\documentclass{article}

\usepackage{arxiv}
\usepackage[utf8]{inputenc} 
\usepackage[T1]{fontenc}    
\usepackage{hyperref}       
\usepackage{url}            
\usepackage{booktabs}       
\usepackage{amsfonts}       
\usepackage{nicefrac}       
\usepackage{microtype}      
\usepackage{lipsum}		
\usepackage{graphicx}
\usepackage{natbib}
\usepackage{doi}
\usepackage{multirow}
\usepackage{booktabs}
\usepackage{makecell} 
\usepackage{array}    
\usepackage{geometry} 
\usepackage{pdflscape}      
\usepackage{graphicx}       
\usepackage{multirow}       
\usepackage{colortbl}       
\usepackage{xcolor}
\usepackage{threeparttable}
\usepackage{tikz}
\usetikzlibrary{arrows.meta, positioning, shapes.geometric, fit, backgrounds}
\usepackage{placeins}
\usepackage{amsmath}

\newcolumntype{C}[1]{>{\centering\arraybackslash}p{#1}}

\newcommand{\yes}{\checkmark}
\newcommand{\no}{\textperiodcentered}

\title{Imputing Missing Data in Longitudinal Life-Logging Athlete Data}

\author{Saaz Kaur Sahdra \\
	Faculty of Computer Science\\
	University of Sydney\\
	Sydney, Australia \\
	\texttt{saaz.sahdra@sydney.edu.au} \\
	\And
	Kalina Yacef \\
	Faculty of Computer Science\\
	University of Sydney\\
	Sydney, Australia \\
	\texttt{kalina.yacef@sydney.edu.au} \\
	\AND
    HaeJin Lee \\
	Faculty of Computer Science\\
	University of Sydney\\
	Sydney, Australia \\
	\texttt{haejin.lee@sydney.edu.au} \\
    \AND
    Judy Kay \\
	Faculty of Computer Science\\
	University of Sydney\\
	Sydney, Australia \\
	\texttt{judy.kay@sydney.edu.au} \\
}

\renewcommand{\shorttitle}{Sahdra et al.}

\hypersetup{
pdftitle={A template for the arxiv style},
pdfsubject={q-bio.NC, q-bio.QM},
pdfauthor={David S.~Hippocampus, Elias D.~Striatum},
pdfkeywords={First keyword, Second keyword, More},
}

\begin{document}
\maketitle

\begin{abstract}
Long-term, diverse personal informatics data enables athletes to understand and act on their health and performance goals, whether assessing readiness right now, tracking how soreness relates to training load over time, or predicting how sleep affects next-day performance. However, the inherent missingness in longitudinal data, if handled inappropriately, can bias these analyses. To our knowledge, no prior work has systematically benchmarked classical and deep learning imputation methods on long-term self-tracking data from athlete populations. We evaluate six imputation methods across four missingness mechanisms at missingness rates spanning 10-80\%. Using seven diverse long-term self-tracking datasets from 231 individuals, we assess both point accuracy and the preservation of temporal and multivariate structure. Our results show that no single method dominates across all conditions and reveal a critical trade-off: while PAIN is superior on point-wise accuracy, deep learning methods (BRITS, GRU-D, and GRIN) are better at preserving multivariate and temporal structure, particularly under higher missingness. Our findings offer a practical, goal-conditioned framework that guides researchers in selecting imputation strategies aligned with both the underlying missingness mechanism and the specific analytical goal of long-term self-tracking studies.
\end{abstract}


\section{Introduction}

\subsection{Motivation}

Self-tracking wearable devices, such as Fitbit or WHOOP, have significantly changed how individuals view and understand their health \cite{pi_review, tangkay2017, karapanos_opportunities_2021, persuasive_reflect}. By building relational models from daily-level self-tracking data, individuals can better understand long-term trends in their health, identify where interventions may help, and evaluate how effective those interventions are. Within this broader landscape, athletes represent a distinct population. Their training, recovery, and injury-prevention goals differ fundamentally from general self-trackers \cite{rapp_self-tracking_2020}, often shifting seasonally or directed by coaches alongside an athlete's own objectives \cite{elvitigala_grand_2024}. They generate unusually rich multi-source data (wearable, self-report, and coach-logged), and their data carries direct, high-stakes value for performance and injury decisions \cite{tenan2023, impute_bball}. This distinctiveness extends to how their data goes missing. A missing day, for instance, may not indicate a lapse in tracking at all, but rather a rest day or a competition day, as some sports, such as soccer, prohibit wrist-worn wearables during games \cite{ifab_law4_2026}. How this missing data is handled can therefore bias resulting estimates of training-load, recovery, or performance trends \cite{tenan2023, impute_bball, imputation_matters}. Yet prior work has not linked individual analytical goals to imputation choice, leaving a gap between how data is handled and what it is ultimately used for.

This gap matters because an athlete's goals take several forms: a point-in-time assessment (e.g., "how ready is my body for a long run today?"), a relational temporal pattern (e.g., "did my soreness track with training load and sleep this week?"), or a predictive insight (e.g., "does consistent 8+ hours of sleep make tomorrow's training feel easier?"). The second and third of these require data spanning a meaningful duration. We define this as long-term monitoring ($\geq$30 daily time points), sufficient to capture weekly and monthly behavioural patterns. Longer collection windows, however, increase the risk of missingness through device non-wear, synchronisation failures, or battery depletion \cite{imputation_matters}.

Missingness is also not uniform across athletes. It can be person- and context-specific \cite{spatio_temporal_imputation}. One athlete may remove their device during high-intensity sessions, another during sleep, another during low-mood periods. Since these patterns are themselves informative about the outcome of interest \cite{idionomics2026}, applying a single imputation method across all athletes risks distorting person-specific inferences. While contextual, group-based models have outperformed one-size-fits-all approaches elsewhere \cite{kunzler2020}, the effect of imputation choice on person-specific, multivariate insight remains underexplored for athletes specifically.

\subsection{Handling Missing Data}

Numerous studies across disciplines have highlighted the importance of addressing missing data, with growing evidence that imputation strategy choice can meaningfully alter downstream conclusions \cite{gomer2023, imputation_matters, tenan2023, imputation_longterm_athlete, tangkay2017, bentley2013, karapanos_opportunities_2021, mi_comparison}. For example, consider estimating the association between sleep duration and next-day resting heart rate from months of wearable data. If non-wear nights disproportionately follow late-night social events or travel, simple list-wise deletion or mean imputation can attenuate or even invert the estimated effect, whereas context-aware deep learning (DL) imputation could potentially recover it by capitalising on missing or self-report data. As more DL imputation methods emerge, researchers are beginning to compare Multiple Imputation with DL approaches under different missingness conditions \cite{imputation_longterm_athlete}. However, these studies have not yet linked individual analytical goals to imputation choice. Since imputation outputs directly feed into personalised feedback and trend interpretation, managing missing data is not merely a pre-processing task but a decision shaping the reliability and interpretability of insights for an individual's goals. Choosing an appropriate method is not straightforward, depending on the nature, prevalence, and recoverability of missingness \cite{beyond_rubin}. We extend this view by proposing that individual analytical goal is another dimension that should inform imputation choice.

\subsection{Research Questions}

Building on the person-specific missingness challenges described above, we ask how imputation methods support analysis of long-term, multivariate wearable data across different missingness mechanisms (MCAR, MAR, MNAR, and Block/Gap, defined in Table~\ref{tab:missingnessmechanisms}) and missingness proportions. We examine not only which method is most accurate, but which best preserves the cross-variable and temporal relationships essential to athlete-monitoring analyses.

\textbf{RQ1: } How accurate are different imputations under different missingness mechanisms and percentages?

\textbf{RQ2: } How do different imputation methods preserve the multivariate relationships and temporal shape of the dataset under different missingness mechanisms and percentages?

\textbf{RQ3: } Which imputation best serves which types of individual analytical goals?

\subsection{Contributions}

Our work makes three contributions. First, a comprehensive benchmark identifying which methods achieve the highest accuracy and best preserve multivariate temporal structure across missingness scenarios. Second, a goal-conditioned decision framework mapping accuracy–structure trade-offs to specific analytical goals (point-in-time, relational, predictive), offering concrete guidance on which metric matters most for a given goal. Third, evaluation across diverse real-world datasets (including WHOOP and Fitbit data), supporting generalisable, cross-device conclusions. This work extends prior research by Choube et al (2025) \cite{imputation_matters}, Tang \& Kay (2017) \cite{tangkay2017}, Künzler et al. (2019) \cite{kunzler2020}, Tenan (2023) \cite{tenan2023}, and Benson et al. (2021) \cite{imputation_longterm_athlete} filling a gap in systematic missing-data benchmarking for athlete populations.

\section{Related Work}

The analysis of missing data and adherence in longitudinal wearable health and behaviour sensing studies is an area of interest within Human-Computer Interaction (HCI) and health disciplines. Missingness in life-logging and wearable datasets can arise from various factors, including but not limited to device non-wear, charging gaps, and user disengagement or re-engagement. Missingness patterns can follow different and complex mechanisms. Despite its ubiquity, the treatment of missing data has been largely overlooked in wearable health literature, with many studies failing to address the implications of their chosen imputation strategy \cite{imputation_matters, simulation_imputation}. 

\subsection{Wearable and Self-Report Data from Athletes}

An important body of HCI research on wearable and life-logging devices has focused on encouraging physical activity in inactive or general populations, or supporting chronic health condition management (e.g., \cite{epstein_mapping_2020, self-tracking-review2023}). Such studies often aim to encourage physical activity through personalised recommendation-driven approaches to increase activity levels or address inactivity-driven health risks. While important, it differs deeply from the potential role of wearables for very active people, such as athletes.

Athletes generate unique data patterns, often characterised by high variability in physiological states and activity levels, and distinctive missingness patterns. For example, a missing day may indicate not just a rest day but a competition day—some as sports, such as soccer, prohibit wrist-worn wearables during games \cite{ifab_law4_2026}. These individuals also require distinct analytical approaches, as their goals, such as optimising performance, preventing injury, or monitoring recovery, differ substantially from those of sedentary populations \cite{rapp_self-tracking_2020}. Furthermore, individual goals may change seasonally or be directed by coaches and sport professionals alongside personal short- and long-term goals \cite{elvitigala_grand_2024}. Sport science routinely applies multivariate statistics and advanced modelling techniques to understand complex relationships between variables such as heart rate, workload, and recovery \cite{exel_precision_2024, bailey_longitudinal_2019}. Yet, missing data in self-tracking datasets can distort these inter-varaible relationships and compromise the validity of resulting insights \cite{tenan2023}.

Handling missing data effectively remains a critical gap in data processing and modelling for athlete populations. Several studies have shown that missing data handling methods can directly impact the accuracy and interpretability of downstream analysis. For instance, Choube et al. (2025) \cite{imputation_matters} demonstrated that imputation technique choice can significantly alter analysis results for heterogeneous wearable data streams, while Benson et al. (2021) \cite{imputation_longterm_athlete} found that, in athlete workload monitoring, imputation methods leveraging multivariate dependencies consistently outperform simpler statistical approaches. Together, these studies highlight the need to understand how imputation strategies perform under the specific conditions generated by athlete populations. Investigating this further could have the potential to inform personalised performance tracking, injury prevention strategies, and recovery monitoring, ultimately contributing to the broader field of wearable health analytics and SportHCI.

\subsection{Characterising Missingness}

HCI has discussed missingness through the lenses of adherence, blocks, gaps, and streaks \cite{tang_adherence2018, tangkay2017, streaks-jochenmeyer}. Other disciplines, including health and psychology, have also used a statistical framework to characterise missingness. Rubin's (1976) \cite{rubin1976} seminal framework on missingness mechanisms categorises three types of missingness. Table \ref{tab:missingnessmechanisms} presents these, MCAR, MAR and MNAR, in its first three rows, with definitions (Col 2) and examples (Col 3). Each mechanism has different statistical approaches for appropriate handling and imputation to avoid bias in downstream analysis. Table \ref{tab:missingnessmechanisms} also summarises their significance and methods for handling them.

Rubin's missingness mechanisms remain a cornerstone for understanding missingness in modern research. Interestingly, real-world multivariate data often exhibit mixtures of these mechanisms, which complicates efforts to apply standard imputation methods \cite{beyond_rubin}. Many studies within and outside of HCI have identified the challenges with incomplete data and adherence in long-term life-logging and wearable tracking data (see \cite{tang_adherence2018, tangkay2017, streaks-jochenmeyer, imputation_longterm_athlete, impute_bball,imputation_longterm_athlete}). Breaks/gaps, both micro and macro, seem inevitable and have diverse patterns across individuals, making it difficult to model for groups and populations \cite{tangkay2017, karapanos_opportunities_2021}. Nevertheless, the presence of missing data itself can also contribute to a user's reflection on behavioural patterns and the system's ability to monitor adherence and provide adaptive feedback \cite{karapanos_opportunities_2021, tangkay2017, bentley2013}. This dual role, where missingness is both an analytical obstacle and a meaningful behavioural signal, underscores the need for methods that can accommodate rather than `correct' for diverse, individual-specific patterns of missingness. 

There is increasing agreement among scholars that long-term analysis results should be contextualised and personalised to individuals and groups for meaningful insights. Contextualised models account for user receptivity and context, outperforming generalised models in mobile health interventions \cite{tangkay2017, kunzler2020}. Tenan (2023) \cite{tenan2023} and Sharma et. al. (2026) \cite{impute_bball} both demonstrated how datasets in sport science commonly deviate from the MCAR assumption, with missingness patterns linked to athlete fatigue, non-wear during recovery, or poor performance days. Therefore, handling missingness within heterogeneous data requires careful consideration of missingness mechanisms, amount, and goal of analysis. Similarly, Gomer and Yuan (2025) \cite{gomer2023} highlighted that a mixture of mechanisms within a single dataset could significantly distort statistical inferences, particularly under large proportions of MNAR values. Like Choube et al. (2025) \cite{imputation_matters}, we emphasise the need for explicit testing and reporting of missingness mechanisms in wearable and life-logging studies, and advocate for participant-level imputation evaluation. This study therefore aims to develop a robust pipeline of imputation strategies tailored to the specific missingness patterns inherent in wearable and self-tracking datasets. 

Epstein et al. (2021) \cite{karapanos_opportunities_2021} discuss the challenges of long-term tracking including changing devices due to time, changing goals, emotional bonds, and the inevitability of gaps or breaks. Researchers need to develop tools and processes to deal with heterogeneous data from disparate sources with missingness, because the gaps in itself could tell a story on their own \cite{karapanos_opportunities_2021}. We agree with Epstein et al. (2021) that gaps should be considered a normal part of the data. However, to the best of our knowledge, we have not seen a fully developed pipeline that equips researchers with a framework to handle gaps and missingness of different mechanisms. Since missingness, such as gaps, can be diverse between users, a generalised model may not adequately capture this person-specific heterogeneity. Building on Xu et al. (2021) \cite{chikersal2021}, in this study we build the foundation for personalised behaviour modelling for within-person and sample-specific approaches, as we believe this method will best characterise individual patterns and useful insights. 

\begin{table}[h]
    \centering
    \caption{Missingness Mechanisms: This table presents the type and definition of each mechanism as described in the literature, operationalises it with an example, and connects it to its significance in PI/HCI. Based on imputation methods and studies identified in our literature review, we also indicate which handling approach is theoretically best suited to each mechanism. However, these methods and mechanisms have not been systematically benchmarked together on athlete.}
    \label{tab:missingnessmechanisms}
    \small
    \begin{tabular}{p{2.7cm} p{3.2cm} p{3.0cm} p{3.0cm} p{3.1cm}}
        \toprule
        Missingness Mechanism & Definition & Example & Significance to PI/HCI & Applicable Handling Methods \\
        \midrule
        MCAR (Missing Completely at Random) & Probability of missingness is unrelated to any observed or unobserved data. & A sensor randomly fails to record data due to a transient hardware glitch, independent of user behavior. & Simplest case; analysis remains unbiased, but reduced sample size may affect power and generalizability. & Listwise deletion, pairwise deletion, mean imputation, multiple imputation, maximum likelihood. \\
        \addlinespace
        MAR (Missing at Random) & Probability of missingness depends only on observed variables, not on the missing values themselves. & Users with lower engagement scores are less likely to answer follow-up questions, but engagement scores are recorded. & Common in PI/HCI logs; must model missingness using observed covariates to avoid bias. & Multiple imputation with covariates, inverse probability weighting, maximum likelihood methods. \\
        \addlinespace
        MNAR (Missing Not at Random) & Missingness depends on unobserved values, including the missing value itself. & Users with very high stress are least likely to complete self-report surveys about stress. & High risk of biased conclusions about sensitive or extreme states; directly impacts design decisions and user modeling. & Sensitivity analyses, selection models, pattern-mixture models, explicit modeling of missingness mechanism. Deep Learning methods such as BRITS.\\
        \addlinespace
        Mechanism Unknown / Mixed & Underlying missingness mechanism is unclear or combines MCAR, MAR, and MNAR. & Complex longitudinal app data with device issues, optional questions, and avoidance of sensitive prompts. & Typical in real-world PI/HCI deployments; requires careful diagnostics and robustness checks. & Exploratory diagnostics, multiple imputation under MAR, scenario-based sensitivity analysis. Deep learning methods such as BRITS and PAIN. \\
 \addlinespace
        Block/Gap Pattern Missingness& Can be either MAR or MNAR missingness but for longer periods of time (i.e., days).& User decides to take off wearable due to high stress (observed or unobserved) for a week.& Typical in real-world PI/HCI where users abandon and re-engage with wearables.& Deep learning methods such as BRITS and PAIN. \\
    \end{tabular}
\end{table}

\subsection{Imputation Strategies}

Multivariate analysis often requires complete-case datasets (i.e., datasets with no missing data). For example, standard linear regression cannot natively handle missing data and will fail or crash if a dataset contains NA values. This missingness must be handled before running the model to avoid biased results or loss of power. Common ways of handling such missingness are list-wise deletion, mean imputation, or advanced imputation methods \cite{imputation_matters}. The type of missingness mechanism can help us determine which handling method is most appropriate (see Table \ref{tab:missingnessmechanisms}).

When analysing in-the-wild studies or real-world datasets with missingness, traditional "easy" methods of handling missing data (i.e., list-wise deletion or mean imputation) can be inappropriate for the complex temporal and multivariate nature of the data. In recent years, researchers have explored both statistical and machine learning based approaches to imputation. For example, Benson et. al. (2021) \cite{imputation_longterm_athlete} evaluated multiple imputation techniques in the context of athlete workload monitoring, finding that machine learning models outperformed regression-based methods by leveraging the inherent relationships between variables. Sharma et. al. (2026) \cite{impute_bball} propose a layered imputation pipeline to ensure that athlete-specific inter-individual variability is captured and that latent constructs are considered. Their method also outperformed traditional models such as Multiple Imputation (MI). 

More DL methods are emerging as promising alternatives for handling missing data in multivariate datasets. Some models leverage temporal dependencies in time series data, while others adapt to incorporate temporal context during imputation \cite{spatio_temporal_imputation}. More recently, graph-based methods have integrated network structures into the imputation process, enabling the preservation of both temporal and inter-variable relationships \cite{GAIN2018, gain2024}. These structure-aware approaches are particularly relevant for wearable datasets, where physiological signals often exhibit strong correlations across time and variables. This has the potential to aid users in understanding their multivariate and temporal data.

As previously mentioned, Choube et al. (2025) \cite{imputation_matters} emphasised the importance of systematically evaluating imputation strategies, particularly for datasets with heterogeneous and interdependent streams of data, as found in mobile and wearable sensing studies. They proposed selecting imputation strategies based on the underlying data structure and the missingness mechanism, which align closely with the goals of this study. 

Recent systematic reviews have compared several of these imputation methods in health data contexts and found that deep learning methods generally outperform statistical approaches for complex, longitudinal health data, though issues with interpretability and fairness remain \cite{liu2023}. These reviews find that MICE is most widely used \cite{ren2024} and MICE, BRITS, GRU-D, and GAIN are well-studied in health data \cite{kazijevs2023, liu2023}. PAIN and GRIN seem promising deep learning methods but have seen limited application in individual health and wearable datasets. Although deep learning methods are recommended, no single method is universally optimal, and many methods like PAIN and GRIN remain under-explored and under-utilised in personal health data \cite{kazijevs2023, ren2024}. Collectively, these systematic reviews \cite{kazijevs2023, liu2023, ren2024} identify a need for more benchmarking of advanced methods in individual health contexts. This gap, according to the available literature, has not yet been addressed in conjunction with analytical goals for athlete populations. 

Building on this gap, we selected six different imputation methods used on time series data (excluding mean imputation) that can handle mixed data types: Multiple Imputation by Chained Equations (MICE), Bidirectional Recurrent Imputation for Time Series (BRITS), Gated Recurrent Unit with Decay (GRU-D), Generative Adversarial Imputation Network Wasserstein VAE hybrid (WGAIN-VAE), Precision Adaptive Imputation Network (PAIN), and Graphical Recurrent Imputation Network (GRIN). This selection spans traditional statistical and deep learning approaches, and deliberately includes under-explored methods (i.e., PAIN and GRIN) alongside well-established ones (i.e., MICE), directly addressing the benchmarking gap identified above. Each method is further defined under Section~\ref{sec:impmethods} Methods.

\subsection{Scope and Hypotheses}

To the best of our knowledge, no study has evaluated different imputation methods systematically by considering not only the missingness mechanism, but also the user analytical goal for downstream analysis. This paper directly addresses this gap by evaluating 6 imputation methods across 4 missingness mechanisms at missingness rates spanning 10-80\%, using 7 diverse long-term self-tracking datasets comprising data from 231 individuals. Such variety in datasets can provide cross-device insights. However, validation of the measurements from the devices used for data collection in these public datasets is out of scope for this study. The data we use are also limited to daily level aggregation. 

We focus on evaluating the imputation methods on the basis of point accuracy and their ability to preserve temporal and multivariate structure of the data. These metrics allow us to map which type of imputation is optimal for a user's analytical goal. We anticipated that each dataset will have users with different missingness characteristics. We also hypothesise that deep learning methods compared to MI will perform better across MAR, MNAR, and Blocks/Gaps conditions across all missingness percentages (H1 for RQ1 and RQ2). We expected the quality of imputation to degrade at 40\%-50\% missingness or higher, based on previous literature \cite{40missing, mice50missing}. Finally, we expected that no one imputation method would give best across point accuracy and preserve temporal and multivariate structure of the data (H2 for RQ3). Rather, we aim to build a decision framework that will guide researchers in determining the best path for their goals and datasets.

\section{Data and Study Design}

\subsection{Operationalising Long-Term}

Our study is based on analyses of at least 4 weeks of data. This threshold is not arbitrary relative to the three analytical goals introduced. A point-in-time assessment (e.g., "how ready is my body for a long run today?") requires only a single day's data and is unaffected by our threshold. A relational temporal pattern (e.g., "did my soreness track with training load and sleep this week?") requires at least one full weekly cycle to be observable. A predictive insight (e.g., "Does consistent 8+ hours of sleep make tomorrow's training feel easier?") requires repeated exposure to the pattern being tested, which a single week cannot provide, but a month can begin to establish. One month is therefore the shortest window in which all three goal types, including the two that depend on duration, can be meaningfully evaluated, while remaining short enough that non-wear, device failure, and athlete dropout remain comparatively rare compared with longer windows. We treat this as a practical lower bound for this benchmark rather than a claim that one month is sufficient for all athlete-monitoring purposes. Goals tied to slower physiological adaptations (e.g., detecting a sustained change in resting heart rate or cardiovascular fitness) are known to require longer observation windows than the relational and predictive goals we benchmark here, and we return to this as a direction for future work in Section~\ref{sec:limitations}.

\subsection{Datasets}

Using 7 diverse long-term self-tracking datasets comprising data from 231 individuals, we evaluated 6 imputation methods across 4 missingness mechanisms at missingness rates spanning 10–80\% (see Appendix~\ref{app:datasets} for more information on which athletes are included in the analysis). The raw public datasets were each cleaned and consecutive days were forced. After cleaning, for each user in each dataset, the longest available complete-case (or near complete-case) subset was extracted. These complete or near-complete subsets are treated as ground-truth, as is common practice in the literature \cite{liu2023, ren2024}. Using the complete subsets, we simulated missingness under 4 mechanisms at 8 different proportions. Finally, a suite of imputation methods were applied to these simulated missingness datasets.

\subsubsection{Ethics Statement}

Six of the seven datasets are public, obtained from Kaggle \cite{Fitbit_public, athlete_public_dataset, WHOOP_public} and research repositories (PMData \cite{PMData}, Human Rhythms Dataset \cite{Human-Rhythms-Dataset}, SoccerMon \cite{soccermon}), released under terms permitting research reuse. As secondary analysis of de-identified public data, this does not require IRB/Ethics review. We rely on each dataset's documented consent and licensing terms and did not independently verify original consent procedures. The seventh dataset is the first author's personal WHOOP data, as self-experimentation with no other data subjects, formal IRB/Ethics review was not sought.

\subsubsection{Extracting complete-case subsets}

We computed each user's complete subset by first enforcing consecutive days and then identifying the longest continuous observation window with minimal missingness ($\leq$5\%) across all target variables. For larger datasets with longer timelines, a greater than 60-day window was applied. Users with less than 60 days were excluded. For shorter few-person datasets, shorter timelines were accepted to provide edge cases. Up to 5\% residual missingness with gaps of at most one day is tolerated and type-aware imputed (numeric: median; logical: modal; categorical: mode). We acknowledge that this windowing choice, while methodologically necessary, may limit generalisation to the broader athlete population. Full extraction parameters are detailed in Appendix~\ref{app:datasets}.

\subsubsection{Dataset Summaries}
\label{sec:dataset-summaries}

Table \ref{tab:combined-span-vars} summarises the span of days and target variable counts per dataset. 
Table \ref{tab:vars} summarises the metrics available in each dataset's complete subset. Full narrative descriptions, collection context, and device/sensor details for each dataset are provided in Appendix Appendix~\ref{app:datasets}. In brief:

\begin{itemize}
    \item \textbf{Human Rhythms \cite{Human-Rhythms-Dataset}:} 166 college students (including 1 NCAA athlete) tracked via Fitbit and daily surveys over 17 months, focused on circadian/ultradian rhythms. Complete subset retained 146 users.
    \item \textbf{SoccerMon \cite{soccermon}:} 77 elite Norwegian women's football players, combining PmSys \cite{pmsys} self-report wellness/load data with GPS/sensor data over 2 seasons. Complete subset retained 45 users.
    \item \textbf{Fitbit \cite{Fitbit_public}:} 33 consumer Fitbit users (23 with sleep data) over 31 days. This dataset is an edge case under our 30-day threshold, included to test generalisability across shorter timelines and consumer-grade devices. Complete subset retained 23 users.
    \item \textbf{PMData \cite{PMData}:} 16 users over 5 months, integrating Fitbit, a sports-logging app, and Google Forms. Complete subset retained 14 users.
    \item \textbf{Personal WHOOP:} First author's own 2-year WHOOP export (700+ days, 500+ observed), collected via the WHOOP Developer API v2 \cite{WHOOP_api}. Complete subset retained 181 days.
    \item \textbf{Public WHOOP \cite{WHOOP_public}:} Single-user Kaggle WHOOP export with user-annotated context, spanning 51 days. An edge case under our 60-day threshold.
    \item \textbf{Athlete \cite{athlete_public_dataset}:} Single NFL player's self-published multi-device data (WHOOP, Polar, Oura, MyFitnessPal) over 94 days.
\end{itemize}

Tables \ref{tab:combined-span-vars} and \ref{tab:vars} 
show how these data sets are heterogeneous in terms of size, populations, and the nature of the data collected.

\begin{table}[htbp]
\centering
\caption{Span of days and number of target variables for each dataset across users.}
\label{tab:combined-span-vars}
\begin{tabular}{lr|rrrr|rr}
\toprule
& \multicolumn{1}{c}{} & \multicolumn{4}{c}{\textbf{Span of Days (Complete Subset)}} & \multicolumn{2}{c}{\textbf{Target Variables}} \\
\cmidrule(lr){3-6} \cmidrule(lr){7-8}
Dataset & $n$ complete & Min & Max & Median & Mean & Raw & Complete \\
\midrule
Human Rhythms  & 146 & 55 & 251 & 105 & 127.10 & 20 & 13 \\
Soccermon      & 45  & 80 & 360 & 294 & 257.07 & 24 & 26 \\
Fitbit         & 23  & 21 & 31  & 31  & 29.43  & 31 & 14 \\
PMData         & 14  & 63 & 152 & 131 & 122.36 & 63 & 83 \\
Personal WHOOP & 1   & --- & 181 & --- & ---   & 70 & 24 \\
Public WHOOP   & 1   & --- & 51  & --- & ---   & 26 & 25 \\
Athlete        & 1   & --- & 94  & --- & ---   & 43 & 40 \\
\bottomrule
\end{tabular}
\begin{flushleft}
\footnotesize
\textit{Note.} Span is measured in days across users within each dataset. For single-user datasets ($n=1$), only the observed span is reported; minimum, median, and mean are undefined (denoted ---). Target variables are counted removing all id columns, admin/metadata columns, and columns created during cleaning from the original dataset. As some columns were created during cleaning from the original variable, the complete subset has more columns than the raw for some datasets.
\end{flushleft}
\end{table}

\begin{table}[htbp]
\centering
\caption{Types of metrics in the complete subset across datasets}
\label{tab:vars}

\footnotesize
\textit{Legend:} PW = Personal WHOOP; UW = Public WHOOP; AT = Public Athlete; 
FB = Public Fitbit; PM = Public PMData;\newline SM = Public SoccerMon; HM = Public Human Rhythms.
\vspace{4pt}

\begin{tabular}{lrrrrrrr}
\toprule
Metric in Complete Subset & PW & UW & AT & FB & PM & SM & HM \\
\midrule
Sleep Duration                                   & \yes & \yes & \yes & \yes & \yes & \yes & \yes \\
Sleep Stages                                     & \yes & \yes & \yes & \no  & \yes & \no  & \no  \\
In-bed Duration                                  & \no  & \yes & \yes & \yes & \yes & \no  & \yes \\
Blood Oxygen                                     & \yes & \yes & \no  & \no  & \no  & \no  & \no  \\
Skin Temp                                        & \yes & \yes & \no  & \no  & \no  & \no  & \no  \\
HRV                                              & \yes & \yes & \yes & \no  & \no  & \no  & \no  \\
RHR                                              & \yes & \yes & \yes & \no  & \yes & \no  & \no  \\
Travel                                           & \yes & \no  & \no  & \no  & \no  & \no  & \no  \\
Respiratory Rate                                 & \no  & \yes & \yes & \no  & \no  & \no  & \no  \\
Number of Workouts                               & \yes & \no  & \no  & \no  & \no  & \no  & \no  \\
Heart Rate Zones Time                            & \yes & \no  & \no  & \no  & \yes & \no  & \no  \\
Max HR                                           & \no  & \yes & \yes & \no  & \yes & \yes & \no  \\
Min HR                                           & \no  & \no  & \no  & \no  & \yes & \no  & \no  \\
Average HR                                       & \no  & \yes & \yes & \no  & \yes & \no  & \no  \\
SD HR                                            & \no  & \no  & \no  & \no  & \yes & \no  & \no  \\
Activity Type                                    & \yes & \no  & \no  & \no  & \yes & \no  & \no  \\
Activity Distance                                & \no  & \no  & \no  & \yes & \yes & \no  & \no  \\
Activity Time                                    & \no  & \no  & \no  & \yes & \yes & \no  & \no  \\
Activity Calories                                & \no  & \no  & \no  & \no  & \yes & \no  & \no  \\
Activity HR                                      & \no  & \no  & \no  & \no  & \yes & \no  & \no  \\
Team Activity \& Type                            & \no  & \no  & \no  & \no  & \yes & \no  & \no  \\
Individual/Team Performance                      & \no  & \no  & \no  & \no  & \no  & \yes & \no  \\
Training Load                                    & \no  & \no  & \no  & \no  & \no  & \yes & \no  \\
Speed                                            & \no  & \no  & \no  & \no  & \no  & \yes & \no  \\
Steps                                            & \no  & \no  & \no  & \yes & \yes & \no  & \yes \\
Injury                                           & \no  & \no  & \no  & \no  & \yes & \yes & \no  \\
Illness                                          & \no  & \no  & \no  & \no  & \no  & \yes & \no  \\
Weight                                           & \no  & \no  & \yes & \yes & \yes & \no  & \no  \\
RPE                                              & \no  & \no  & \yes & \no  & \yes & \no  & \no  \\
Scripted Intensity                               & \no  & \no  & \yes & \no  & \no  & \no  & \no  \\
Calories in/out                                  & \no  & \no  & \yes & \emph{out} & \emph{out} & \no  & \no  \\
Macros/Micros Nutrition                          & \no  & \no  & \yes & \no  & \no  & \no  & \no  \\
Water Intake                                     & \no  & \no  & \yes & \no  & \yes & \no  & \no  \\
Self-report feeling (i.e., recovery, mood, etc.) & \no  & \yes & \yes & \no  & \yes & \yes & \yes \\
Recovery activity                                & \no  & \no  & \yes & \no  & \no  & \no  & \no  \\
\bottomrule
\end{tabular}
\end{table}

\subsection{Naturally Occurring Missingness Mechanisms in Datasets}
\label{sec:naturalmiss}

To characterise missingness in our longitudinal datasets, we implement a staged diagnostic pipeline operationalising the MCAR/MAR/MNAR taxonomy plus a Block/Gap pattern, while remaining conservative about mechanism identifiability. The pipeline sequentially tests for MCAR (pattern-wise independence between missingness and variable values), MAR (whether missingness can be predicted from other observed variables), and MNAR (via run-length and tail-clustering diagnostics testing whether missingness depends on the unobserved value itself), escalating through these tests via a priority ladder to assign a single verdict per user-variable combination. 
Appendix~\ref{app:mechanismdetection} has full statistical procedures, thresholds, and correction methods.

Crucially, these mechanisms cannot be identified from data alone, as distinguishing MAR from MNAR requires knowledge of why values are missing. We therefore use "Undetermined" liberally rather than over-claiming a mechanism the data cannot adjudicate.

\subsection{Simulated Missingness}
\label{sec:amputation}

We construct a controlled benchmark in which the true missingness mechanism is known by design, following the formal framework of Rubin (1976) \cite{rubin1976} and the HCI-defined Block/Gap pattern. We deliberately impose known missingness onto each user's complete subset, a process commonly termed "amputation," the standard approach for missing data simulation studies \cite{gomer2023, imputation_matters, liu2023, ren2024}. Each complete subset per user serves as the ground-truth reference against which all subsequent amputations and downstream imputations are evaluated. We inject NA values into each per-user complete subset under four distinct mechanisms (MCAR, MAR, MNAR, Block/Gaps) at eight severity levels (10\%–80\% missingness proportions). We acknowledge that these amputations are not an exhaustive list of how missingness can manifest in athletes' real-life data, and discuss this as a limitation in Section~\ref{sec:limitations}.

MCAR masks cells uniformly at random. MAR and MNAR condition missingness on observed covariates (weekend flags, auxiliary high-value indicators) and on each variable's own extreme values, respectively, so that missingness probability depends on observed data (MAR) or on the value itself (MNAR). Block/Gap imposes contiguous per-user gaps of scaling size, simulating realistic device non-wear periods. All amputations use fixed seeds for reproducibility, and the realised NA rate (overall and per-variable) is recorded alongside each amputated file. Full weighting schemes, seed values, and block-placement algorithms are provided in Appendix~\ref{app:simulatedconstruct}.

\section{Imputation Methods}
\label{sec:impmethods}

Six imputation methods are run against every amputated file: 

\begin{enumerate}
    \item \textbf{MICE: } Multiple Imputation by Chained Equations. \cite{mice_how}
    \item \textbf{BRITS: } Bidirectional Recurrent Imputation for Time Series. \cite{BRITS2018}
    \item \textbf{GRU-D: } A gated recurrent unit with decay mechanisms for irregular series. \cite{gru_d}
    \item \textbf{WGAIN-VAE: } Generative Adversarial Imputation Network Wasserstein VAE hybrid. \cite{GAIN2018, gain2024}
    \item \textbf{PAIN: } Precision Adaptive Imputation Network. \cite{PAIN2025}
    \item \textbf{GRIN: } Graphical Recurrent Imputation Network. \cite{GRIN2022}
\end{enumerate}

This selection spans the methodological spectrum from classical statistical MI to recurrent, adversarial, and graph-based DL, allowing us to evaluate how each fares against mechanisms with known ground truth.

Every method imputes one user at a time, with parallelism occurring across users (N\_CORES=16) and never within a user. For each amputated file, data are split by user, date-sorted, and dispatched to a worker. MICE runs in pure R, while the deep learning (DL) workers are called through Python. The full pipeline runs twice, governed by two window strategies: "full," which imputes each user's entire date-sorted sequence at once (requiring $\geq$10 rows), and "win30," which partitions each user into tumbling 30-day windows, imputes each independently, and re-binds results (users with <30 rows are skipped). The win30 strategy serves as a sensitivity check against non-stationarity, where a user's per-chunk results aggregate into an overall "ok," "partial," or "failed" status. Notably, there is no mean/median fallback. Failed imputations are skipped and logged rather than crudely filled, preserving ground-truth comparison integrity. Floors are enforced before any model runs (minimum 10 user rows for MICE/DL, minimum 3 observed values per column), and chunks with no missing cells are returned unchanged. Full implementation details (hyperparameters, architecture specifics, and shared training infrastructure) are provided in Appendix~\ref{app:imputationdetails}.

\textbf{MICE} fills missing values by repeatedly predicting each incomplete variable from the others, producing several "complete" dataset versions that capture estimation uncertainty. It performs best under MAR, where the probability of missingness can be explained by observed variables, but does not natively account for time, leaving temporal patterns unmodelled, precisely the assumption this study examines.

\textbf{BRITS} uses a neural network that reads sequences both forwards and backwards in time, reconstructing missing values from patterns at both earlier and later time points \cite{BRITS2018}. It should perform well under MCAR, MAR, and block missingness, and has also been applied to MNAR settings and to healthcare/physical-activity data, suggesting suitability for wearable data, though at higher computational cost and interpretability trade-offs than MICE.

\textbf{GRU-D} extends a standard recurrent network with built-in decay mechanisms modelling how the influence of past observations fades as time gaps grow, explicitly treating missingness patterns as meaningful rather than fixing decay rates in advance \cite{gru_d}.

\textbf{WGAIN-VAE} builds on GAIN's adversarial framework, in which a generator fills missing values while a discriminator tries to detect which values were originally missing, pushed toward realism via a "hint" mechanism \cite{GAIN2018, gain2024}. The original GAIN formulation outperforms MICE under MNAR conditions but is not designed for temporal modelling. The hybrid combines a Wasserstein-GAN with a variational autoencoder (VAE) to add temporal and distributional stability.

\textbf{PAIN} adapts dynamically to different data types, distributions, and missingness patterns via a three-step process: combining statistical methods, random forests, and autoencoders; adaptively weighting these based on data characteristics; and refining imputed values to reduce noise \cite{PAIN2025}. It supports mixed data types and shows promise for health data, though it remains comparatively under-explored.

\textbf{GRIN} combines a recurrent network for temporal patterns with a graph neural network for variable relationships, learning representations via bidirectional message passing \cite{GRIN2022}. It makes no assumptions about the distribution of missing values but assumes underlying patterns remain stable over time, narrowing its effective use toward MAR scenarios.

All methods handle mixed data types via consistent pre-processing (logical/categorical variables coerced to factors or numeric), and the three recurrent models (BRITS, GRU-D, GRIN) share common training infrastructure (gradient clipping, time-limited early stopping). A shared post-processing routine standardises numeric values before modelling and constrains results within observed-data ranges afterward. Together, these conventions ensure all six methods are evaluated under a uniform, reproducible protocol, isolating differences in imputation quality to the methods themselves. Full architectural and hyperparameter detail for each method is provided in Appendix~\ref{app:imputationdetails}.

\section{Evaluation Methods}

Our benchmark imposes missingness on complete subsets under known mechanisms, so every imputed value can be scored against ground truth. All accuracy metrics are computed on truly masked cells. The evaluation is organised around our research questions: point-level reconstruction accuracy, preservation of multivariate (between-variable) structure, preservation of temporal (within-variable) structure, and preservation of distributional shape.

A methodological note applies throughout: since standard metrics are typically averaged only over successfully evaluated users, methods that silently fail on harder users (e.g., shorter series, sparse columns) can post a deceptively low error. We therefore introduce a coverage-penalised family of metrics (prefixed cp-) that deflate each raw metric by the fraction of attempted users actually evaluated, ensuring low error on a subset is never rewarded over consistent accuracy across the full sample. At full coverage, each coverage-penalised metric equals its raw counterpart. Pooled (cross-dataset) versions use sample-size weighting so larger datasets and those with higher coverage contribute proportionally more. Full derivations for all metrics are provided in Appendix~\ref{app:metricevaluation}.

\subsection{Point-level reconstruction accuracy}

For numeric and ordinal targets, we report normalised RMSE and MAE (NRMSE, NMAE) as primary point-accuracy measures, enabling comparison across variables and datasets of different scales (Appendix~\ref{app:point-accuracy_eq}, Eq.~\eqref{eq:nmae}). Bias is recorded as the mean signed residual to distinguish noisy-but-centred methods from those that systematically over- or under-impute, a failure mode RMSE alone conceals. These three measures (cpNRMSE, cpNMAE, cpNBias) are each coverage-penalised following the scheme above (Appendix~\ref{app:point-accuracy_eq}, Eqs.~\eqref{eq:cpnrmse-pool}–\eqref{eq:contribution}).

\subsection{Preserving multivariate (between-variable) structure}

Low point-error does not guarantee inter-variable relationships are preserved. A method could minimise RMSE while flattening covariance structure, which matters particularly for network-model analyses. We compute two diagnostics. First, cpCorr is the Pearson correlation between imputed and true values on masked cells, capturing whether rank and shape are preserved even when absolute level is off. Since correlation is undefined for constant masked vectors, we track a correlation-specific coverage term alongside standard coverage (Appendix~\ref{app:multivariate_eq}, Eqs.~\eqref{eq:corrdefine}–\eqref{eq:cpcorrpool}). Second, cpFrob and cpPartialFrob compare full and partial correlation matrices (across all numeric/ordinal targets jointly) using the Frobenius norm of their difference, with partial correlation isolating direct pairwise relationships after conditioning on all other variables (Appendix~\ref{app:multivariate_eq}, Eqs.~\eqref{eq:corr-frob}–\eqref{eq:cpfrob}). Because these matrix metrics are computed over each user's full series, they are only interpretable at higher missingness levels (50–80\%), where a meaningful share of the matrix is genuinely reconstructed rather than copied through unchanged.

\subsection{Preserving temporal (within-variable, over-time) structure}

A method can score well on point accuracy while destroying persistence, dynamics, or rhythm. We assess three full-series temporal diagnostics, computed per user~$\times$~variable on the complete reconstructed series, since autocorrelation, differencing, and spectral content are not well-defined on a scattered subset of timepoints. For all three, lower values indicate better preservation, with 0 denoting perfect fidelity:

\begin{itemize}
    \item \textit{Autocorrelation error}: mean absolute difference in lag-1–3 autocorrelation between true and reconstructed series, testing whether persistence (e.g., a high day following a high day) is preserved (Appendix~\ref{app:temporal_eq}, Eq.~\eqref{eq:acf}).
    \item \textit{Range-normalised first-difference RMSE}: RMSE between first-differenced true and reconstructed series, testing whether the magnitude of day-to-day swings is preserved rather than smoothed away (Appendix~\ref{app:temporal_eq}, Eqs.~\eqref{eq:firstdiff-raw}–\eqref{eq:firstdiff-norm}).
    \item \textit{Spectral difference}: total absolute difference between normalised power spectral densities, testing whether periodic (e.g., weekly) structure is preserved (Appendix~\ref{app:temporal_eq}, Eq.~\eqref{eq:spectral}).
\end{itemize}

We additionally compare performance between full-series and 30-day windowed (win30) imputation strategies. Stability across the two would indicate robust temporal handling, while degradation under win30 could indicate reliance on regime-specific context. All temporal metrics use the same coverage-weighted pooling scheme as above (Appendix~\ref{app:temporal_eq}).

\subsection{Preserving distributional shape}

Unlike the temporal metrics, which are order-dependent and diagnostic only at higher missingness, distributional metrics are order-agnostic and masked-cell-only, remaining diagnostic at every missingness level. This matters because a method can preserve marginal distribution while scrambling temporal order, or vice versa. Both families together are needed to separate these failure modes.

\begin{itemize}
    \item \textit{Variance ratio}: the ratio of imputed to true variance over masked cells, directly exposing the common "regress-to-the-mean" failure that NRMSE cannot detect; a value of 1 indicates perfectly reproduced spread, below 1 indicates variance collapse, and above 1 indicates inflated variability (Appendix~\ref{app:distributional_eq}).
    \item \textit{Kolmogorov-Smirnov (KS) statistic}: the maximum gap between true and imputed empirical CDFs, capturing whole-distribution mismatch (Appendix~\ref{app:distributional_eq}).
    \item \textit{1-Wasserstein distance}: the full displacement between true and imputed distributions, complementing KS by capturing mismatches spread across the full range rather than concentrated at a single point (Appendix~\ref{app:distributional_eq}).
\end{itemize}

KS and Wasserstein distance are likewise coverage-penalised (cpKS, cpWasserstein1) and pooled across datasets using the same scheme (Appendix~\ref{app:distributional_eq}, Eqs.~\eqref{eq:cpks}–\eqref{eq:cpwasserstein1}).

\section{Results}

This section first reports the results for accuracy (RQ1) then, in terms of preservation of multivariate structure, temporal structure and distributional shape (RQ2). Finally, it compares
point accuracy with temporal/distributional structure and shape (RQ3). This provides a foundation for the discussion in Section~\ref{sec:discussion} which summarises the key ideas in a flow diagram (Figure~\ref{fig:decision-flow}) for selecting suitable imputations, depending on the goal and data properties.

\subsection{RQ1: Imputation Accuracy}

\subsubsection{Fail Counts}

Each user (n=231) was run per method 32 times (4 mechanisms × 8 missingness proportions), giving 7,392 total runs per method. MICE failed for 7,144 runs, versus only 384 total failures across all other methods combined (Appendix~\ref{app:failcounts}, Figure~\ref{fig:failcountplot}). This imbalance is why the coverage-penalised metrics below are necessary. A method's apparent accuracy on its successfully-imputed users could otherwise misrepresent its reliability across the full sample. Reasons for MICE's large failure rate are discussed in Section~\ref{sec:mice}.

\subsubsection{cpNRMSE}
\label{sec:cpnrmse}

\begin{figure}[htbp]
    \centering
    \includegraphics[width=1\textwidth]{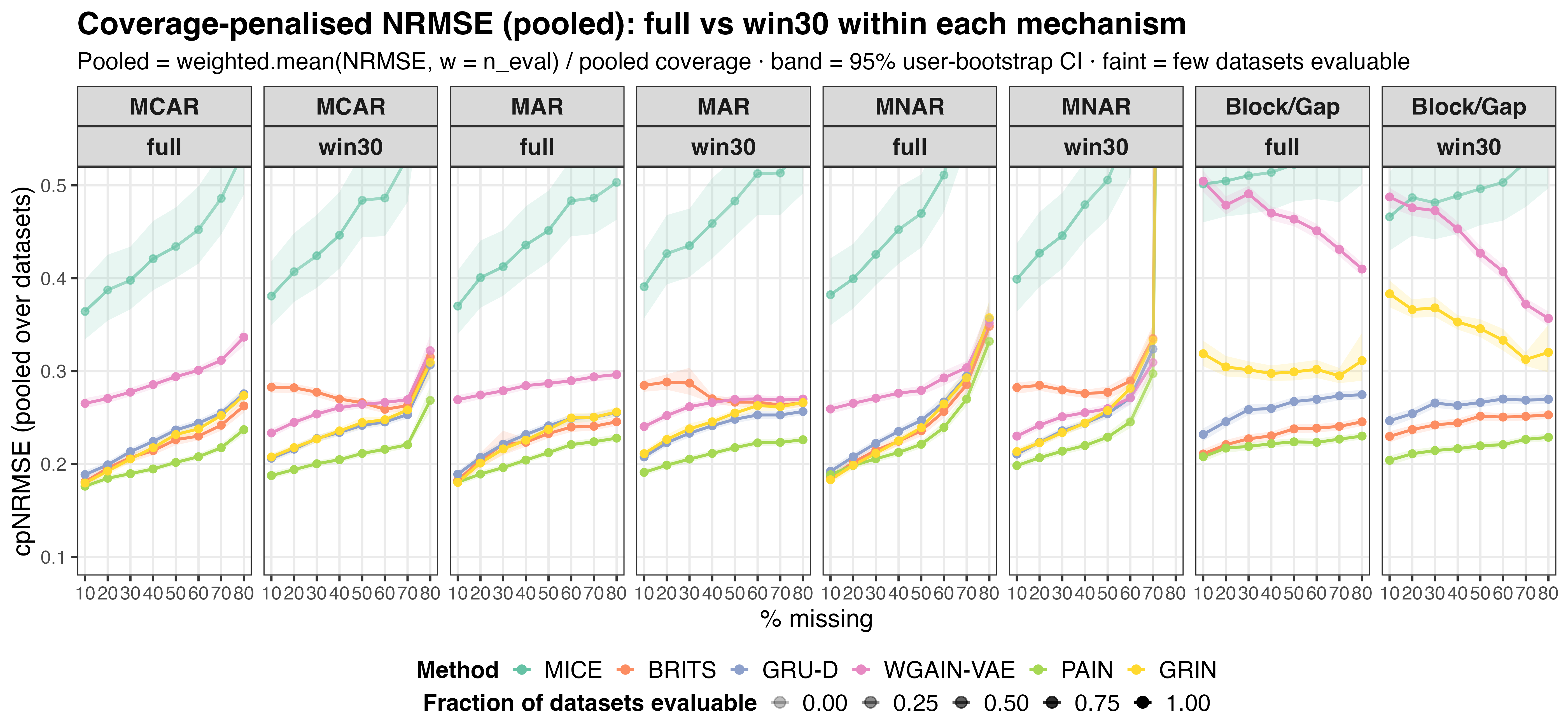} 
    \caption{cpNRMSE pooled across datasets (n\_users\_attempted=231), with 95\% user-bootstrap confidence bands, across missingness mechanisms and window strategies.}
    \label{fig:cpNRMSEzoompooled}
\end{figure}

Figure~\ref{fig:cpNRMSEzoompooled} summarises cpNRMSE pooled across datasets, by mechanism and window strategy. Per-dataset trends are in Appendix~\ref{app:cpnrmse-supplementary} (Figure~\ref{fig:cpNRMSEzoom}). MICE exhibits higher cpNRMSE than all DL methods with wider confidence bands, indicating higher and less stable error. All DL methods remain within 0.1–0.5 cpNRMSE with narrow, overlapping bands.

Dispersion among DL methods is greatest under Block/Gap, where WGAIN-VAE performs worst for both window strategies. PAIN attains the lowest pooled cpNRMSE overall, ranking first in 94–100\% of user-bootstrap resamples across nearly all mechanism–strategy combinations (Figure~\ref{fig:rankstability}, Appendix~\ref{app:cpnrmse-supplementary}). The only contested conditions are MAR/MNAR (full window, 10\% missingness), where GRIN claims a share of top ranks, and Block/Gap (full window, 10–20\%), where BRITS claims second.

Methods fail less and perform better under the full window strategy, especially at higher missingness. Under MNAR win30, all DL methods spike at 80\% missingness to join MICE beyond 0.5 cpNRMSE. No DL method fails on any dataset below 70\% missingness. DL failures emerge only under MNAR at 70–80\% (full window, 0.5–0.8 of datasets evaluable) and under MCAR at 80\%/MNAR at 60–80\% (win30, 0.3–0.83 evaluable). MICE struggles across all four mechanisms (0.2–0.6 evaluable throughout). Win30 is markedly more fragile than the full window, especially under MNAR at high missingness.

\subsubsection{cpNMAE}

Comparing cpNRMSE and cpNMAE diagnoses error-tail heaviness (Appendix~\ref{app:cpnmae}, Figure~\ref{fig:rmsemaeratio}). MICE, BRITS, GRU-D, PAIN, and GRIN cluster together (ratios $\approx$1.21–1.37) across most conditions. WGAIN-VAE is a consistent outlier with a lighter error tail ($\approx$1.08–1.26) despite higher average error. This is most pronounced under Block/Gap, where its ratio stays 0.1–0.2 below the next-lowest method while the others converge toward 1.30–1.37. This gap disappears under MNAR win30, where all methods converge toward 1.21–1.28 as evaluable datasets decline.

\subsubsection{cpNBias}

\begin{figure}[htbp]
    \centering
    \includegraphics[width=1\textwidth]{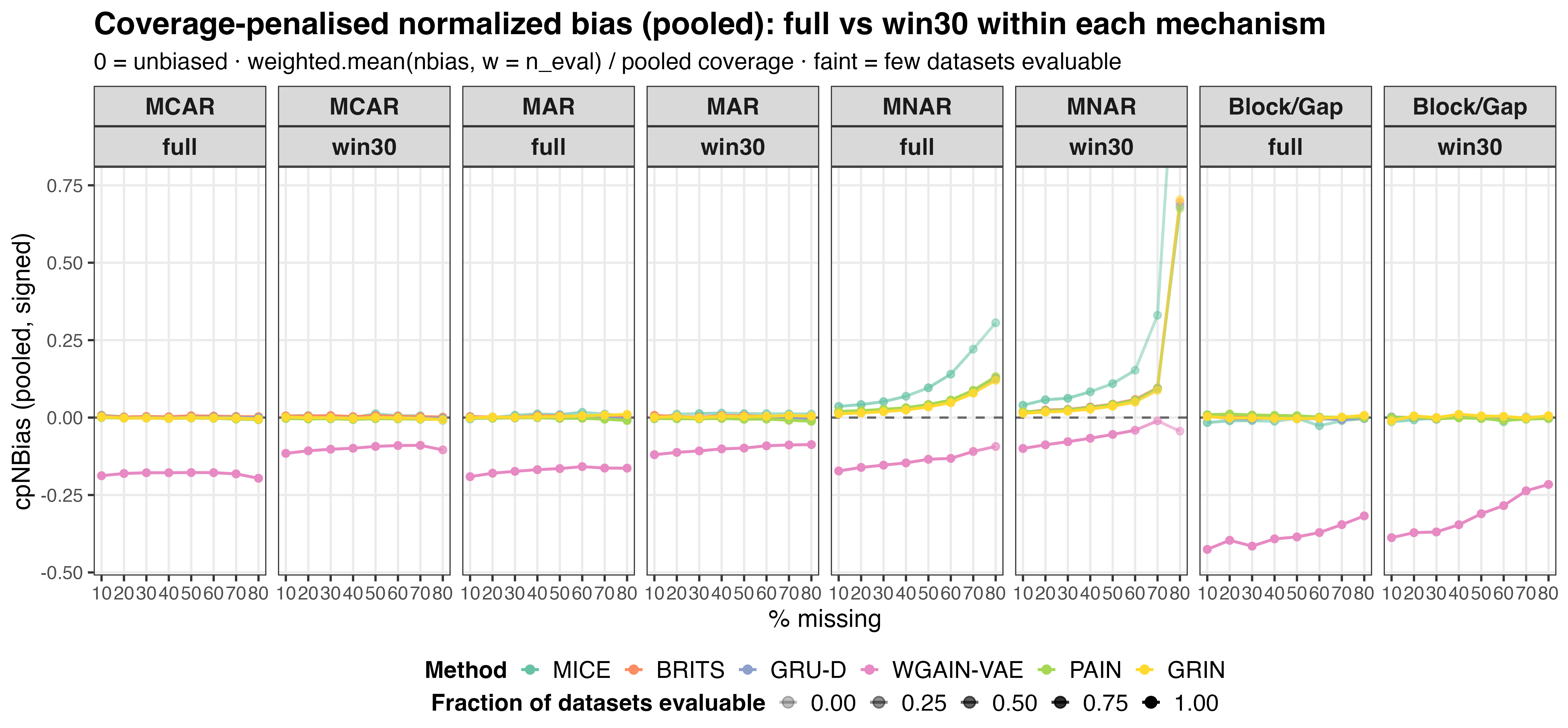} 
    \caption{Coverage-penalised normalised bias (cpNBias), pooled across datasets, by missingness mechanism and window strategy. Line fading indicates the fraction of datasets evaluable at each missingness percentage; $\times$ denotes complete failure, $\circ$ denotes no numeric/ordinal target available.}
    \label{fig:cpnbiaspooled}
\end{figure}

Figure~\ref{fig:cpnbiaspooled} presents cpNBias pooled across datasets. Pairwise comparisons against PAIN and per-dataset breakdowns are in Appendix~\ref{app:cpnbias} (Figures~\ref{fig:biasgapvspain}-\ref{fig:cpnbiasdatasets}). WGAIN-VAE is the only method with a consistent directional bias, under-imputing across nearly every condition. This shows most severely under Block/Gap ($\approx$-0.4 at 10\% missingness, $\approx$-0.2/-0.3 at 80\%), BH-significant (\textit{p} < .001) throughout.

MICE, BRITS, GRU-D, PAIN, and GRIN remain near-unbiased (within $\pm$0.02) under MCAR, MAR, and Block/Gap. Under MNAR, bias grows sharply with missingness for all of these methods, but at a markedly more extreme rate for MICE, worsening further under win30.

\subsubsection{Runtime-Accuracy trade-off}

Appendix~\ref{app:runtime} (Figure~\ref{fig:runtimefull} and Figure~\ref{fig:runtimewin30}) presents per-user runtime against NRMSE. Runtime spread is greater under win30 ($\approx$1–50s) than full ($\approx$1–10s). PAIN and MICE form the tightest runtime-accuracy clouds under full window, but PAIN achieves the same profile with substantially greater coverage, so we rank PAIN above MICE. GRIN matches PAIN/GRU-D's runtime but shows a higher-error subset (NRMSE 0.6–0.8) under all mechanisms. BRITS shows the same pattern except under Block/Gap, where it stays within 0.1–0.4 — a reasonable second choice there. GRU-D is the strongest overall second choice, matching top runtimes while maintaining a tight 0.1–0.4 NRMSE band across all mechanisms. WGAIN-VAE performs worst on this joint criterion, with the widest NRMSE spread and no runtime advantage.

\subsection{RQ3: Preserving multivariate structure}

\subsubsection{cpCorr}

\begin{figure}[htbp]
    \centering
    \includegraphics[width=1\textwidth]{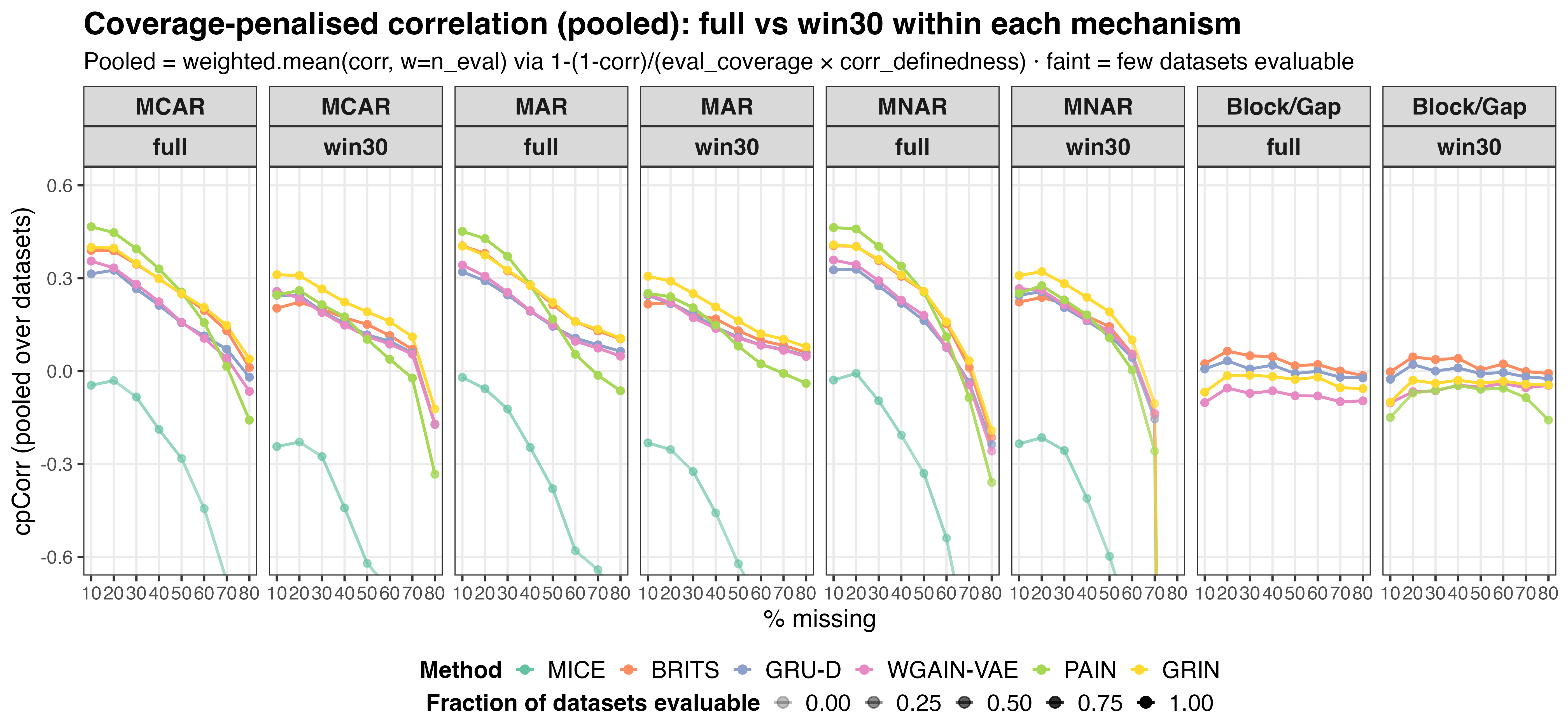} 
    \caption{Coverage-penalised correlation (cpCorr), pooled across datasets, by missingness mechanism and window strategy. Line opacity indicates the fraction of datasets evaluable at each missingness percentage.}
    \label{fig:cpcorrpooled}
\end{figure}

Figure~\ref{fig:cpcorrpooled} presents cpCorr pooled across datasets. Per-dataset detail is in Appendix~\ref{app:cpcorr} (Figure~\ref{fig:cpcorrdatasets}). Under MCAR, MAR, and MNAR (full window), PAIN leads at low missingness (10–20\%, $\approx$0.45) but declines more steeply than other DL methods, crossing over at $\approx$40–50\% and becoming the lowest-performing DL method by 70–80\%. GRIN and BRITS remain comparatively stable, attaining the highest cpCorr among DL methods at high missingness under these three mechanisms.

Under Block/Gap, this crossover does not occur: PAIN fails dramatically under the full window (Appendix~\ref{app:cpcorr}, Figure~\ref{fig:cpcorrdatasets}), while GRIN and WGAIN-VAE remain flat near zero. BRITS and GRU-D maintain a small, consistent advantage. MICE performs worst overall, declining near-monotonically and falling below zero in almost all conditions, with the greatest cross-dataset variability.

\subsubsection{cpFrob and cpPartialFrob}

We report coverage-penalised Frobenius error for full (cpFrob) and partial (cpPartialFrob) correlation matrices (Appendix~\ref{app:cpfrob}, Figures~\ref{fig:cpfrobpooled} and~\ref{fig:cpfrobdatasets}), where lower indicates better-preserved multivariate structure. A coverage caveat applies: three of seven datasets (Personal-WHOOP, Public-PMData, Public-SoccerMon) are structurally undefined across nearly every condition due to zero-variance variables. This is a dataset property, not an imputation-quality signal. We therefore treat the pooled summary as illustrative, carried effectively by only four datasets (Public-Athlete, Public-Fitbit, Public-HumanRhythms, Public-WHOOP).

The pooled plot shows a sharp rise under win30 MNAR at 70–80\% missingness, driven almost entirely by Public-Fitbit and Public-HumanRhythms. This is a coverage artefact rather than evidence of worse covariance structure. Public-Athlete and Public-WHOOP show a stable, flat baseline across methods. These findings indicate correlation and multivariate structure preservation are not fully redundant. PAIN's cpCorr advantage at low missingness does not translate to a cpFrob advantage, where it performs comparably to MICE.

\subsection{Preserving temporal structure}

\begin{figure}[htbp]
    \centering
    \includegraphics[width=1\textwidth]{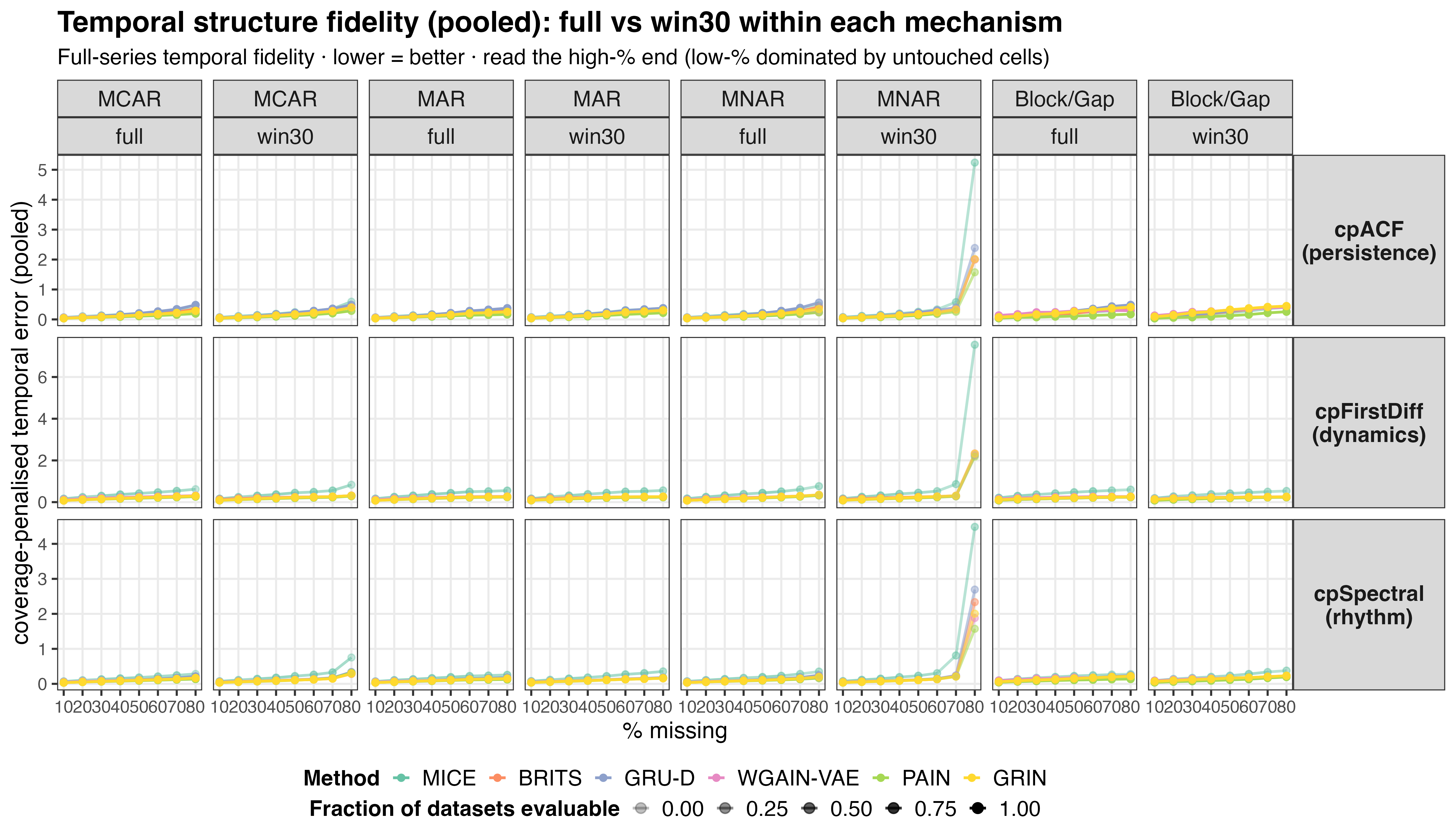} 
    \caption{Coverage-penalised cpACF, cpFirstDiff, and cpSpectral, pooled across datasets, by missingness mechanism and window strategy. Lower indicates better-preserved temporal structure. Line/point size indicates the fraction of datasets evaluable at each missingness percentage.}
    \label{fig:cptemporalpooled}
\end{figure}

Figure~\ref{fig:cptemporalpooled} shows cpACF (persistence), cpFirstDiff (dynamics), and cpSpectral (rhythm) pooled across datasets, where 0 denotes perfect preservation. Per-dataset detail is in Appendix~\ref{app:temporal} (Figures~\ref{fig:cpacfdatasets}-\ref{fig:cpspectraldatasets}. As full-series metrics, higher missingness is the more informative comparison.

Temporal structure is well preserved by DL methods until the most extreme missingness. MICE is again a dominant outlier, rising away from the DL cluster at higher missingness, while DL methods remain flat near-floor, except under MNAR win30 at 70–80\% missingness, where all DL methods rise ($\approx$2.3–2.6), though less than MICE. Per-dataset results show both patterns are driven heavily by Public-Fitbit, which has the richest day-to-day autocorrelation structure. MICE is a consistent outlier there across cpACF/cpFirstDiff from $\approx$30\% missingness onward. cpSpectral shows a similar pattern, with all methods rising together under MNAR win30 at 60–80\% missingness, indicating periodic structure becomes unrecoverable at higher missingness regardless of imputer in these dataset-mechanism-strategy interactions.

\subsection{Preserving distributional shape}

\subsubsection{Variance Ratio}

\begin{figure}[htbp]
    \centering
    \includegraphics[width=1\textwidth]{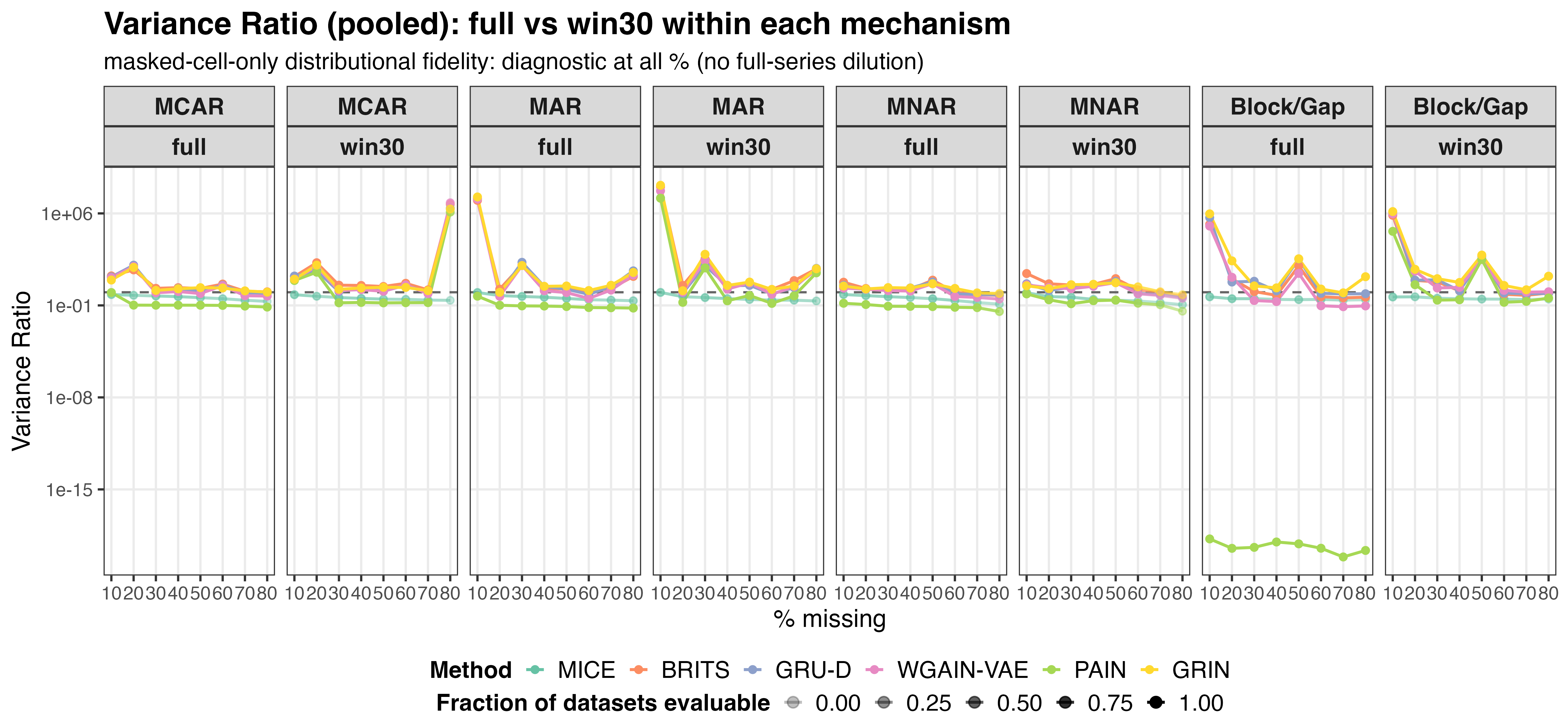} 
    \caption{Variance ratio (imputed/true variance over masked cells), pooled across datasets, by missingness mechanism and window strategy. 1 (dashed line) = ideal; below 1 = collapse; above 1 = over-dispersion. Line/point size indicates the fraction of datasets evaluable at each missingness percentage.}
    \label{fig:varratiopooled}
\end{figure}

\begin{figure}[p]
    \centering
    \includegraphics[width=0.73\textwidth]{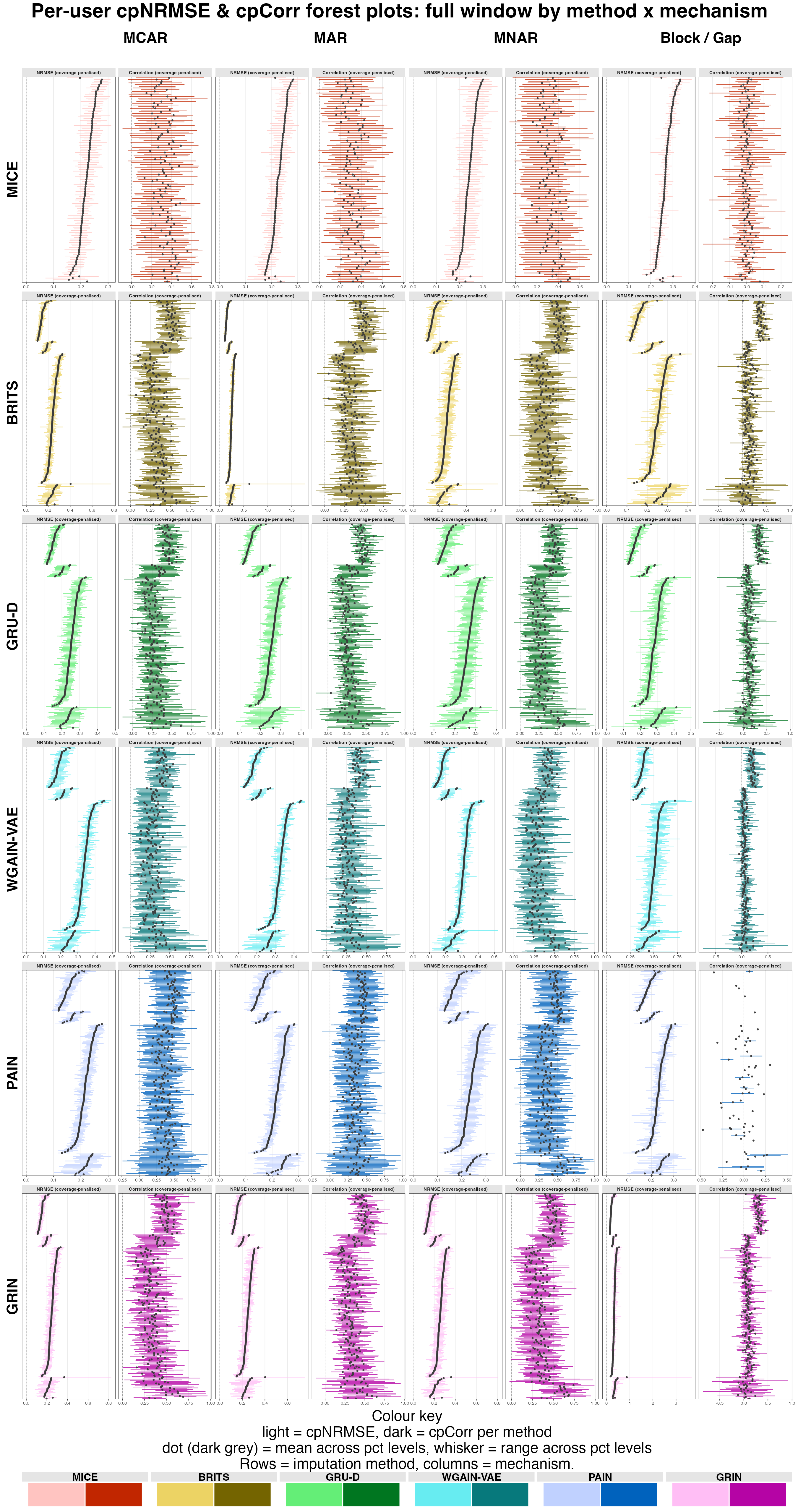} 
    \caption{Per-user cpNRMSE and cpCorr forest plots, full window strategy, by method (rows) and missingness mechanism (columns). Within each mechanism, the \textbf{light panel} shows per-user cpNRMSE (point accuracy, lower is better) and the \textbf{dark panel} shows per-user cpCorr (structure preservation, higher is better). Each line is one user; the point marks their mean across 8 missingness percentages, and the whisker spans their min--max range across those percentages (longer whisker = more sensitive to missingness severity). \textbf{Take-away:} PAIN matches top methods on cpNRMSE across all mechanisms, but its cpCorr panels collapse into near-zero with wider wiskers (i.e., greater sensitivity to missingness severity). PAIN's disordered scatter specifically under Block/Gap, clearly exposes an accuracy/structure trade-off.}
    \label{fig:forest_full}
\end{figure}

Figure~\ref{fig:varratiopooled} shows the pooled variance ratio (imputed/true variance; 1 = ideal). Per-dataset detail is in Appendix~\ref{app:distributional} (Figure~\ref{fig:varratiodatasets}). Under MCAR and MNAR, all methods cluster near ideal across most of the range. Under MAR, BRITS, GRU-D, WGAIN-VAE, and GRIN show isolated spikes to $10^5$-$10^6$, reflecting occasional over-dispersion rather than a systematic trend. MICE and PAIN sit consistently just below ideal, indicating mild, stable under-dispersion.

PAIN is the exception under Block/Gap (full window). It collapses to a near-zero variance ratio across the entire missingness range. This is the signature of imputing a near-constant value across contiguous gaps. Every other method remains near or above ideal. This collapse is consistent across six of seven datasets (Appendix~\ref{app:distributional}, Figure~\ref{fig:varratiodatasets}).

\subsubsection{cpKS and cpWasserstein1}

Under all mechanisms, MICE is the dominant outlier, while DL methods sit flush against the ideal line (Appendix~\ref{app:distributional}, Figures~\ref{fig:cpKSpooled}-\ref{fig:cpW1datasets}). The MNAR win30 facet is a partial breakdown regime: BRITS, GRU-D, PAIN, and GRIN rise alongside MICE at 70–80\% missingness, though by a smaller margin.

\subsection{Point Accuracy vs. Temporal/Distributional Structure and Shape}

Figure~\ref{fig:forest_full} presents per-user cpNRMSE and cpCorr as forest plots (full window). The win30 equivalents are in Appendix~\ref{app:forest} (Figure~\ref{fig:forest_win30}). Each horizontal line represents one user, ordered by mean cpNRMSE within dataset. The dark grey point marks that user's mean across the 8 missingness percentages, and the whisker spans their full range (min-max) across those percentages. A long whisker indicates a user whose accuracy is highly sensitive to missingness severity, while a short whisker indicates stability across the missingness range for that individual.

Combined with PAIN's leading cpNRMSE, low cpNBias, and favourable runtime-accuracy trade-off, PAIN delivers strong point accuracy at low cost. GRIN, GRU-D, and BRITS are comparable on accuracy alone. MICE's poor cpNRMSE, large positive cpNBias, and high failure count mark it as the least reliable method overall. WGAIN-VAE shows systematic under-imputation with a comparatively light error tail, at higher computational cost.

Under MCAR, MAR, and MNAR, PAIN's per-user whiskers resemble other DL methods. Under Block/Gap (full window), however, PAIN's cpCorr panel scatters near zero without smooth ordering, with wide, asymmetric whiskers, though it performs better under win30 for this mechanism (Appendix~\ref{app:forest}). BRITS, GRU-D, WGAIN-VAE, and GRIN retain smoother, low-dispersion cpCorr ordering under Block/Gap. This is corroborated by the variance-ratio results (Figure~\ref{fig:varratiopooled}). PAIN's leading cpNRMSE co-occurs with severe variance collapse under Block/Gap ($\approx10^{-13}$), while every other method remains near ideal. Across all mechanisms, MICE's whiskers are longer and shifted higher than any DL method's.

Appendix~\ref{app:tradeoffs} (Figures~\ref{fig:acc_vs_multivariate} and~\ref{fig:acc_vs_variance}) presents point accuracy jointly with multivariate structure and variance preservation. All methods cluster near-ideal on both axes under all mechanisms for the full window, except under MNAR win30 at high missingness, where methods separate from the ideal cluster simultaneously. MICE deviates most, while BRITS, GRU-D, WGAIN-VAE, and GRIN rise together by a smaller margin, and PAIN rises by a comparably smaller margin on cpFrob.

\section{Discussion}
\label{sec:discussion}

The central contribution of this work is a goal-conditioned decision framework for imputing missing data in long-term, multivariate, athlete-specific wearable datasets.
Figure~\ref{fig:decision-flow} summarises this in a flowchart to guide the selection of the appropriate imputation mechanism,
based on our experiments. Rather than asking "which imputation method is best?", we show that the right question is "best \textit{for what}?" and that the answer depends systematically on whether a researcher's analytical goal is point-in-time assessment, relational/temporal pattern detection, or predictive insight. In this section, we first present this framework and the reasoning behind each of its recommendations i.e., RQ3 (Section~\ref{sec:RQ3}), then substantiate it with the underlying evidence: how missingness manifests in athlete data (Section~\ref{sec:charmiss}), and how methods perform on accuracy and on multivariate/temporal structure across mechanisms and missingness rates i.e., RQ1 and RQ2 (Section~\ref{sec:RQ1RQ2}). Together, these results give practitioners and researchers a practical, evidence-backed guide for handling missing data in similar datasets.

\subsection{A Goal-Conditioned Decision Framework for Imputation (RQ3)}
\label{sec:RQ3}

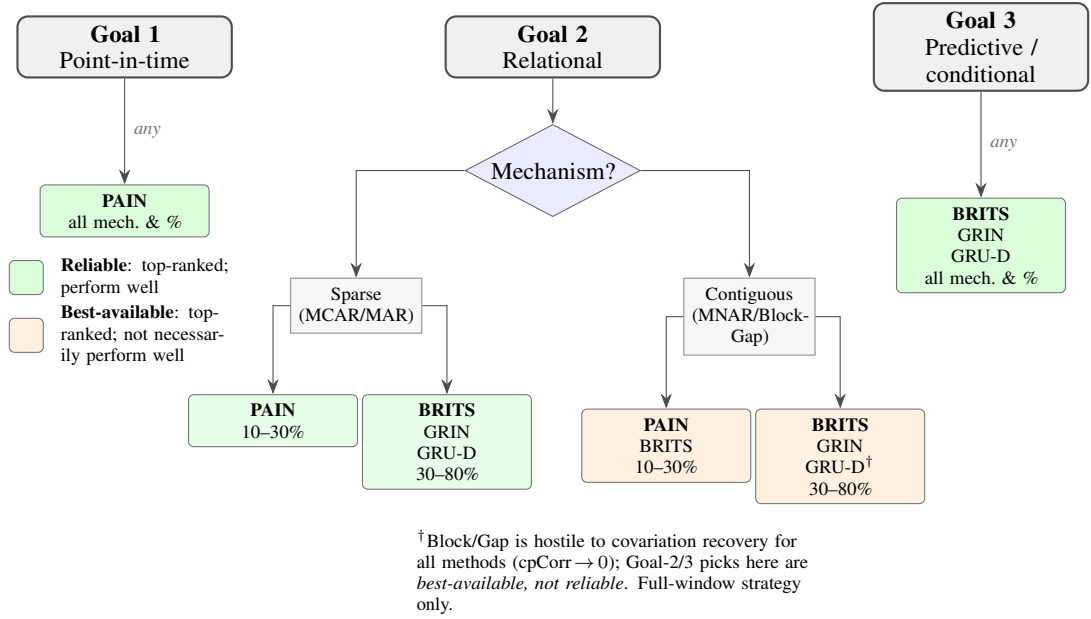
\begin{figure*}[t]
  \centering
  \begin{tikzpicture}[
    font=\footnotesize,
    node distance = 6mm and 9mm,
    goal/.style   = {rectangle, rounded corners, draw=black!70, thick,
                     fill=black!6, align=center, minimum height=8mm,
                     text width=26mm},
    dec/.style    = {diamond, aspect=1.9, draw=black!60, fill=blue!8,
                     align=center, inner sep=1pt, minimum height=7mm,
                     text width=17mm},
    band/.style   = {rectangle, draw=black!45, fill=black!3, align=center,
                     minimum height=6mm, text width=15mm, font=\scriptsize},
    leaf/.style   = {rectangle, rounded corners=2pt, draw=black!55,
                     fill=green!10, align=center, minimum height=6mm,
                     text width=20mm, font=\scriptsize},
    line/.style   = {draw, -{Stealth[length=2mm]}, black!70},
    lbl/.style    = {font=\scriptsize\itshape, black!55},
    legbox/.style = {rectangle, rounded corners=2pt, draw=black!55,
                     minimum height=4.5mm, minimum width=4.5mm}
  ]

  \node[goal] (g1) {\textbf{Goal 1}\\Point-in-time};
  \node[goal, right=28mm of g1] (g2) {\textbf{Goal 2}\\Relational};
  \node[goal, right=28mm of g2] (g3) {\textbf{Goal 3}\\Predictive /\\conditional};

  \node[leaf, below=14mm of g1, fill=green!14] (l1)
       {\textbf{PAIN}\\\scriptsize all mech.\ \& \%};
  \path[line] (g1) -- node[lbl,right]{any} (l1);

  \node[leaf, below=14mm of g3, fill=green!14] (l3)
       {\textbf{BRITS}\\GRIN\\GRU-D\\\scriptsize all mech.\ \& \%};
  \path[line] (g3) -- node[lbl,right]{any} (l3);

  \node[dec, below=6mm of g2] (m2) {Mechanism?};
  \path[line] (g2) -- (m2);

  \node[band, below=8mm of m2, xshift=-26mm] (sp) {Sparse\\(MCAR/MAR)};
  \node[band, below=8mm of m2, xshift=26mm] (ct) {Contiguous\\(MNAR/Block-Gap)};
  \path[line] (m2) -| (sp);
  \path[line] (m2) -| (ct);

  \node[leaf, below=8mm of sp, xshift=-11mm] (sp_lo)
       {\textbf{PAIN}\\\scriptsize 10--30\%};
  \node[leaf, below=8mm of sp, xshift=12mm] (sp_hi)
       {\textbf{BRITS}\\GRIN\\GRU-D\\\scriptsize 30--80\%};
  \path[line] (sp) -| (sp_lo);
  \path[line] (sp) -| (sp_hi);

  \node[leaf, below=7mm of ct, xshift=-11mm, fill=orange!12] (ct_lo)
       {\textbf{PAIN}\\BRITS\\\scriptsize 10--30\%};
  \node[leaf, below=7mm of ct, xshift=12mm, fill=orange!12] (ct_hi)
       {\textbf{BRITS}\\GRIN\\GRU-D$^{\dagger}$\\\scriptsize 30--80\%};
  \path[line] (ct) -| (ct_lo);
  \path[line] (ct) -| (ct_hi);

  \node[legbox, fill=green!14, below=24mm of g1, xshift=-13mm] (leg1) {};
  \node[font=\scriptsize, right=1mm of leg1, align=left, text width=22mm]
       {\textbf{Reliable}: top-ranked; perform well};
  \node[legbox, fill=orange!12, below=3mm of leg1] (leg2) {};
  \node[font=\scriptsize, right=1mm of leg2, align=left, text width=22mm]
       {\textbf{Best-available}: top-ranked; not necessarily perform well};

  \node[align=left, font=\scriptsize, text width=52mm,
        below=4mm of sp_hi, xshift=22mm] (note)
       {$^{\dagger}$Block/Gap is hostile to covariation recovery for all
        methods (cpCorr $\!\to\!0$); Goal-2/3 picks here are
        \emph{best-available, not reliable}. Full-window strategy only.};

  \end{tikzpicture}
  \caption{Decision flowchart for goal-conditioned imputation-method selection (\emph{full-window strategy}). Read top-to-bottom: start at your analysis goal (Goal 1–3), follow the branches through missingness mechanism and percentage where shown, and select the listed method(s) at each leaf. Green leaves denote Reliable recommendations (consistently top-ranked across multiple evaluation metrics and performed well). Orange leaves denote Best-available recommendations (top-ranked among tested methods, but not necessarily performing well). Point-in-time (Goal 1) and predictive/conditional (Goal 3) tasks require no mechanism- or percentage-specific branching, so PAIN and [BRITS, GRIN, GRU-D] are recommended uniformly across all mechanisms and missingness levels, respectively. For relational tasks (Goal 2) under contiguous (MNAR/Block-Gap) missingness, all methods' between-variable correlation recovery collapses toward zero (cpCorr $\!\to\!0$), therefore the Goal-2 picks shown here are best-available rather than reliable, and apply to the full-window strategy only.}
  \label{fig:decision-flow}
\end{figure*}

Choube et al.~(2025)~\cite{imputation_matters} argue that "researcher[s] should choose right evaluation metrics to evaluate imputation strategies depending on their end goal." We operationalise this principle for athlete datasets by identifying three recurring types of analytical goal: point-in-time assessment (e.g., "how ready is my body for a long run today?"), relational temporal pattern detection (e.g., "did my soreness track with training load and sleep duration this month?"), and predictive insight (e.g., "does sleeping 8+ hours and eating more carbs make next-day training feel easier?"). Consistent with Choube et al.~\cite{imputation_matters}, we find that no single imputation method dominates across all three goals. This directly confirms H2: no single method is simultaneously optimal for point accuracy and for preserving temporal/multivariate structure (Section~\ref{sec:dl-methods} alongside Figure~\ref{fig:cpNRMSEzoompooled} and Figure~\ref{fig:cpcorrpooled} provide the clearest evidence, where PAIN leads on cpNRMSE but collapses on cpCorr under Block/Gap). Instead, the appropriate choice depends on which property of the reconstructed signal the downstream question actually interrogates (i.e., point-wise accuracy, marginal/multivariate structure, or conditional structure). Figure~\ref{fig:decision-flow} summarises these recommendations. The remainder of this subsection explains the reasoning behind each branch, and Sections~\ref{sec:charmiss} and~\ref{sec:RQ1RQ2} provide the supporting evidence.

Across all three goals, two cross-cutting recommendations hold. First, our coverage-penalised metric suite is what makes these per-goal recommendations possible. The coverage-penalised metric suite disentangles failure modes (bias, variance collapse, structural distortion) that raw error metrics alone would conflate, letting us pinpoint where and why each method fails rather than simply that it does (evidence in Section~\ref{sec:RQ1RQ2}). Second, across datasets and conditions, the win30 windowing strategy degrades performance most under MNAR missingness. We recommend against this mechanism-window pairing for all three goal types, and treat the full-window strategy as the default (Section~\ref{sec:mice}).

\subsubsection{Goal 1: Point-in-time assessment.}
For questions about a single value at a specific point in the time series (e.g., an athlete or coach asking how training went in Friday's strength session) the relevant property to optimise is point-wise reconstruction accuracy. We recommend PAIN. PAIN attains the lowest cpNRMSE under almost all conditions, shows minimal systematic bias, and has a predictable, low computational cost (Section~\ref{sec:dl-methods}). Its accuracy is not an artefact of rare catastrophic errors being averaged away, and it offers the most attractive runtime-accuracy trade-off among the deep learning methods, particularly under the full-window strategy. GRIN, GRU-D, and BRITS are comparable alternatives for practitioners who prefer to standardise on a single structure-preserving method across goals. We do not recommend WGAIN-VAE. Its conservative, mean-reverting behaviour and high computational cost are not offset by any accuracy advantage.

\subsubsection{Goal 2: Relational temporal pattern detection.}
For questions about how signals co-vary over time (e.g., whether an athlete's carbohydrate intake correlates with sleep duration) the priority shifts to multivariate structure preservation, examining how cross-channel covariation, temporal dynamics, and marginal spread are retained (Section~\ref{sec:dl-methods}). The recommendation here is more conditional than for Goal 1. For missingness below 40\% under MCAR, MAR, and some MNAR conditions with a full-window strategy, PAIN remains a reasonable choice if runtime is also a priority. However, PAIN degrades sharply at higher missingness rates, systematically collapsing variance, so we strongly advise against using PAIN under Block/Gap missingness specifically, where this collapse is most severe. Above 40\% missingness, BRITS and GRIN are the more dependable structure-preservers, with a flatter performance decay, though both degrade faster under MNAR. This 40\% breakpoint is consistent with our expectation, grounded in prior literature \cite{40missing, mice50missing}, that imputation quality would degrade meaningfully at or above this threshold. For Block/Gap missingness specifically, we recommend BRITS as the best-available option, with the caveat that "best available" is not the same as "reliable". Like other DL methods (except PAIN), BRITS cpCorr still clusters just above the zero line, reflecting a general difficulty recovering between-variable relationships under this mechanism.

\subsubsection{Goal 3: Predictive / conditional insight.}
Conditional "if X and Y, then Z" questions (e.g., what affects an athlete's next-day readiness) concern direct dependencies among channels after accounting for others. We recommend BRITS and GRIN, the most stable methods across conditions for this goal, again as a best-available rather than fully reliable recommendation. GRU-D is a comparable method as well. We consider PAIN, WGAIN-VAE, and MICE inappropriate for this goal given PAIN's structure-preservation deficits (Goal 2), WGAIN-VAE's systematic under-imputation, and MICE's outright failures (Section~\ref{sec:mice}).

\subsubsection{Combining goals in practice.}
In practice, researchers often pursue multiple goals simultaneously, and deployment on athlete- or coach-facing systems requires balancing reconstruction quality against computational cost, especially at scale. 

For the combination of point-in-time and relational goals (e.g., an accurate readiness estimate today and a faithful account of how readiness has covaried with sleep and load), no single method optimises both. PAIN wins on accuracy but collapses spread and structure, while BRITS and GRIN preserve structure without leading on accuracy. BRITS is the most defensible compromise under Block/Gap missingness (lower cpNRMSE, higher cpCorr than GRIN). GRIN is a defensible competitor under MCAR, MAR, and MNAR (lower cpNRMSE, higher cpCorr than BRITS). GRU-D is a reasonable third fallback, performing comparably on cpNRMSE and cpCorr while offering a tighter runtime-accuracy profile than GRIN.

For the combination of relational and predictive goals, both of which depend on recovering covariation (marginal and conditional, respectively), the recommendation is unambiguous. The same trio, b BRITS, GRU-D, and GRIN, dominates both axes. As above, we recommend the full-window strategy over win30.

For the combination of point-in-time and predictive goals (an accurate value today plus trustworthy conditional structure), the tension is sharpest, since structure-preservation is volatile across datasets and users. Again, our recommendation is best-available rather than most-reliable. BRITS, GRU-D, and GRIN perform similarly and show relative stability across missingness rates, so we suggest differentiating between them by runtime-accuracy trade-off. Given that BRITS has previously performed well on health records and biomedical sensor data~\cite{kazijevs2023}, a domain athlete-specific data may resemble, we lean toward recommending BRITS here.

These recommendations are collectively reflected in Figure~\ref{fig:decision-flow}. The remainder of this section presents the evidence underlying them.

\subsection{Evidence I: Characterising Missingness Mechanisms}
\label{sec:charmiss}

The decision framework above hinges on first identifying which missingness mechanism is present in a dataset, since this determines which branch of Figure~\ref{fig:decision-flow} applies. This subsection establishes that athlete data exhibits genuinely distinct missingness mechanisms worth distinguishing, and that doing so is non-trivial.

There are a variety of reasons why data can be missing, and the nature of that missingness in athlete-specific datasets is itself informative~\cite{imputation_longterm_athlete, impute_bball}. Missing data can be a strong behavioural signal in its own right~\cite{imputation_matters}. Yet the nature of missingness remains under-reported in the literature~\cite{imputation_matters}. We build on prior work to characterise missingness across four mechanisms: Missing Completely at Random (MCAR), Missing at Random (MAR), Missing Not at Random (MNAR), and Block/Gap. The traditional MCAR/MAR/MNAR taxonomy is useful but not always the most meaningful and useful categorisation~\cite{beyond_rubin}, which is why HCI researchers have additionally focused on "gaps" as a distinct, practically meaningful pattern~\cite{tang_adherence2018}. Across the 231 users in our study, we observe a wide range of missingness patterns (tested using methods outlined in Section~\ref{sec:naturalmiss},) with clear differences in downstream performance across all four mechanisms (Section~\ref{sec:RQ1RQ2}). There is heterogeneity in the manifestation of missingness patterns, and testing what mechanism is present for an individual's data can help our decision-making. This confirms that illuminating what kind of missingness is present in a dataset is a necessary first step for choosing an imputation strategy suited to a researcher's end goal, the premise on which Figure~\ref{fig:decision-flow} is built.

\subsection{Evidence II: Accuracy and Structural Performance Across Methods (RQ1, RQ2)}
\label{sec:RQ1RQ2}

Having established that missingness mechanisms are meaningfully distinct, we now turn to how imputation methods perform against them, the evidence base for the per-goal recommendations in Section~\ref{sec:RQ3}. We ask how imputation accuracy varies across methods, mechanisms, and missing percentages (RQ1), and how well imputations preserve multivariate and temporal structure under the same conditions (RQ2). The headline finding is that there is no one-size-fits-all answer. While clear failure modes exist for athlete long-term, multi-modal, multivariate data, choosing the best method is nuanced and goal-dependent. Point accuracy is strongly governed by missingness mechanism, while multivariate-structure preservation is governed by both mechanism and the amount of missingness, consistent with prior athlete-imputation studies~\cite{impute_bball, imputation_longterm_athlete}.

\subsubsection{Why MICE fails so often.}
\label{sec:mice}
MICE is the clear worst coverage-penalised performer on both point accuracy and multivariate structure, indicating pervasive rather than mechanism-specific unreliability. Its high failure count quantifies how often it produced no valid, evaluable reconstruction at all. This is unsurprising: MICE was designed for cross-sectional inference and pooled multiple-imputation estimates, not long per-user time series, and is validated primarily under MCAR/MAR assumptions, whereas our evaluation was strictly per-user and spanned varying mechanisms. Over 80\% of logged MICE failures were labelled `residual\_na > 0`, meaning MICE ran but left cells unimputed. Although it typically succeeded on most columns, each failing chunk saw one or more target columns skipped entirely, whose original missing cells remained NA (we deliberately did not implement a fallback to capture these failures across all imputation methods). Columns were skipped for one of three reasons: (1) too little data to learn from (fewer than 3 non-missing values remaining in the chunk); (2) the column appeared constant; or (3) failure on one column crashed the full chunk. Athletes often show many such near-constant variables, as tracked daily habits can be highly consistent. More broadly, sport science data exhibits non-linear patterns, substantial individual variability (visible in our forest plots, Figure~\ref{fig:forest_full} and Figure~\ref{fig:forest_win30}), and temporal dependencies~\cite{impute_bball, imputation_longterm_athlete}. These are the conditions under which MICE's distributional assumptions are hard to validate~\cite{impute_bball}, so its failure here is not unexpected.

Interestingly, the win30 windowing strategy reduces MICE's failure rate and tames its error tail, suggesting its breakdown is partly a function of context length. By constraining reconstruction to a bounded 30-day window, the long-range extrapolation responsible for MICE's most catastrophic misses is removed, at the cost of longer and more variable runtime. Overall, these results point to clear practical limitations of applying MICE to long-term multivariate athlete data, and motivates our recommendation against it for any of the three analytical goals (Section~\ref{sec:RQ3}).

\subsubsection{Comparing deep learning methods.}
\label{sec:dl-methods}
PAIN's reliability on point accuracy, combined with its low computational cost and corroborating cpNMAE results, suggests its strong cpNRMSE performance is genuine rather than masking rare catastrophic errors. This strengthens the case for its point-accuracy reliability, underpinning its Goal 1 recommendation above. WGAIN-VAE, by contrast, shows a combination of systematic under-imputation and a light error tail consistent with a conservative, mean-reverting imputer. Its high computational cost is not offset by a corresponding accuracy advantage, which is why we exclude it from all three goal recommendations. Our results partially support H1. Deep learning methods in our evaluation do outperform MICE on both point accuracy and structure preservation across all four mechanisms, including the hypothesised MAR, MNAR, and Block/Gap conditions. However, PAIN breaks the pattern under Block-Gap.

PAIN's strong aggregate point accuracy does not hold uniformly across mechanisms, which is precisely why its Goal 2 (relational) recommendation above is conditional rather than unqualified. Under Block/Gap missingness, individual users' cpCorr swings between reasonable and near-zero (or negative) correlation depending on missingness percentage, rather than degrading gradually. This is a pattern specific to PAIN among the methods examined. Read alongside the variance-ratio collapse, PAIN's Block/Gap failure is not confined to the pooled variance-ratio floor. We see even at the individual-user level, correlation with ground truth becomes unstable and frequently indistinguishable from noise. BRITS, GRU-D, and GRIN, by contrast, preserve user-level stability in both metrics throughout. This decouples point accuracy from shape preservation specifically for PAIN under Block/Gap. Therefore, the method with the best reconstruction accuracy is simultaneously the method with the most severe spread collapse, which is why BRITS, GRU-D, and GRIN are the more broadly trustworthy choice wherever user-level stability in both point accuracy and variable structure matters (Goals 2 and 3).

Beyond PAIN and WGAIN-VAE, BRITS, GRU-D, and GRIN emerge as serious competitors wherever preserving distributional and multivariate structure jointly with point accuracy matters, even though their raw point accuracy alone does not always match PAIN's. This is the basis for their Goal 2/3 recommendations. Our finding on BRITS coincides with Kazijevs and Samad (2023)~\cite{kazijevs2023}, who applied BRITS to health records and biomedical sensor data, a domain athlete datasets can resemble. Our results support their finding that BRITS performs well under MNAR missingness (of which Block/Gap is a form), across time series lengths, datasets, and high missingness rates. We also observe heterogeneity in how DL methods perform across individual users, but all achieve broadly satisfactory results, supporting Liu et al.~(2023)~\cite{liu2023}.

\subsubsection{Limitations}
\label{sec:limitations}

Several limitations bound the generality of our conclusions. First, our MICE implementation pools its five completed datasets into a single point estimate by averaging numeric values and taking the modal category, in order to support point-wise comparison against ground truth across all methods on a common footing. This collapses the between-imputation variance that is MICE's principal statistical advantage. Although we note that MICE naturally captures the uncertainty in its estimates through its multiple-imputation design, our evaluation pipeline discards this uncertainty to produce a single completed dataset. This is a defensible simplification for our point-accuracy and structure-preservation comparisons, but it means our results do not reflect MICE's performance when used as intended, i.e., with its multiple imputations retained and pooled downstream via Rubin's rules rather than averaged upfront. Researchers specifically interested in uncertainty quantification, rather than point reconstruction, should not treat our MICE results as representative of the method's full capability, and may find MICE more competitive under an evaluation protocol that preserves its between-imputation variance.

Second, our two breakdown regimes constitute threats to validity for any goal that touches structure or dynamics. Under MNAR/win30 every method spikes simultaneously across temporal-structure and distributional metrics at 70-80\% missingness. Furthermore, Block/Gap collapses cpCorr to a narrow band around zero for all methods. Since these are precisely the mechanisms most likely to arise in real wearable and self-report data, our goal-conditioned recommendations should be read as conditional on missingness remaining below these extremes. 

Our analysis operates at a daily aggregation level and does not examine minute-to-minute signals, so conclusions about temporal-structure preservation pertain to day-scale rhythms rather than intra-day dynamics where reconstruction difficulty and relative ordering of methods could differ. We also did not incorporate device wear time per day as an explicit covariate or masking driver. As non-wear itself is a primary generator of MNAR micro gaps in life-logging data, excluding it means our synthetic mechanisms may understate the difficulty of real deployments. Future studies can build on our daily aggregated pipeline to include daily non-wear time and develop a minute-to-minute pipeline. Our analysis also assumed the DL methods to have implicitly captured lagged dependencies through their sequence modelling, which is consistent with the full window strategy's superior robustness. However, future studies can explicitly evaluate lagged observations within the win30 imputation window as a means of anchoring inter-series modelling.

Although real datasets typically can exhibit multiple co-occurring missingness mechanisms, we imputed under one mechanism at a time. Interactions between mechanisms with a single time-series are therefore not captured and the observed method rankings may shift under mixed regimes. Future work can explore these interactions explicitly to supplement our work. A wider more systematically varied set of missingness patterns would be needed to bootstrap and stress-test our findings. We identify this as the most important direction for future work before the goal-conditioned recommendations can be treated as broadly generalisable.

\section{Conclusion}

Missing data in wearable and life-logging datasets from athletes has received comparatively little attention in HCI, despite this population's distinctive behavioural, physiological, and goal-driven data patterns. We characterised missingness across 231 users under four mechanisms and systematically benchmarked six imputation methods, spanning statistical and deep learning approaches, on point accuracy, multivariate structure, temporal structure, and distributional shape.

No single method dominates across all analytical goals, mechanisms, and missingness levels. PAIN is the strongest choice for point-wise reconstruction under most conditions, while BRITS, GRU-D, and GRIN are more broadly dependable wherever multivariate or temporal structure must be preserved alongside accuracy, particularly under Block/Gap and at higher missingness rates. We recommend the full window strategy over win30, and caution against MNAR/win30 regimes specifically, where reconstruction quality degrades sharply across all methods.

These findings contribute a practical, goal-conditioned decision framework for researchers handling missing data in long-term, multivariate, athlete wearable and life-logging datasets, an area where such guidance has been largely absent, and demonstrate the value of evaluating imputation quality along multiple complementary axes rather than accuracy alone. As noted in our limitations (Section~\ref{sec:limitations}), these recommendations are bounded by the missingness patterns we simulated, our daily aggregation level, and the assumption of single, non-interacting mechanisms. Extending this work to mixed missingness regimes, finer-grained temporal resolutions, and explicit non-wear modelling are the most important next steps toward generalising these findings for real-world SportHCI deployment.

\bibliography{references}  

\appendix

\section{Methodological Transparency \& Reproducibility Appendix (META)}

\subsection{Extended Dataset Descriptions}
\label{app:datasets}

This appendix provides full descriptive detail for each of the 7 datasets summarised in Section~\ref{sec:dataset-summaries}, supporting reproducibility and dataset-selection transparency.

\subsubsection{Complete-Case Extraction Parameters}

We computed each user's complete subset by first enforcing consecutive days and then identifying the longest continuous observation window with minimal missingness ($\leq$5\%) across all target variables. This is a necessary methodological choice; we acknowledge that selecting such windows may limit generalisation to the rest of the athlete population. For larger datasets with longer timelines (e.g., Human Rhythms), a greater than 60-day window was applied, and users with less than 60 days were excluded. For shorter few-person datasets (e.g., Public WHOOP), shorter timelines were accepted to make the most of the data at hand and to provide edge cases. Up to 5\% residual missingness with gaps of at most one day is tolerated. A type-aware imputation is applied for these minimal gaps: numeric columns use the median, logical columns use the modal false/true value, and categorical columns use the mode. These complete subsets define the simulation ground truth used throughout the study.

\subsubsection{SoccerMon}

The SoccerMon dataset \cite{soccermon} is a large-scale longitudinal sports science public dataset focused on elite women's football (soccer). The dataset combines two main categories of athlete-monitoring data collected from 77 professional female players in the Norwegian top division: (1) subjective self-report data such as wellness and training load questionnaires collected via PmSys \cite{pmsys}, (2) objective sensor-based data such as GPS tracking and other load indicators. The raw dataset has over nine thousand parquet and csv files containing minute level data from sensor-based data and daily level data for subjective data. For this study, we aggregated all data to the daily level, which provided 38,000+ observations, across all individuals over 2 seasons (51.45\% missing data). The complete subset extract retained 45 out of the 77 individuals. 

\subsubsection{Human Rhythms}

The Human Rhythms Dataset \cite{Human-Rhythms-Dataset} is a longitudinal public dataset designed to study human behavioural rhythms and physiological patterns. It consists of Fitbit, mobile, and daily survey data collected from 166 college students between August 2021 and January 2023. 1 participant was reported to be an elite athlete (NCAA), and the dataset could contain unreported student-athletes (non-NCAA). This dataset captures multivariate metrics such as activity levels, sleep, and mood. The purpose of this dataset is unique as it focuses on circadian and ultradian rhythms, providing temporal perspective on human behaviour. The dataset also includes contextual metrics, such as COVID data, weather, and university events. For this study, we retain the Fitbit and survey streams across 166 processed participants with 36,000+ observations (29.33\% missing data). The complete subset yielded 146 users.

\subsubsection{Fitbit}

Fitbit is one of the most widely used consumer-grade wearables, offering a broad range of health and fitness tracking features. Fitbit devices measure metrics such as steps, heart rate, calories burned, and sleep patterns, making them popular for both general health monitoring and research applications. We use a public Fitbit dataset available on Kaggle \cite{Fitbit_public}. 23 of the 33 users had sleep-tracking data, with a total of 600+ observations across 31 days and 31 variables (19.54\% missing data). The complete subset extracted 23 users. This dataset provides an edge case scenario of less than 30 days of data of wearable users. We are unsure whether these users are athletesfo which could be contributing to this datasets "outlier" like results where MICE and other methods deviated most. Nevertheless, these are daily wearable users. Combining such a dataset in this study enables evaluation of generalisability across different wearable devices, time frames, and a wider, heterogeneous consumer sample.

\subsubsection{PMData}

PMData is a public sport-logging dataset of 16 persons spanning five months, that integrates data from multiple life-logging sources, including Fitbit wearables, a sports logging app, and Google Forms \cite{PMData}. It is designed to support research on personal health and behaviour, offering a rich, multivariate dataset with diverse data streams. PMData includes metrics such as physical activity, injuries, sleep, heart rate, nutrition, and mood. It is uniquely suited to studying the interplay between physiological and contextual factors. The multi-source nature of PMData introduces heterogeneous resolutions and missingness patterns, which are critical for stress-testing imputation and network analysis robustness. The raw dataset has 63 variables with 2000+ observations across 152 days (30.16\% missing days). The complete subset extracted retained 14 users. 

\subsubsection{Personal WHOOP}

The first author's personal WHOOP data, collected over two years, provides a single-user longitudinal dataset with 500+ days of history and 16 naturally occurring data gaps. WHOOP is a wearable device designed for athletes and fitness enthusiasts, offering advanced tracking of physiological and behavioural metrics. Unlike general-purpose wearables like Fitbit or Apple Watch, WHOOP focuses on optimising human performance and recovery by providing detailed insights into physiological cycles \cite{WHOOP_howitworks}. Physiological cycles are defined as a Sleep-to-Sleep cycle to provide recovery insights based on natural sleep patterns rather than a fixed 24-hour clock. WHOOP collects measures such as heart rate variability (HRV), sleep stages, strain, and recovery scores. The device also allows users to log behavioural data daily on the app interface journal feature, such as alcohol consumption, stress levels, nutrition, and subjective recovery \cite{WHOOP_howitworks}. WHOOP is particularly suitable for multivariate analysis. These features make WHOOP highly relevant for this study, as it provides both physiological and behavioural data streams that are prone to missingness. 

Physiological and activity data were extracted programmatically using the official WHOOP Developer API v2 with OAuth 2.0 authentication \cite{WHOOP_api}, which provides authenticated, granular access to the cycles, sleep, recovery, and workout endpoints. The resulting dataset was structured at the daily (cycle) level, with each row containing aligned sleep, recovery, and activity variables. The richness and longitudinal span of this dataset (700+ days with 500+ observed) enabled extraction of a clean complete-case window over 181 days.

\subsubsection{WHOOP Dataset}

The Public WHOOP Dataset \cite{WHOOP_public} is a Kaggle-hosted, real-world, single-user WHOOP export that includes WHOOP derived physiological metrics alongside contextual annotations entered by the user. The raw dataset downloaded from Kaggle had 53 observations across 51 days. This dataset is another edge-case that is less than 60-days.

\subsubsection{NFL Athlete Dataset}

This long-term public dataset, available on Kaggle \cite{athlete_public_dataset}, contains daily performance data for a National Football League (NFL) player. NFL is a premier professional American football league in the United States. This player self-published their data and integrates data from WHOOP, Polar, Oura Ring, and MyFitnessPal. The dataset including metrics for sleep, nutrition, training intensity, and recovery, providing a detailed view of the athlete's physiological and performance trends. A complete-case subset was extracted spanning 94 days. This dataset is particularly valuable for studying the application of wearable data analytics in high-performance sports. 

\subsection{Mechanism Detection Pipeline}
\label{app:mechanismdetection}

This appendix provides the full statistical procedure underlying the staged diagnostic pipeline summarised in Section~\ref{sec:naturalmiss}, used to characterise naturally occurring missingness mechanisms (MCAR, MAR, MNAR, Block/Gap) in each of our seven datasets.

\subsubsection{MCAR testing}
We first test MCAR using a pattern-wise chi-square test, binning each variable into terciles and testing independence between missingness pattern and bin, with Bonferroni correction across variables. MCAR is rejected at p$<$0.05, consistent with Little's (1988) \cite{little1988test} classic formulation, under which missingness is unrelated to observed or unobserved values. Since such tests are sample-size sensitive, we gate the result behind a power heuristic: it is trusted only when n$\geq$500, at least three distinct missingness patterns exist, and minimum expected cell count $\geq$5. We additionally compute co-missingness (correlation among variables' missingness indicators), flagging |r|$>$0.3 as possible device-level dropout rather than genuine MAR.

\subsubsection{MAR testing}
When MCAR is rejected, we probe for MAR by fitting logistic regressions to predict each variable's missingness from other variables, synthesised temporal predictors, and user-supplied covariates, using missing-indicator augmentation. Benjamini-Hochberg (BH)–adjusted likelihood-ratio p-values below the alpha flag a MAR signal is used to operationalise missingness' dependence on observed data.

\subsubsection{MNAR testing}
We assess MNAR via two tests: (1) a long-run signal (run-length encoding of missing streaks, flagged when $\geq$50\% of missing days fall in runs $\geq$5), and (2) a tail-clustering test, evaluated against a 200-shuffle permutation null with BH correction, checking whether values adjacent to gaps disproportionately fall in distribution extremes (bottom/top 10\%). This aims to capture missingness that could be dependent on the unobserved value itself.

\subsubsection{Verdict assignment}
A single verdict is assigned via a priority ladder: unavailable MCAR test → "Undetermined (test unavailable)"; MCAR not rejected with adequate power → "MCAR (not rejected)"; MCAR not rejected but underpowered → "Undetermined (low power)". If MCAR is rejected, the pipeline escalates through tail-clustering, combined MAR+MNAR, long-run MNAR, MAR, and finally co-missingness. Crucially, these mechanisms cannot be identified from data alone. Distinguishing MAR from MNAR requires knowledge of why values are missing. We use "Undetermined" liberally rather than over-claiming a mechanism the data cannot adjudicate.

\subsection{Simulated Missingness Construction}
\label{app:simulatedconstruct}

This appendix provides the full construction procedure underlying the four simulated missingness mechanisms summarised in Section~\ref{sec:amputation}.

To ensure MAR and MNAR mechanisms depart from pure randomness in a controlled way, we condition on three signals. First, a weekend logical flag marks Saturday and Sunday. Second, one or two auxiliary variables are chosen as the highest-variance numeric columns or the most balanced categorical ones, paired with a flag marking rows where the auxiliary is "high" (top quartile for numeric variables, modal category otherwise). Third, per-target self-low and self-high flags mark each variable's own low values (bottom quartile) and high values (top quartile). Logical or binary columns use TRUE/FALSE or 0/1, and categorical columns distinguish rare from common categories.

For every dataset, missingness is injected at proportions of 10, 20, 30, 40, 50, 60, 70, and 80\%, each using a fixed seed for reproducibility (SEED\_MCAR = 2121, SEED\_MAR = 4242, SEED\_MNAR = 6363, SEED\_BLOCK = 8484, offset by the proportion). The four mechanisms are implemented as follows:

\textit{MCAR:} We compute the total number of cells to mask (rows × targets × proportion), then sample that many cell positions uniformly at random across the target matrix and set them to NA, so that missingness is unrelated to any values.

\textit{MAR:} We build a weight matrix in which weekend rows receive weight 3 across all targets, and rows where an auxiliary variable is "high" receive weight 2 on the other targets. These weights are scaled so the overall NA fraction reaches the target proportion (capped at 0.95 per cell), and a per-column draw then determines which cells become NA. The pipeline logs the resulting weekend-versus-weekday missing rates to confirm the intended dependence on observed values.

\textit{MNAR:} The weight rules key off each variable's own values. A row receives weight 4 where that variable falls in its low range (self-low) and weight 3 where it falls in its high range (self-high), so that extreme values are preferentially deleted, inducing missingness that depends on an artificial "unobserved" value.

\textit{Block/Gap:} These represent realistic contiguous gaps in the data (e.g., a device not being worn). Operating per user, we convert the target proportion into a number of missing days, partition these into blocks whose size scales with severity (3 days at $\leq$20\%, 5 at $\leq$40\%, 7 at $\leq$60\%, else 10), and use a rolling-sum scan to place non-overlapping blocks at random valid positions, blanking all target variables across the contiguous rows. More specifically, for each block size $b$ (processed largest-first), we compute a length-$b$ moving sum over a binary occupancy vector, select all window positions with zero sum (i.e., fully free), sample one uniformly at random, and mark it occupied before placing the next block, blanking all target variables across the resulting contiguous rows.

For each of the 8 proportions × 4 mechanisms, we write the amputated CSV (e.g., MAR\_30pct.csv) and record both the realised overall NA rate (actual versus target percentage) and per-variable NA rates. Each dataset directory thus receives 32 amputated files, 4 metadata files, and cross-dataset summaries.

\subsection{Imputation Method Implementation Details}
\label{app:imputationdetails}

This appendix provides full implementation detail for each of the six imputation methods summarised in Section~\ref{sec:impmethods}, supporting reproducibility.

\subsubsection{Window strategies and shared infrastructure}

The defining architectural principle is that every method imputes one user at a time, so parallelism occurs across users (N\_CORES=16) and never within a user. For each amputated file, the data are split by the user column, each user's rows are date-sorted, and each user is dispatched to a worker. MICE runs in pure R, whereas the deep-learning (DL) workers are called through Python. The full pipeline runs twice, governed by two window strategies. The first, "full," imputes the entire per-user, date-sorted sequence at once, providing the richest retrospective context. This strategy requires users to have at least 10 rows. The second, "win30," partitions each user into tumbling 30-day windows, imputes each window independently, and re-binds the results. Users with fewer than 30 rows are skipped and logged. The win30 strategy serves as a sensitivity and robustness check against non-stationarity. A user's per-chunk results are aggregated into an overall status of "ok," "partial," or "failed," depending on how many chunks succeeded. Notably, there is no mean or median fallback. If imputation fails for a user, that user is skipped and logged rather than crudely filled, preserving the integrity of the downstream ground-truth comparison. Before any model runs, several floors are enforced: a minimum of 10 user rows for MICE and DL methods, and a minimum of 3 observed values per column. A chunk containing no missing cells is returned unchanged.

\subsubsection{MICE}

Multiple Imputation by Chained Equations (MICE) \cite{mice_how} fills missing values by repeatedly predicting each incomplete variable from the others. For each variable with missing data, a model is built from observed data, and missing values are filled by drawing from that model's predictions. This cycle repeats across all variables several times, producing several different "complete" versions of the dataset for analysis. Because MICE generates several predictions per missing value rather than one, it naturally captures the uncertainty its estimates. It works best when data are Missing at Random (MAR), meaning the chance a value is missing can be explained by other observed variables, making it widely used in health and wearable studies. However, because it assumes MAR, MICE does not naturally account for time. It can be adapted using lagged variables (past values as predictors), but this raises computational cost for large datasets, and the standard version still treats each time point independently, leaving time-based patterns unmodelled, precisely one assumption this study examines.

In our implementation, MICE generates 5 complete datasets over 10 cycles, seeded at 2121 for reproducibility. A setup pass automatically selects a prediction method and predictor columns for each variable. We assign methods by column type: predictive mean matching (filling a value using an observed value from a similar data point) for numeric columns; logistic regression for true/false columns; and polytomous regression (predicting one of several categories) for multi-category columns. Each target column uses at most 20 predictors, or one-third of available rows, whichever is smaller. The five completed datasets are then combined: numeric values are averaged, and categorical values are set to the most common prediction, with true/false columns temporarily treated as categories during this step. If any value remains missing, that portion is marked failed. This mirrors the standard MICE approach of generating several completed datasets and pooling them into one.

\subsubsection{BRITS}

Bidirectional Recurrent Imputation for Time Series (BRITS) is a deep learning method using a neural network that reads sequences both forwards and backwards in time to fill missing values \cite{BRITS2018}. It treats each missing value as something to be learned and refined, using patterns from both earlier and later time points to reconstruct what is missing. BRITS should perform well under MCAR, MAR, and block missingness and has also been applied to MNAR settings. The original paper tested it on healthcare and physical-activity data, suggesting suitability for wearable data. It captures complex patterns across time and variables, but requires substantial data and is harder to interpret and more computationally demanding than MICE. To handle mixed data types, BRITS requires manual preparation: logical and categorical variables are coerced to factors or numeric values.  

In our implementation, two recurrent network units (one reading forward, one backward) each maintain an internal "memory" that fades at a rate learned by the model, based on the time gap since the last observation. At each step, the model makes two estimates (from recent history and from other variables) and blends them using a learned weighting. Originally observed values are always kept as-is. The model is trained to minimise errors on observed values across both estimates, while encouraging the forward and backward readings to agree.

\subsubsection{GRU-D}

GRU-D extends a standard recurrent neural network with built-in decay mechanisms modelling how the influence of past observations fades as time gaps grow \cite{gru_d}. As with BRITS, mixed data types require the same manual preparation. It fills each missing value using a weighted mix of the last observed value and the overall column average, combined with a learned temporal pattern and decay rates for both inputs and memory, accounting for irregular timing. GRU-D explicitly models time gaps through masking, and is described as an "effective solution to characterise the missing patterns of not missing-completely-at-random time series data with masking and time interval" \cite{gru_d}. While the original paper does not explicitly test it where missingness depends on the unobserved value itself, it treats missingness patterns as meaningful and learns decay rates from data rather than fixing them in advance \cite{gru_d}. Our implementation uses a single forward-reading network with two decay mechanisms, one pulling missing inputs toward the column average, and one decaying the model's memory. A final linear model predicts values from this decayed memory, trained only on observed values.

\subsubsection{WGAIN-VAE}

Our adversarial imputer builds on GAIN, which uses two competing neural networks: a generator that fills missing values, and a discriminator that tries to detect which values were originally missing \cite{GAIN2018, gain2024}. This competition, guided by a "hint" mechanism giving partial information about which values were missing, pushes the generator toward realistic imputations. The same manual preparation of categorical variables applies. The original GAIN paper found it outperforms MICE under MNAR conditions \cite{GAIN2018, gain2024}. However, standard GAIN is not designed for temporal modelling and may underperform time-aware models under MNAR especially.

Our implementation combines a Wasserstein-GAN (a more stable training variant) with a variational autoencoder (VAE, a model that learns compressed data representations) \cite{gain2024}. The generator reads sequences using a recurrent VAE, paired with a bidirectional critic network trained using a hint mask, a gradient penalty for stability, and three critic updates per generator update. Data are rescaled to 0–1 per column, the critic step is skipped for very short sequences (fewer than 8 points), and at prediction time the model averages 10 stochastic samples with the deterministic pass.

\subsubsection{PAIN}

PAIN is a recent method designed to adapt dynamically to different data types, distributions, and missingness patterns \cite{PAIN2025}. It uses a three-step process: (a) combining statistical methods, random forests (an ensemble of decision trees), and autoencoders (neural networks learning compressed data representations); (b) adaptively weighting these methods based on data characteristics; and (c) refining imputed values to reduce noise. PAIN supports mixed data types and shows promise for health data, though further research is needed.

Our implementation is layered and each layer completion is tracked. The first layer blends column mean, median, and nearest-neighbour KNN imputation, weighted by missingness severity. The second uses an iterative random forest, imputing one column at a time from least-to-most missing, for up to three passes. A de-noising autoencoder then refines and averages with this output. The final layer clips values to a realistic range (based on observed 5th–95th percentiles, extended for natural spread) and rounds whole-number columns.

\subsubsection{GRIN}

GRIN combines a recurrent network for time patterns with a graph neural network (a model learning relationships between variables as a network) for variable relationships, learning representations via bidirectional message passing \cite{GRIN2022}. This enables imputation in datasets with multiple interdependent variables and shows strong potential for wearable health data. Although GRIN makes no assumptions about the distribution of missing values, it assumes underlying patterns remain stable over time \cite{GRIN2022}, narrowing its effective use toward MAR scenarios.

Our implementation learns a network of relationships between variables, passing information using the model's best current estimate rather than placeholder zeros, so artificial zeros never distort learned relationships. A recurrent unit carries information through time, and the final prediction combines this temporal information with variable-relationship information.

\subsubsection{Shared Training Infrastructure and Post-Processing}

All models are organised into per-user batches. The three recurrent models (BRITS, GRU-D, GRIN) share further common infrastructure: time gaps are calculated in both directions, and training uses gradient clipping (preventing unstable updates), a 300-second time limit, and early stopping after 10 attempts without improvement. A common post-processing routine applies to all deep learning methods: numeric values are standardised before processing and converted back afterward, and results are constrained within a reasonable range of observed data. Together, these conventions ensure all six methods are evaluated under a uniform, reproducible protocol, isolating differences in imputation quality to the methods themselves.

\section{Evaluation Metric Derivations}
\label{app:metricevaluation}

\subsection{Point-level reconstruction accuracy}
\label{app:point-accuracy_eq}

\textbf{NRMSE.} To account for survivorship bias, we propose a coverage-penalised NRMSE (cpNRMSE). Standard NRMSE is averaged only over users actually evaluated; this silently drops users with shorter series or sparse columns where the method fails, which can produce a deceptively low NRMSE. cpNRMSE deflates apparent accuracy by the coverage achieved:

\begin{equation}
    \text{coverage} = \frac{\text{n\_users\_eval}}{\text{n\_users\_attempted}}
    \label{eq:coverage}
\end{equation}

\begin{equation}
    \text{cpNRMSE} = \frac{\text{NRMSE}}{\text{coverage}}
    \label{eq:cpnrmse}
\end{equation}

cpNRMSE was preferred over a weighted F-score or a $1-\text{NRMSE}$ recasting to preserve interpretability and avoid unnecessary assumptions. At full coverage, cpNRMSE = NRMSE, remaining on the same scale and directly comparable. It diverges only as coverage drops; at coverage = 0, both NRMSE and cpNRMSE are coded NA. cpNRMSE strictly increases as error rises and as coverage falls—the intended ranking behaviour, where low error on everyone outranks low error on an easy subset. Dividing by coverage is the natural first-order selection-bias correction and requires no arbitrary rescaling into $(0,1)$, which an F-style harmonic mean would force.

For each missingness-percentage $\times$ method combination, each dataset's contribution to the pooled failure rate is:

\begin{equation}
    \text{contribution}_d = 100 \times n_d^{\text{failed}} \Big/ \sum_d n_d^{\text{attempted}}
    \label{eq:contribution}
\end{equation}

Using a common denominator means segments stack to a value with real interpretation (the overall pooled failure rate), and larger datasets contribute proportionally to how many users they represent.

Pooled NRMSE is the coverage-weighted (by $n_d^{\text{eval}}$) mean of per-dataset NRMSEs—standard sample-size weighting for combining group means—and pooled coverage is total evaluated $\div$ total attempted:

\begin{equation}
    \text{cpNRMSE}^{\text{pool}}
    = \frac{\text{NRMSE}^{\text{pool}}}{\bar{c}}
    = \frac{\sum_d n_d^{\text{eval}}\, \text{NRMSE}_d}{\sum_d n_d^{\text{eval}}}
      \Bigg/
      \frac{\sum_d n_d^{\text{eval}}}{\sum_d n_d^{\text{att}}}
    \label{eq:cpnrmse-pool}
\end{equation}

which simplifies to:

\begin{equation}
    \text{cpNRMSE}^{\text{pool}}
    = \frac{\sum_d n_d^{\text{att}} \sum_d n_d^{\text{eval}}\, \text{NRMSE}_d}
           {\left( \sum_d n_d^{\text{eval}} \right)^{2}}
    \label{eq:cpnrmse-compact}
\end{equation}

where, summing over datasets $d$:
\begin{itemize}
    \item $n_d^{\text{att}}$ = users attempted,
    \item $n_d^{\text{eval}}$ = users evaluated,
    \item $\text{NRMSE}_d$ = that dataset's mean NRMSE.
\end{itemize}

This preserves partial-missingness information that a naive collapse would discard, so a pooled point resting on 1-of-5 datasets is visibly less trustworthy than one resting on 5-of-5.

\textbf{MAE.} MAE carries the identical scale-dependence problem as RMSE, so we normalise and coverage-penalise identically:

\begin{equation}
    \text{NMAE} = \frac{\text{MAE}}{\text{range}}
    \label{eq:nmae}
\end{equation}

\begin{equation}
    \text{cpNMAE} = \frac{\text{NMAE}}{\text{coverage}}
    \label{eq:cpnmae}
\end{equation}

where range is RMSE divided by NRMSE. NMAE mirrors the NRMSE definition, landing on the same dimensionless scale and feeding the same cross-dataset and window-comparison rollups.

\textbf{Bias.} Raw bias carries the same survivorship bias as NRMSE, so it is averaged only over successfully evaluated users, and a coverage-penalised Bias is calculated analogously. Bias's optimum is 0, it is unbounded, and symmetric about that optimum; dividing the signed value by coverage $<1$ therefore always pushes it away from 0 (worse), regardless of sign—the same clean behaviour cpNRMSE relies on, requiring no $1-x$ recast.

We first normalise bias to the same range used for NRMSE:

\begin{equation}
    \text{range} = \frac{\text{RMSE}}{\text{NRMSE}}
    \quad \Rightarrow \quad
    \text{nbias} = \frac{\text{bias}}{\text{range}}
    = \text{bias} \times \frac{\text{NRMSE}}{\text{RMSE}}
    \label{eq:nbias}
\end{equation}

This places bias on the same normalised scale as NRMSE and NMAE (0 = unbiased; sign = direction). The same coverage term then yields:

\begin{equation}
    \text{cpNBias} = \frac{\text{nbias}}{\text{coverage}}
    \label{eq:cpnbias}
\end{equation}

\subsection{Preserving multivariate (between-variable) structure}
\label{app:multivariate_eq}

\textbf{cpCorr.} Correlation is calculated per user $\times$ variable as the Pearson correlation between complete (ground-truth) and imputed values, computed only over masked cells (the same cells scored by RMSE). Because this is not a correlation over the whole series, correlation inherits the same survivorship bias as NRMSE: a method that quietly drops hard users can post a flattering correlation. Pearson correlation is additionally undefined whenever either the true or imputed vector of masked values is constant within a user $\times$ variable cell—a failure mode not captured by coverage alone, requiring a separate correlation-coverage term:

\begin{equation}
    \text{corr\_definedness} = \frac{n_{corrDefined}}{n_{corrRows}}
    \label{eq:corrdefine}
\end{equation}

\begin{equation}
    \text{corr\_coverage} = \text{coverage} \times \text{corr\_definedness}
    \label{eq:corrcoverage}
\end{equation}

The two terms are multiplied because they are sequential, independent filters on the same pool of cells: a cell must first survive whole-user attrition (coverage), and among surviving cells, some further fraction still fail to yield a defined correlation (corr\_definedness).

Because correlation is bounded in $[-1,1]$, signed, and inverted relative to NRMSE (higher is better), dividing correlation directly by coverage $<1$ does not consistently move it toward "worse." We therefore recast:

\begin{equation}
    \text{cpCorr} = \frac{1-\text{corr}}{\text{corr\_coverage}}
    \label{eq:cpcorr}
\end{equation}

This recast retains the same properties as cpNRMSE: at full coverage, cpCorr = corr, and cpCorr strictly worsens as corr falls or coverage falls, regardless of corr's sign.

Pooled cpCorr follows the same principles as pooled cpNRMSE:

\begin{align}
    \text{corr}^{\text{pool}}
        &= \frac{\sum_d n_d^{\text{eval}}\, \text{corr}_d}
                {\sum_d n_d^{\text{eval}}} \label{eq:corr-pool} \\[10pt]
    \text{coverage}^{\text{pool}}
        &= \frac{\sum_d n_d^{\text{eval}}}{\sum_d n_d^{\text{att}}},
        \qquad
    \text{corr\_definedness}^{\text{pool}}
        = \frac{\sum_d n_d^{\text{corr\_defined}}}
                {\sum_d n_d^{\text{corr\_rows}}} \label{eq:corr-def} \\[10pt]
    \text{corr\_coverage}^{\text{pool}}
        &= \text{coverage}^{\text{pool}} \times
           \text{corr\_definedness}^{\text{pool}} \label{eq:corr-cov} \\[10pt]
    \text{cpCorr}^{\text{pool}}
        &= 1 - \frac{1 - \text{corr}^{\text{pool}}}
                    {\text{corr\_coverage}^{\text{pool}}} \label{eq:cpcorrpool}
\end{align}

where, summing over datasets $d$:
\begin{itemize}
    \item $n_d^{\text{att}}$ = users attempted, $n_d^{\text{eval}}$ = users evaluated—the same counts cpNRMSE's pooling uses;
    \item $n_d^{\text{corr\_rows}}$ = (user $\times$ variable) cells attempted for correlation, $n_d^{\text{corr\_defined}}$ = how many yielded a defined correlation;
    \item $\text{corr}_d$ = that dataset's mean correlation, weighted by $n_d^{\text{eval}}$ in the pooled mean (same weighting as NRMSE\_pooled).
\end{itemize}

Pooling corr\_definedness as a ratio of summed counts (rather than averaging per-dataset fractions) means a dataset contributing only a handful of evaluable cells does not receive the same vote as one contributing thousands.

\textbf{cpFrob.} For each user, we compare the full and partial correlation matrix of the complete data against the same matrix computed on reconstructed data, across every numeric/ordinal target variable jointly, using the Frobenius norm of the difference:

\begin{equation}
    \text{corr\_matrix\_frob}
    = \left\| R^{\text{true}} - R^{\text{imp}} \right\|_F
    = \sqrt{\sum_{i=1}^{p} \sum_{j=1}^{p}
            \left( R^{\text{true}}_{ij} - R^{\text{imp}}_{ij} \right)^{2}}
    \label{eq:corr-frob}
\end{equation}

where $R^{\text{true}}$ and $R^{\text{imp}}$ are the $p \times p$ Pearson correlation matrices of the complete and imputed target variables. A value of 0 means every pairwise relationship was reproduced exactly; the score grows with the number and size of distorted pairs.

\begin{equation}
    \text{partial\_corr\_frob}
    = \left\| P^{\text{true}} - P^{\text{imp}} \right\|_F
    = \sqrt{\sum_{i=1}^{p} \sum_{j=1}^{p}
            \left( P^{\text{true}}_{ij} - P^{\text{imp}}_{ij} \right)^{2}}
    \label{eq:pcorr-frob}
\end{equation}

where $P^{\text{true}}$ and $P^{\text{imp}}$ are partial-correlation matrices, whose off-diagonal entry $P_{ij}$ is the correlation between variables $i$ and $j$ after conditioning on all remaining variables. This is the sharper test: two variables may still appear related in Eq.~\eqref{eq:corr-frob} simply because both depend on a third, well-reconstructed variable, whereas partial correlation isolates the direct relationship.

corr\_matrix\_frob and partial\_corr\_frob are already proper losses (nonnegative, zero at perfect reconstruction, unbounded above, lower-is-better)—the same shape as NRMSE—so, unlike cpCorr, they can be deflated directly:

\begin{align}
    \text{rel\_definedness}
        &= \frac{n_{\text{rel\_defined}}}{n_{\text{rel\_rows}}}
        \label{eq:rel-def} \\[10pt]
    \text{rel\_coverage}
        &= \text{coverage} \times \text{rel\_definedness}
        \label{eq:rel-cov} \\[10pt]
    \text{cpFrob}
        &= \frac{\text{corr\_matrix\_frob}}{\text{rel\_coverage}},
        \qquad
    \text{cpPartialFrob}
        = \frac{\text{partial\_corr\_frob}}{\text{rel\_coverage}}
        \label{eq:cpfrob}
\end{align}

rel\_definedness (Eq.~\eqref{eq:rel-def}) is, of users for which a relationship row exists, the fraction whose row yielded a finite Frobenius norm. It exists for the same reason corr\_definedness did: a relationship row can return NA even for a successfully imputed user—either the cell had fewer than two numeric/ordinal target variables from which to build a matrix, or the correlation/partial-correlation matrix came out non-finite (e.g., near-singular or near-constant columns). This failure mode is invisible to coverage alone, necessitating the additional definedness term.

Unlike cpCorr, these are computed over each user's entire series with observed cells copied through unchanged; only masked cells differ between true and imputed matrices. At low missingness targets, most of both matrices is therefore literally identical data, so a small Frobenius norm reflects "most of the series was untouched," not "structure was preserved." These metrics should be read only at the high missingness end (50–80\%), where a real share of each matrix is genuinely reconstructed.

\subsection{Preserving temporal (within-variable, over-time) structure}
\label{app:temporal_eq}

We use three full-series temporal metrics, computed on each user's entire per-user series (observed cells copied through, only masked cells reconstructed) rather than on scattered masked cells alone, because autocorrelation, differencing, and spectral content are not well-defined on a random subset of timepoints. Each metric is computed at the granularity of one row per user $\times$ variable and then pooled. For a given user $u$ and variable $v$, let $x=(x_1,\ldots,x_T)$ denote the true series and $\hat{x}=(\hat{x}_1,\ldots,\hat{x}_T)$ the reconstruction over $T$ observed time steps. For all three metrics, lower is better (0 denotes perfectly preserved temporal structure).

\textbf{Autocorrelation error.} This directly tests persistence: does a high day tend to be followed by a high day in the reconstruction, as in reality? We compute the lag-$k$ autocorrelation $\rho_k$ of each series and take the mean absolute error over lags 1–3:

\begin{equation}
\label{eq:acf}
\text{acf\_lag1\_3\_err}_{u,v}
= \frac{1}{3}\sum_{k=1}^{3}\bigl|\,\rho_k(x) - \rho_k(\hat{x})\,\bigr|,
\qquad
\rho_k(x) = \frac{\sum_{t=k+1}^{T}(x_t-\bar{x})(x_{t-k}-\bar{x})}
                 {\sum_{t=1}^{T}(x_t-\bar{x})^2}.
\end{equation}

\textbf{Range-normalised first-difference RMSE.} This captures dynamics/roughness: whether the reconstruction reproduces the magnitude of daily swings, or smooths them into a flat line. We take the RMSE between the first-differenced true and reconstructed series, where $\Delta x_t = x_t - x_{t-1}$:

\begin{equation}
\label{eq:firstdiff-raw}
\text{firstdiff\_rmse}^{\text{raw}}_{u,v}
= \sqrt{\frac{1}{T-1}\sum_{t=2}^{T}\bigl(\Delta x_t - \Delta \hat{x}_t\bigr)^2}.
\end{equation}

Since this raw quantity is expressed in each variable's native units, it cannot be pooled across heterogeneous variables and datasets. We therefore apply the same range-recovery normalisation used for cpNBias and cpNMAE, rescaling by the ratio of the normalised to the raw RMSE of the level series, equivalently, dividing by the range $R_{u,v} = \max(x) - \min(x)$:

\begin{equation}
\label{eq:firstdiff-norm}
\text{firstdiff\_rmse}_{u,v}
= \text{firstdiff\_rmse}^{\text{raw}}_{u,v}
  \times \frac{\text{NRMSE}_{u,v}}{\text{RMSE}_{u,v}}
= \frac{\text{firstdiff\_rmse}^{\text{raw}}_{u,v}}{R_{u,v}}.
\end{equation}

This yields a unit-free quantity that is directly comparable across all variables and datasets.

\textbf{Spectral difference.} This captures rhythm: whether weekly or otherwise periodic structure survives, or is washed out. Let $P_x(f)$ and $P_{\hat{x}}(f)$ be the power spectral densities of the true and reconstructed series across frequencies $f$. We quantify the discrepancy in spectral content as the total absolute difference between the two spectra after normalising each to unit total power:

\begin{equation}
\label{eq:spectral}
\text{spectral\_diff}_{u,v}
= \sum_{f}\bigl|\,\tilde{P}_x(f) - \tilde{P}_{\hat{x}}(f)\,\bigr|,
\qquad
\tilde{P}_x(f) = \frac{P_x(f)}{\sum_{f'} P_x(f')}.
\end{equation}

Normalising each spectrum to unit total power isolates shape—which rhythms are present, and in what relative proportion—from overall signal energy.

\textbf{Pooling.} Consistent with our other metrics, each quantity in Equations~\eqref{eq:acf}, \eqref{eq:firstdiff-norm}, and~\eqref{eq:spectral} is computed per user $\times$ variable and then aggregated using the coverage-weighted pooling scheme described in Appendix~\ref{app:point-accuracy_eq}, so that datasets and variables with more evaluable cells contribute proportionally and the reported aggregates remain survivorship-aware. First differences in Equation~\eqref{eq:firstdiff-raw} are taken over consecutive observed time steps, consistent with the missingness mask defined in Section~\ref{sec:amputation}.

\subsection{Preserving distributional shape}
\label{app:distributional_eq}

The temporal metrics above test whether the series moves through time correctly (persistence, dynamics, rhythm), since in a time series the ordering of observations is itself informative. These are computed on the full reconstructed series and therefore carry a low-missingness dilution caveat: at small missingness targets the metric is dominated by untouched observed cells, so it is only trustworthy at the 50–80\% end.

The distributional metrics introduced here are complementary in two deliberate ways. First, they are order-agnostic: they interrogate the marginal distribution (the spread and shape of the reconstructed values) independently of temporal ordering. Second, they are masked-cell-only, evaluated exclusively on held-out targets (like cpCorr), and therefore free of the dilution caveat—diagnostic at every missingness level. Since a method can preserve the marginal distribution perfectly while scrambling temporal order (or vice versa), both families are required to separate these two failure modes.

For a given user $u$ and variable $v$, let $M$ denote the set of masked cells, $x=\{x_i : i\in M\}$ the true masked values, and $\hat{x}=\{\hat{x}_i : i\in M\}$ the corresponding imputed values.

\textbf{Variance ratio.} This directly measures whether the method reproduced the spread of the masked values—precisely the quantity to which NRMSE is blind, since a method that regresses every target toward the mean scores well on MAE/NRMSE while destroying the distribution. It is defined as the ratio of imputed to true variance over the masked cells:

\begin{equation}
    \text{var\_ratio}_{u,v} = \frac{\operatorname{Var}(\hat{x})}{\operatorname{Var}(x)}
= \frac{\tfrac{1}{|M|}\sum_{i \in M}(\hat{x}_i - \bar{\hat{x}})^2}
       {\tfrac{1}{|M|}\sum_{i \in M}(x_i - \bar{x})^2}.
\end{equation}

A value of 1 denotes perfectly reproduced spread; values below 1 indicate variance collapse (over-smoothing), and values above 1 indicate inflated variability. This metric exposes the regress-to-the-mean trap directly, which is why we additionally examine it against NRMSE. Variance ratio is deliberately not coverage-penalised: the central optimum and near-zero collapse make any log-deflation blow up, so it is kept on a raw log scale with an ideal line at 1.

\textbf{Kolmogorov-Smirnov statistic.} The KS statistic is the maximum absolute gap between the two empirical cumulative distribution functions (CDFs), catching whole-distribution mismatch that variance ratio alone would miss:

\begin{equation}
    \text{KS\_stat}_{u,v}
= \sup_{z}\bigl|\,F_x(z) - F_{\hat{x}}(z)\,\bigr|,
\end{equation}

where $F_x$ and $F_{\hat{x}}$ are the empirical CDFs of the true and imputed masked values. It ranges over $[0,1]$, with 0 denoting identical distributions.

\textbf{1-Wasserstein distance.} Where KS reports a single worst-point discrepancy on the CDF, the 1-Wasserstein (earth-mover's) distance integrates the entire displacement, capturing both how far and how much of the distribution must move. For one-dimensional distributions this equals the $L^1$ area between the two CDFs:

\begin{equation}
    \text{Wasserstein1}_{u,v}
= \int_{-\infty}^{\infty}\bigl|\,F_x(z) - F_{\hat{x}}(z)\,\bigr|\, dz.
\end{equation}

This complements the KS statistic: a small KS can coexist with a large Wasserstein distance when the mismatch is spread across the whole range rather than concentrated at a single point. Unlike the unit-free KS statistic, Wasserstein distance is expressed in the variable's own units. For pooling across heterogeneous variables and datasets we apply the same range normalisation used elsewhere, dividing by the target range $R_{u,v} = \max(x) - \min(x)$.

\textbf{cpKS and cpWasserstein1.} Let $c_{u,v} \in (0,1]$ denote the coverage for user $u$ and variable $v$ (the fraction of masked target cells for which the method produced a valid imputation), with base distances computed over those evaluable masked cells. The coverage penalty divides the raw distance by $c_{u,v}$, so that a method reconstructing only a subset of the masked distribution is penalised in proportion to what it left unimputed:

\begin{equation}
\label{eq:cpks}
    \text{cpKS}_{u,v}
= \frac{1}{c_{u,v}}\,\sup_{z}\bigl|\,F_x(z) - F_{\hat{x}}(z)\,\bigr|,
\end{equation}

\begin{equation}
\label{eq:cpwasserstein1}
    \text{cpWasserstein1}_{u,v}
= \frac{1}{c_{u,v}}\cdot\frac{1}{R_{u,v}}
  \int_{-\infty}^{\infty}\bigl|\,F_x(z) - F_{\hat{x}}(z)\,\bigr|\, dz,
\end{equation}

where $F_x$ and $F_{\hat{x}}$ are the empirical CDFs of the true and imputed masked values, and $R_{u,v} = \max(x) - \min(x)$ is the target range applied to render cpWasserstein1 unit-free and poolable across variables. Aggregation across datasets uses the same coverage-weighted pooling scheme as our other metrics (Appendix~\ref{app:point-accuracy_eq}).

\newpage

\section{Supplementary Imputation Accuracy Results}

\subsection{Fail Counts by Method and Missingness Percentage}
\label{app:failcounts}

\begin{figure}[htbp]
    \centering
    \includegraphics[width=1\textwidth]{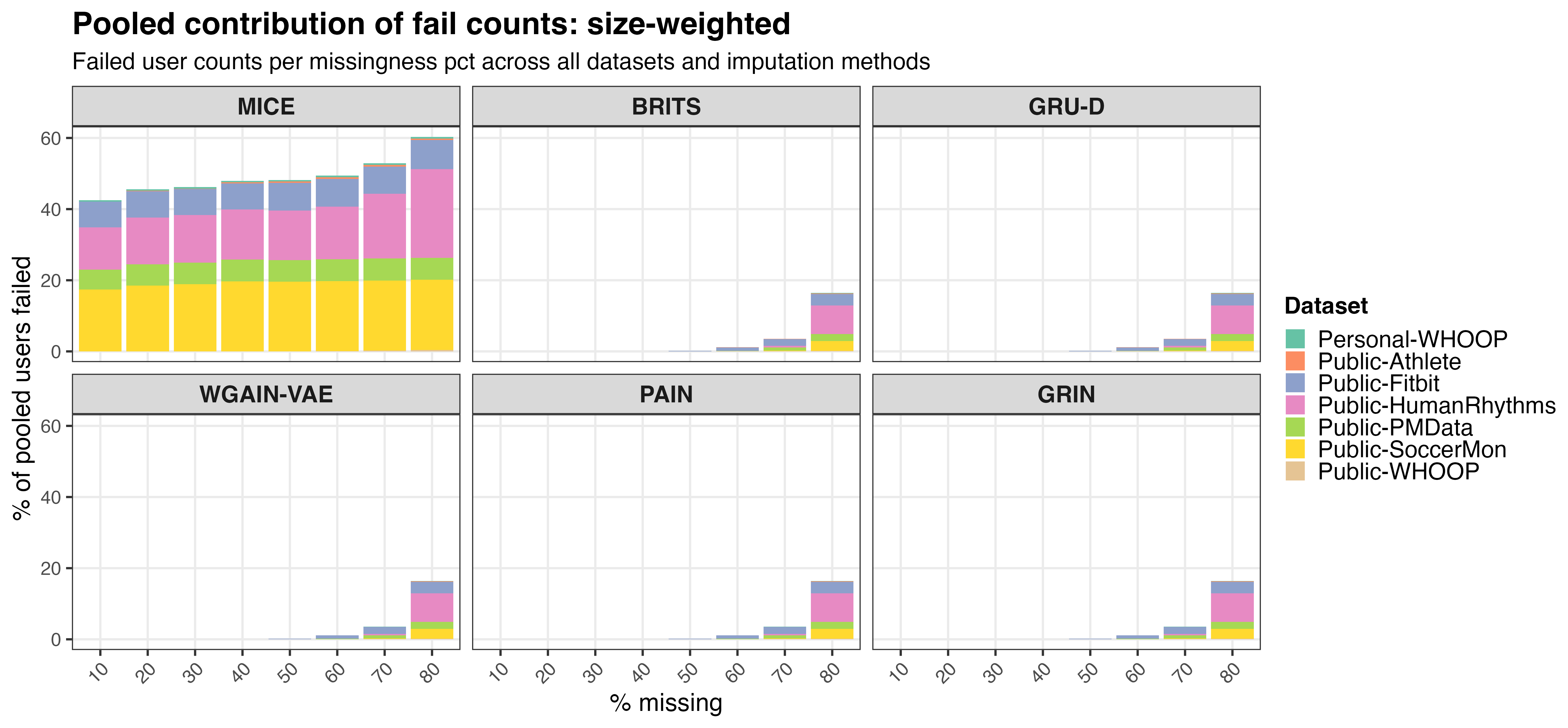} 
    \caption{Failed user counts per missingness percentage across all datasets and imputation methods.}
    \label{fig:failcountplot}
\end{figure}

Figure~\ref{fig:failcountplot} reports the number of users for which each method failed to produce a valid, evaluable reconstruction, broken down by missingness percentage across all datasets and imputation methods. This supports the coverage-penalisation rationale developed in Appendix~\ref{app:point-accuracy_eq}. A method with a low raw error but a high fail count is misrepresenting its reliability unless deflated by coverage.

\FloatBarrier

\subsection{Supplementary cpNRMSE Results}
\label{app:cpnrmse-supplementary}

This appendix provides supplementary cpNRMSE results supporting Section~\ref{sec:cpnrmse} of the main text: per-dataset (non-pooled) trends across all mechanisms and window strategies, the full-range view alongside its zoomed counterpart, and the rank-stability analysis underlying the main text's claims about PAIN's dominance.

\begin{figure}[t]
    \centering
    \includegraphics[width=1\textwidth]{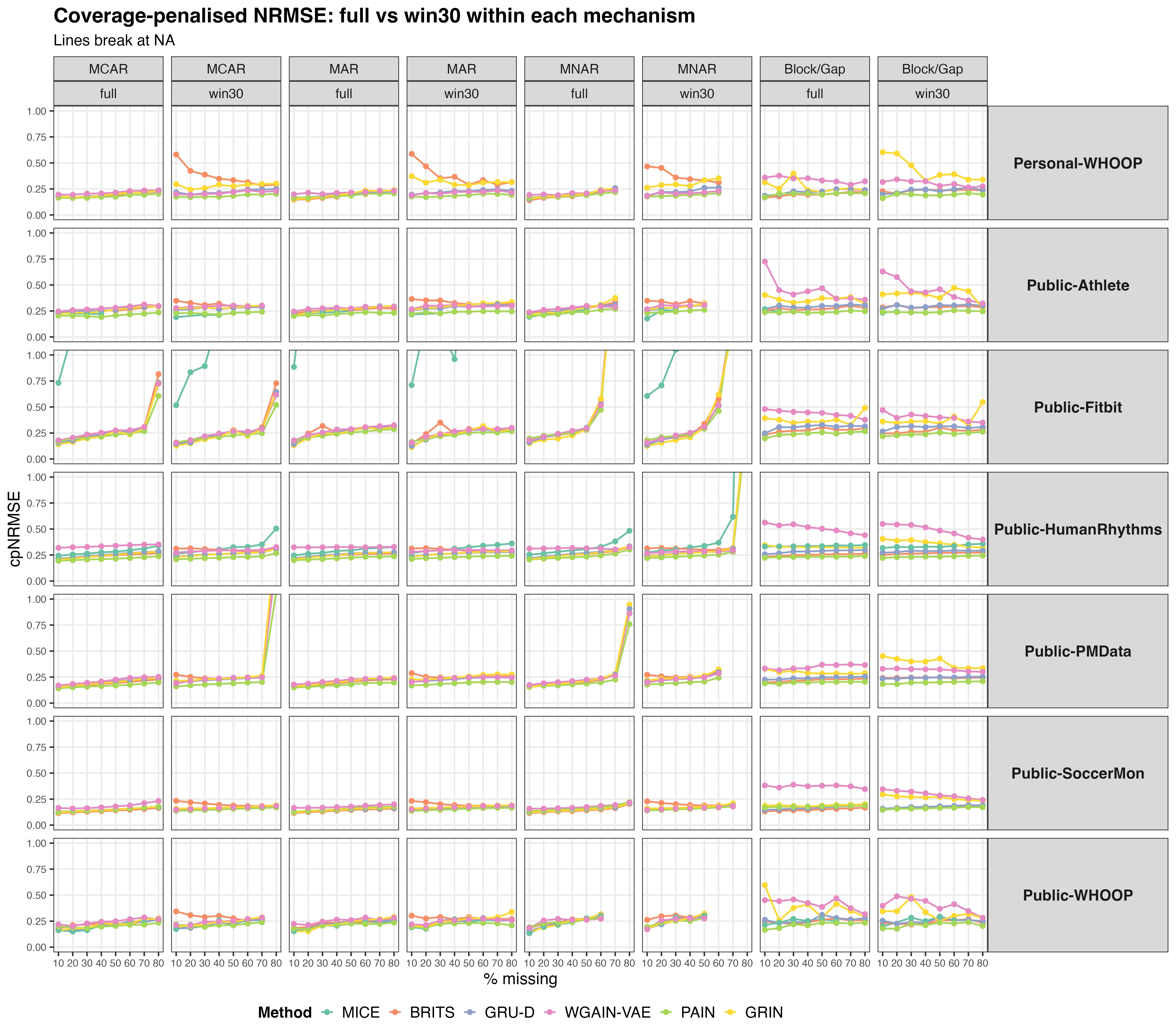} 
    \caption{cpNRMSE per dataset (clipped y-axis [0,1]), across missingness mechanisms and window strategies.}
    \label{fig:cpNRMSEzoom}
\end{figure}

The per-dataset, non-pooled view shown in Figure~\ref{fig:cpNRMSEzoom} complements the pooled Figure~\ref{fig:cpNRMSEzoompooled} in the main text by showing where each dataset sits relative to the overall trend, rather than collapsing all seven into a single line.

\begin{figure}[t]
    \centering
    \includegraphics[width=1\textwidth]{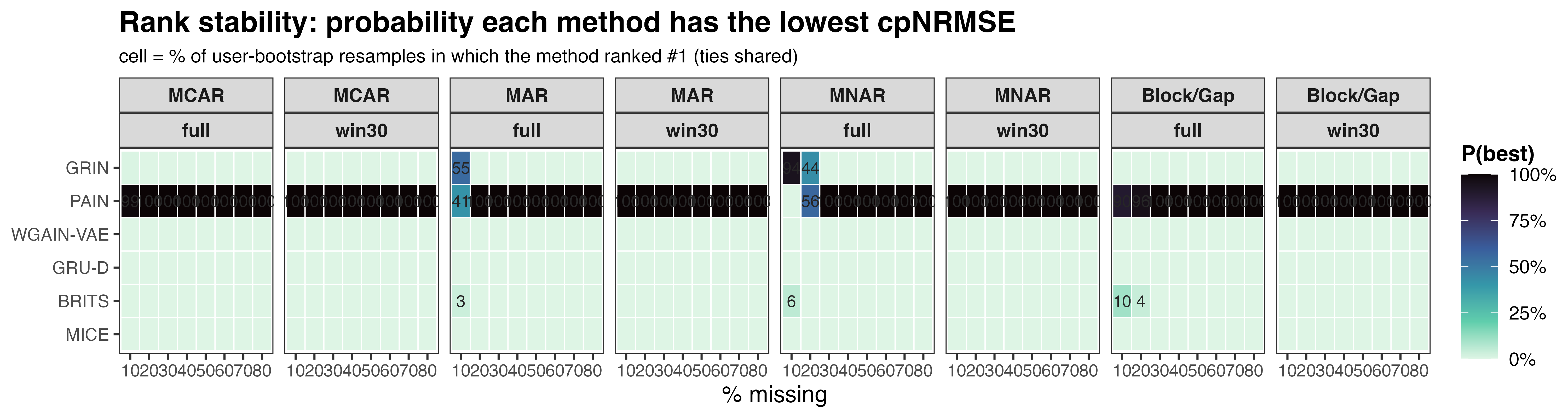} 
    \caption{Rank stability: probability that each method achieves the lowest cpNRMSE, estimated via user-bootstrap resampling, across missingness mechanisms, window strategies, and missingness percentages.}
    \label{fig:rankstability}
\end{figure}

Figure~\ref{fig:rankstability} is the rank-stability analysis which quantifies, via user-bootstrap resampling, how often each method achieves the lowest cpNRMSE, substantiating the main text's claim that PAIN's point-accuracy advantage is stable rather than an artefact of a single sample.

\FloatBarrier

\subsection{Supplementary cpNMAE Results}
\label{app:cpnmae}

\begin{figure}[htbp]
    \centering
    \includegraphics[width=1\textwidth]{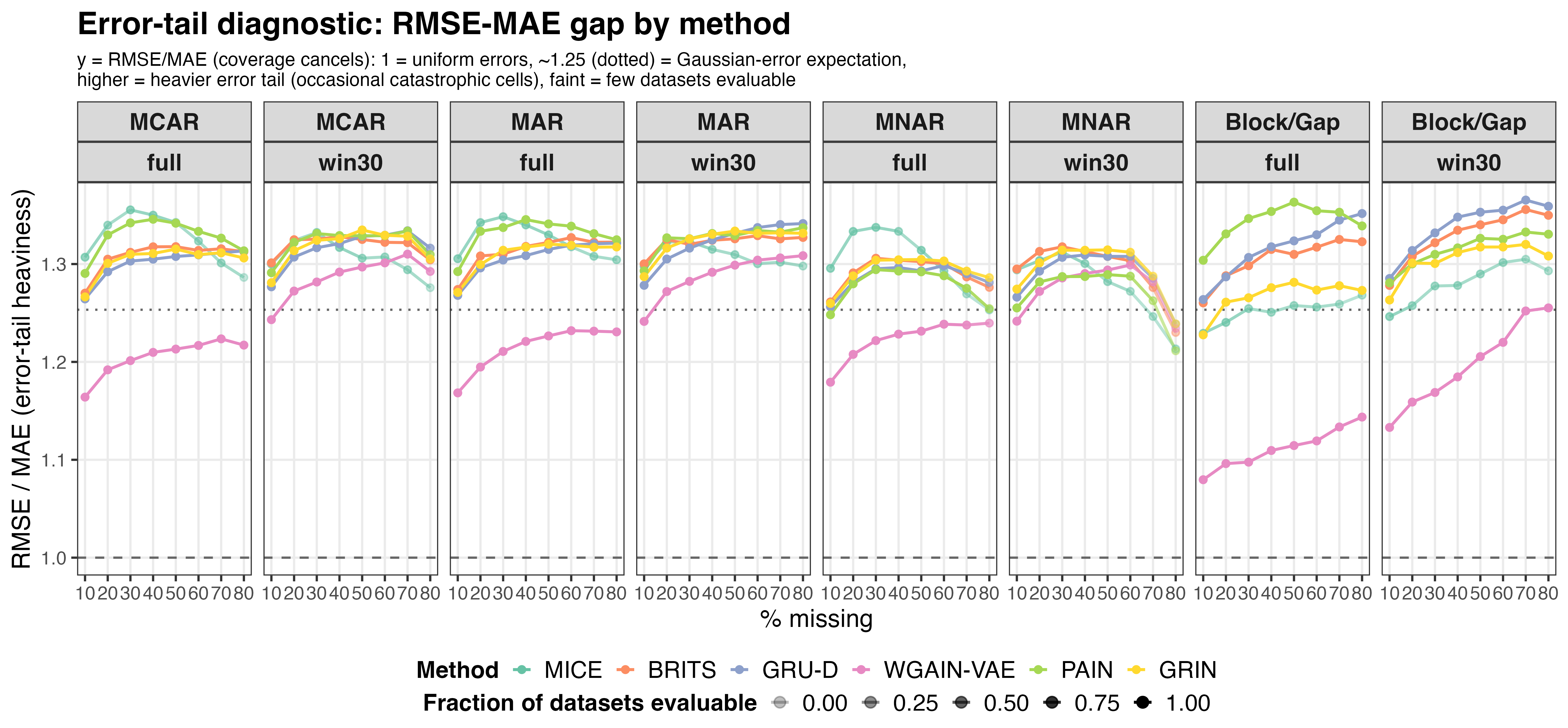} 
    \caption{Error-tail diagnostic: RMSE/MAE ratio by method, missingness mechanism, and window strategy. A ratio of 1 (dashed line) is the theoretical floor, indicating uniformly distributed errors. By Jensen's inequality, RMSE always bounds MAE from above, with equality only when every error has identical magnitude. A ratio of $\approx 1.25$ (dotted line, $\sqrt{\pi/2}$) is the expected value under Gaussian-distributed errors, making it a more realistic baseline than the strict uniform-error floor. Values climbing meaningfully above this line indicate a heavier error tail (occasional large-magnitude misses). Line opacity reflects the fraction of datasets evaluable at each missingness percentage.}
    \label{fig:rmsemaeratio}
\end{figure}

Figure~\ref{fig:rmsemaeratio} shows the RMSE/MAE ratio used as an error-tail diagnostic. Values near 1 indicate uniformly distributed errors, while higher values flag methods whose error is driven by occasional large-magnitude misses rather than consistent small deviations.

\FloatBarrier

\subsection{Supplementary cpNBias Results}
\label{app:cpnbias}

We present supplementary cpNBias results supporting the main text with figures showing the per-user signed-bias gap against PAIN, its statistical significance via paired Wilcoxon tests (BH-corrected), and the per-dataset breakdown underlying the pooled trends. 

\begin{figure}[htbp]
    \centering
    \includegraphics[width=1\textwidth]{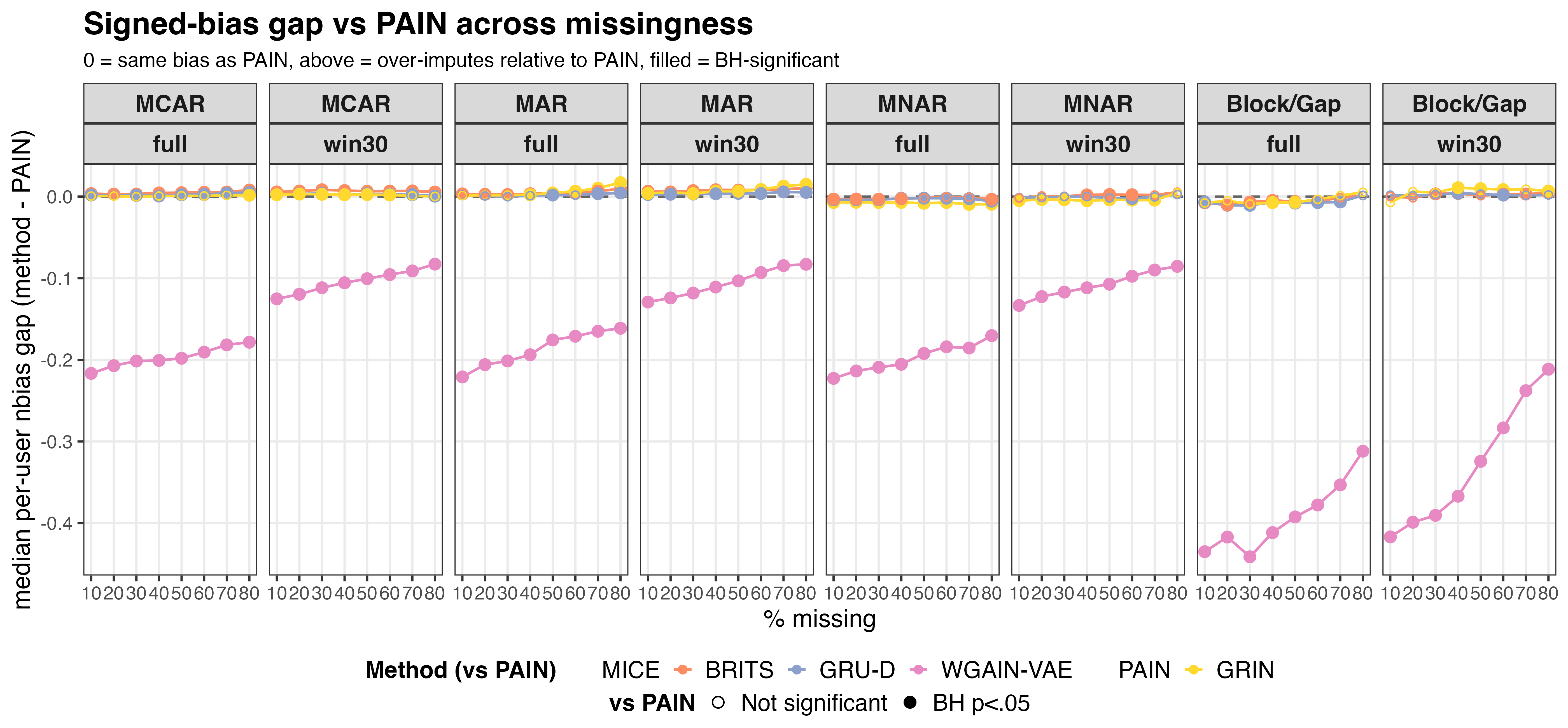} 
    \caption{Median per-user signed-bias gap (method minus PAIN) across missingness mechanisms and window strategies. Filled points denote BH-significant differences ($p<.05$); values above zero indicate the method over-imputes relative to PAIN, below zero indicates under-imputation relative to PAIN.}
    \label{fig:biasgapvspain}
\end{figure}

\begin{figure}[htbp]
    \centering
    \includegraphics[width=1\textwidth]{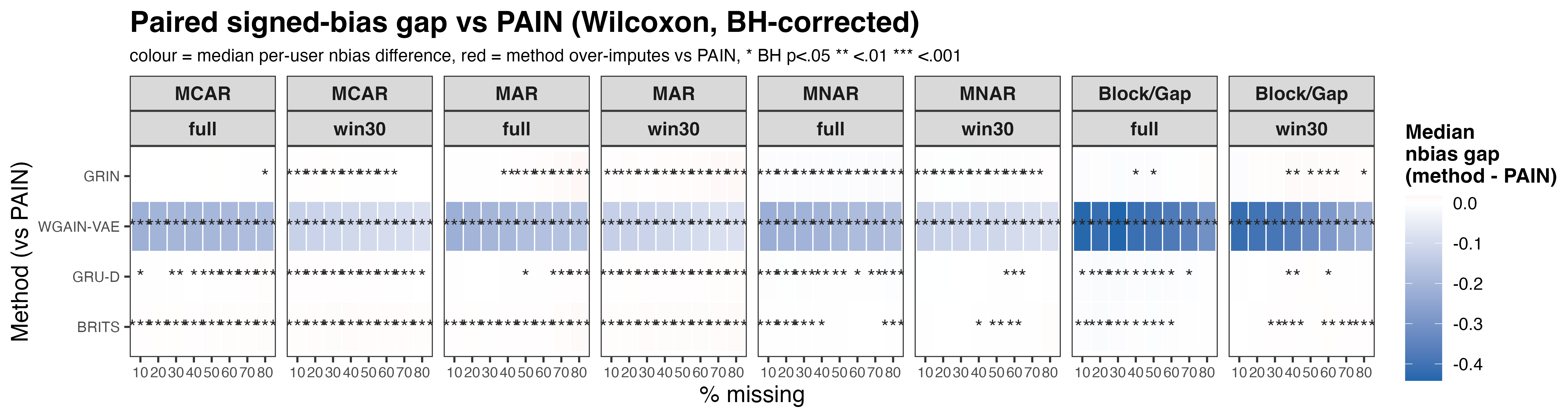} 
    \caption{Paired signed-bias gap vs. PAIN (Wilcoxon, BH-corrected). Colour indicates the median per-user nbias difference (red = over-imputes relative to PAIN, blue = under-imputes); asterisks denote significance (*BH $p<.05$, **$p<.01$, ***$p<.001$).}
    \label{fig:biasgapheatmap}
\end{figure}

\begin{figure}[htbp]
    \centering
    \includegraphics[width=1\textwidth]{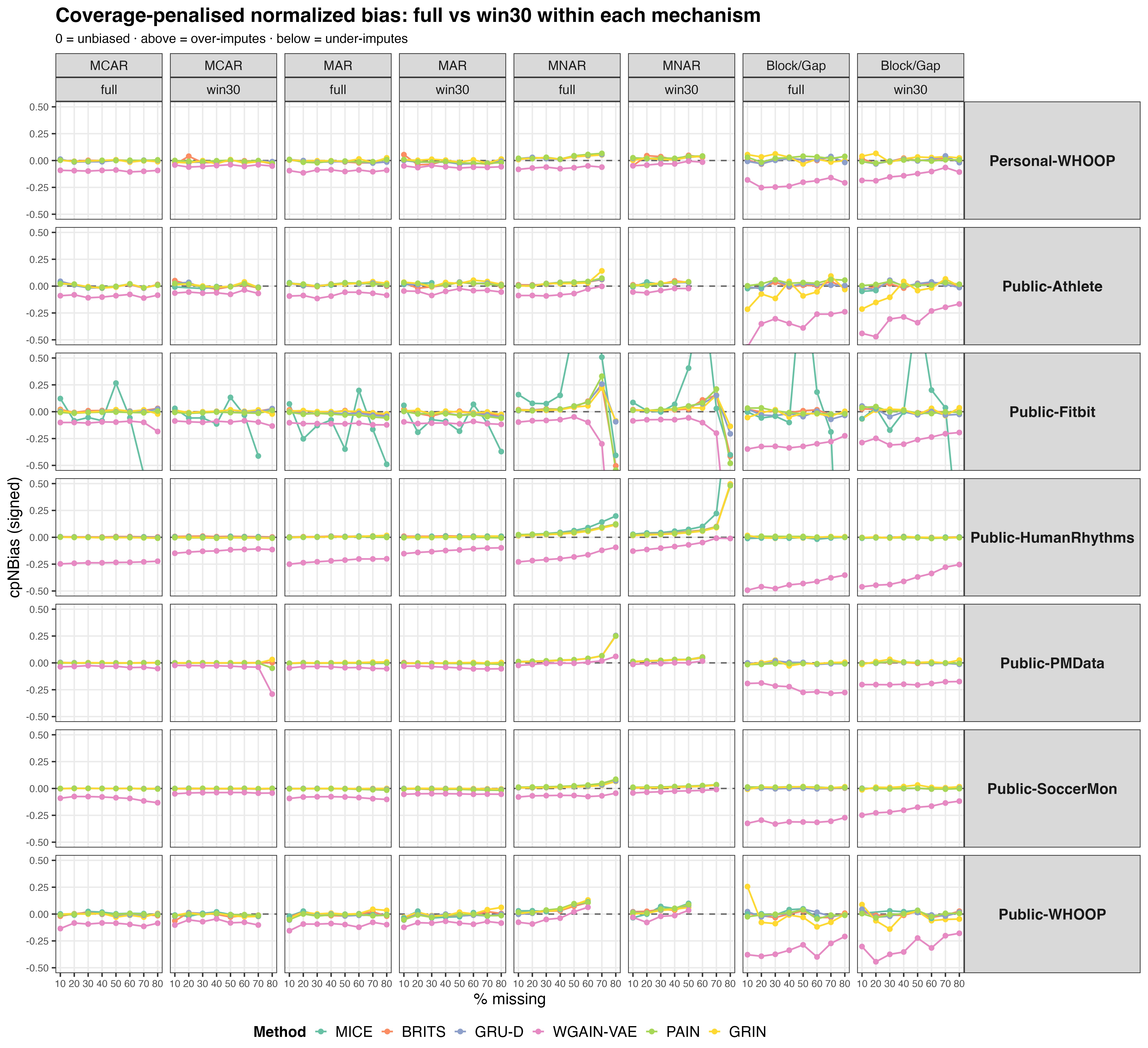} 
    \caption{Coverage-penalised normalised bias (signed), by individual dataset, missingness mechanism, and window strategy. Plot is clipped at [-0.5, 0.5].}
    \label{fig:cpnbiasdatasets}
\end{figure}

The lollipop plot below (Figure~\ref{fig:biasgapvspain}) summarises, for each method, the median per-user signed-bias gap relative to PAIN, with filled points marking statistically significant differences. The tile heatmap (Figure~\ref{fig:biasgapheatmap}) presents the same paired comparison in a complementary format, making it easier to scan significance patterns across the full grid of mechanisms and window strategies at once. Finally, the per-dataset breakdown (Figure~\ref{fig:cpnbiasdatasets}) shows how coverage-penalised signed bias varies by individual dataset, complementing the pooled trends discussed in the main text.

\FloatBarrier

\subsection{Runtime-Accuracy Trade-off: Full Detail}
\label{app:runtime}

The two figures (Figure~\ref{fig:runtimefull} and Figure~\ref{fig:runtimewin30}) below present per-user runtime plotted against per-user NRMSE, faceted by missingness mechanism and method, for the full and win30 window strategies respectively, supporting the runtime-accuracy discussion in the main text.

\begin{figure}[htbp]
    \centering
    \includegraphics[width=1\textwidth]{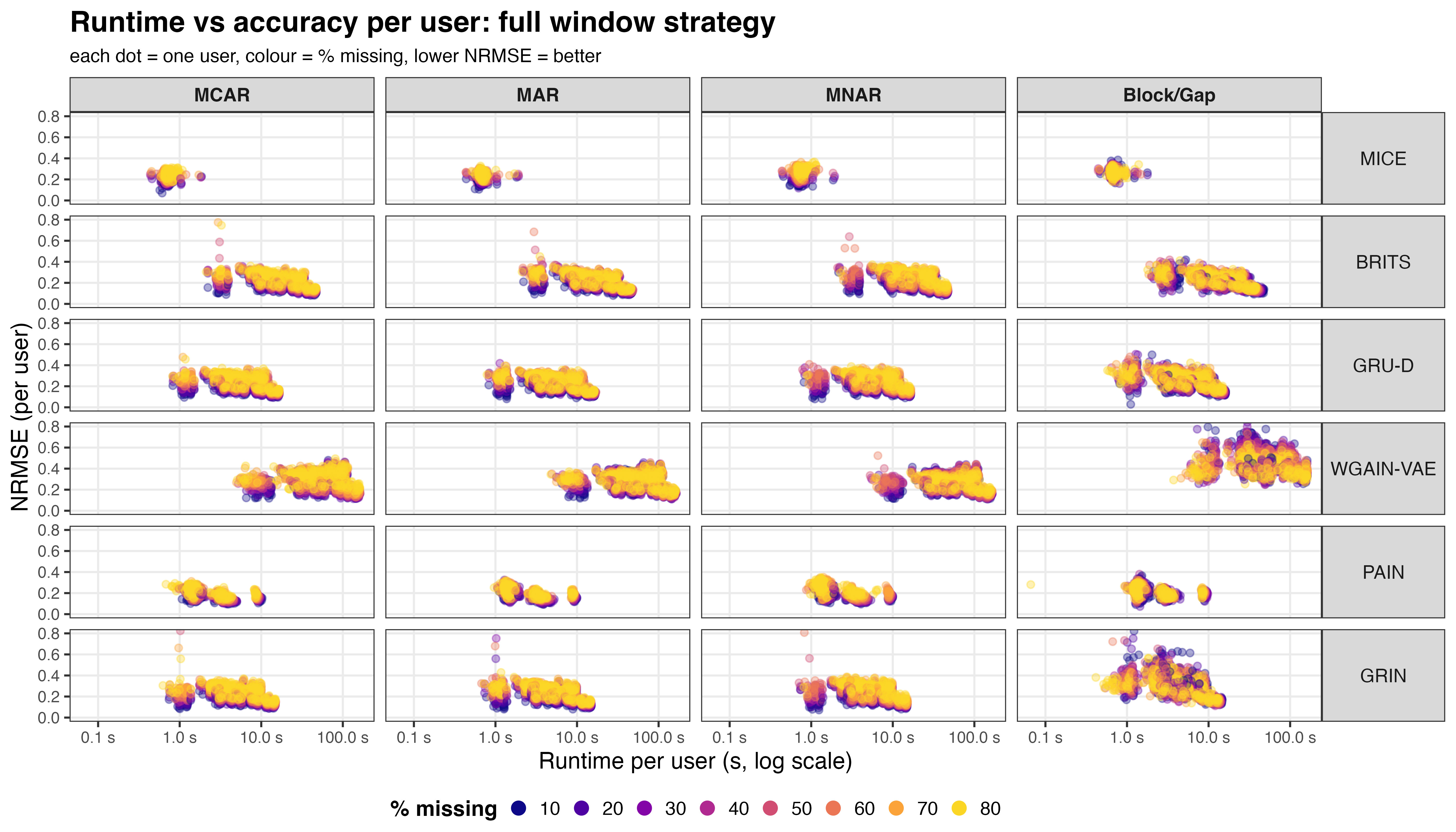} 
    \caption{Runtime vs. accuracy per user, full window strategy, by missingness mechanism and method. Each point is one user; colour indicates \% missing.}
    \label{fig:runtimefull}
\end{figure}

\begin{figure}[htbp]
    \centering
    \includegraphics[width=1\textwidth]{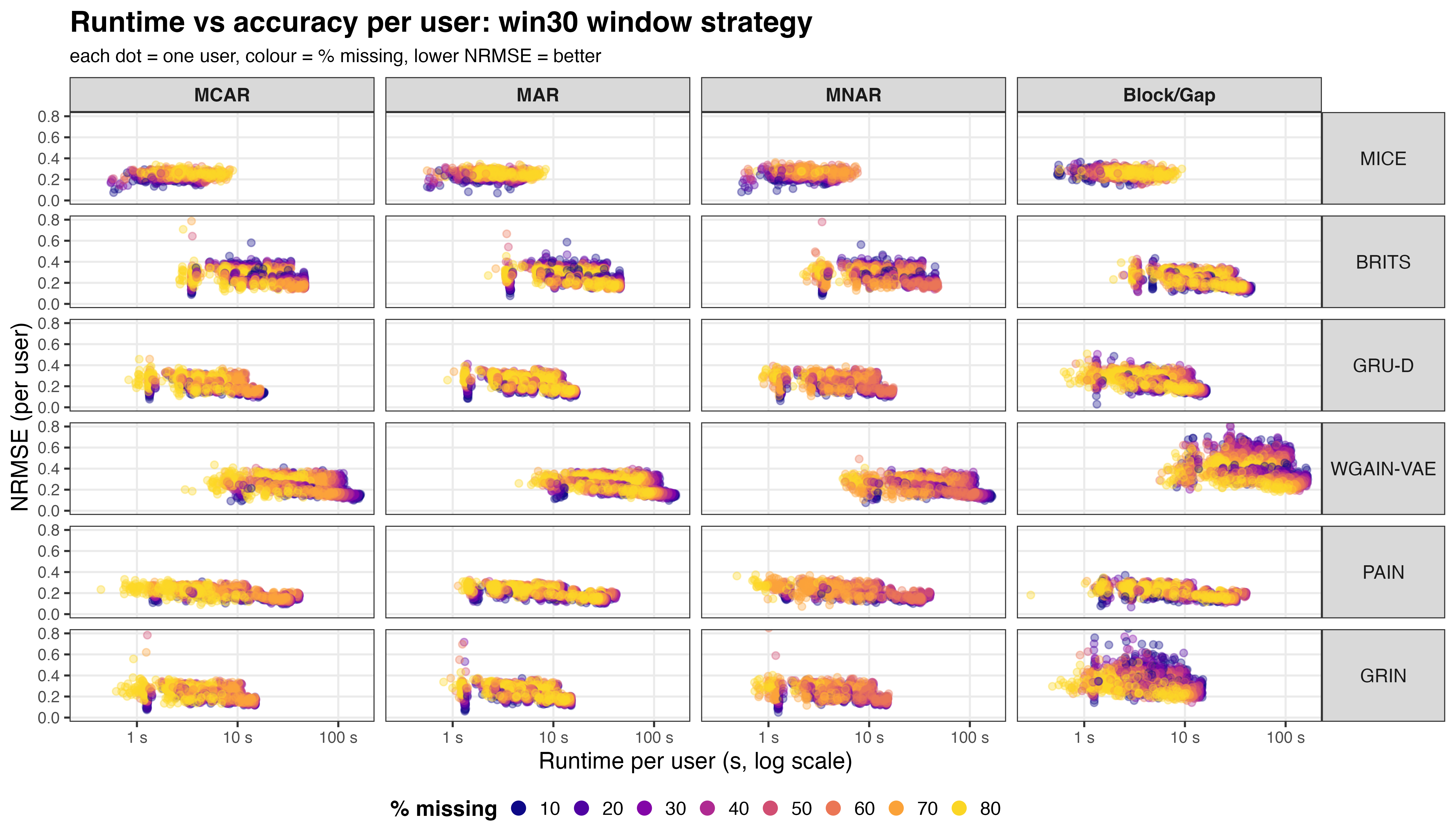} 
    \caption{Runtime vs. accuracy per user, win30 window strategy, by missingness mechanism and method. Each point is one user; colour indicates \% missing.}
    \label{fig:runtimewin30}
\end{figure}

\FloatBarrier

\section{Preserving Multivariate Structure}

\subsection{Supplementary cpCorr Results}
\label{app:cpcorr}

\begin{figure}[htbp]
    \centering
    \includegraphics[width=1\textwidth]{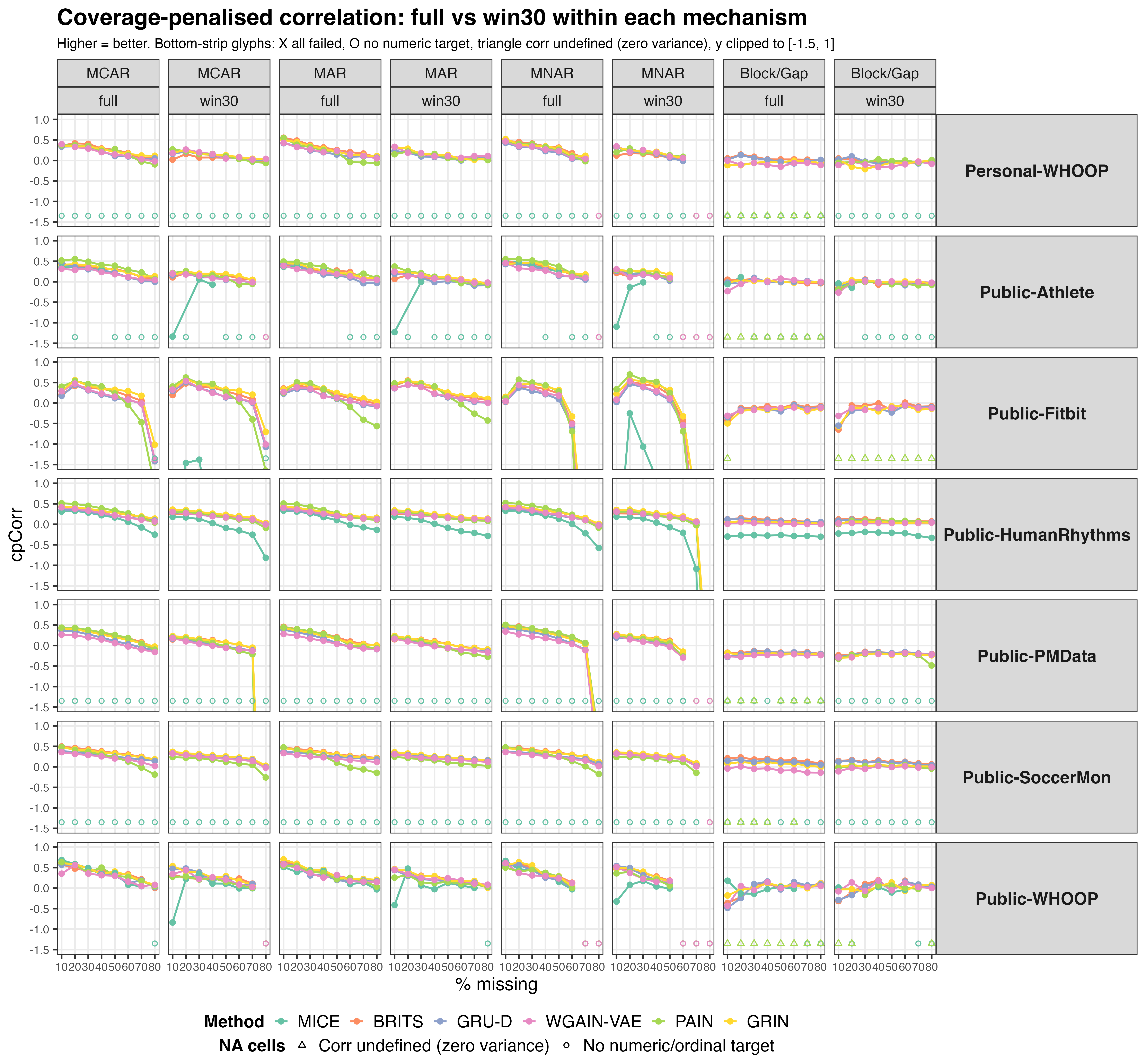} 
    \caption{Coverage-penalised correlation (cpCorr) by individual dataset, missingness mechanism, and window strategy. Bottom-strip glyphs: $\times$ all failed; $\circ$ no numeric/ordinal target; $\triangle$ correlation undefined (zero variance). Y-axis clipped to $[-1.5, 1]$.}
    \label{fig:cpcorrdatasets}
\end{figure}

Figure~\ref{fig:cpcorrdatasets} breaks down coverage-penalised correlation (cpCorr) by individual dataset, missingness mechanism, and window strategy, complementing the pooled view in the main text and making visible the dataset-level divergence underlying PAIN's crossover behaviour at high missingness.

\FloatBarrier

\subsection{Supplementary cpFrob and cpPartialFrob Results}
\label{app:cpfrob}

\begin{figure}[htbp]
    \centering
    \includegraphics[width=1\textwidth]{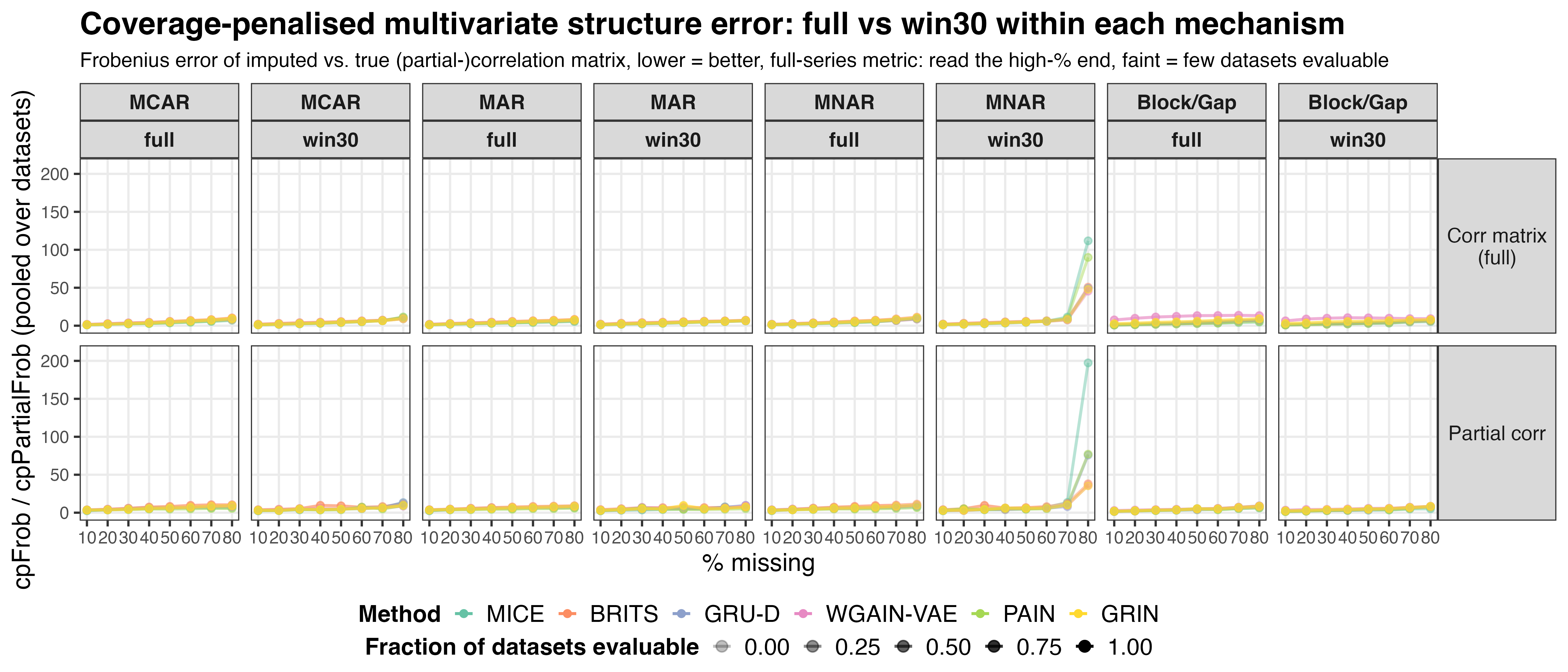} 
    \caption{Coverage-penalised Frobenius error of the imputed vs. true correlation matrix (top row) and partial correlation matrix (bottom row), pooled across datasets, by missingness mechanism and window strategy. Line opacity indicates the fraction of datasets evaluable at each missingness percentage.}
    \label{fig:cpfrobpooled}
\end{figure}

\begin{figure}[htbp]
    \centering
    \includegraphics[width=1\textwidth]{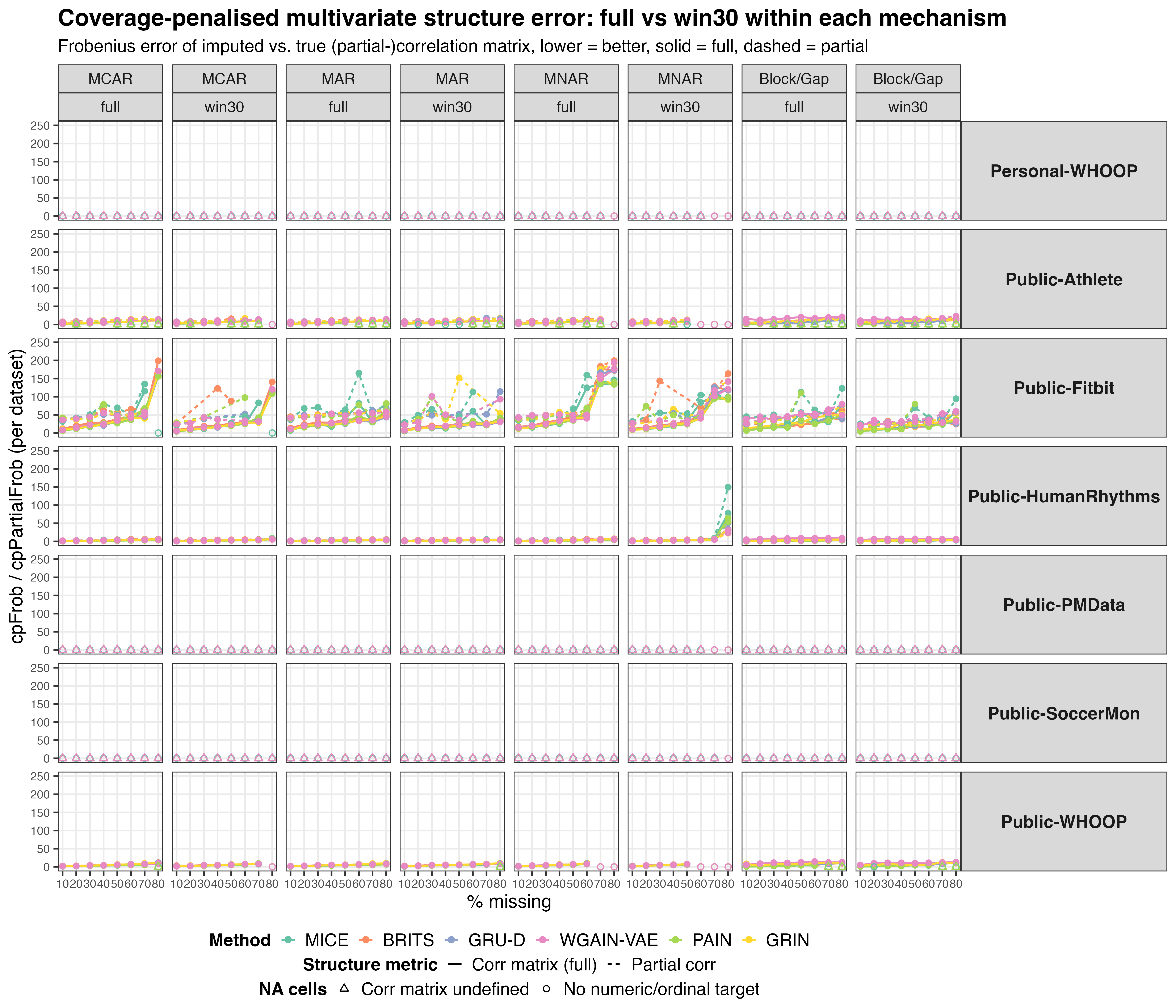} 
    \caption{Coverage-penalised Frobenius error by individual dataset, missingness mechanism, and window strategy. Solid lines: full correlation matrix; dashed lines: partial correlation matrix. Bottom-strip glyphs: $\triangle$ correlation matrix undefined (zero-variance variable); $\circ$ no numeric/ordinal target available.}
    \label{fig:cpfrobdatasets}
\end{figure}

The pooled plot (Figure~\ref{fig:cpfrobpooled}) shows coverage-penalised Frobenius error for both the full correlation matrix and the partial correlation matrix, averaged across datasets. The per-dataset breakdown (Figure~\ref{fig:cpfrobdatasets}) makes explicit which datasets are structurally undefined for this metric (marked by the bottom-strip glyphs) versus which contribute a genuine, evaluable signal. 

\FloatBarrier

\section{Supplementary Preserving Temporal Structure Results}
\label{app:temporal}

\begin{figure}[htbp]
    \centering
    \includegraphics[width=1\textwidth]{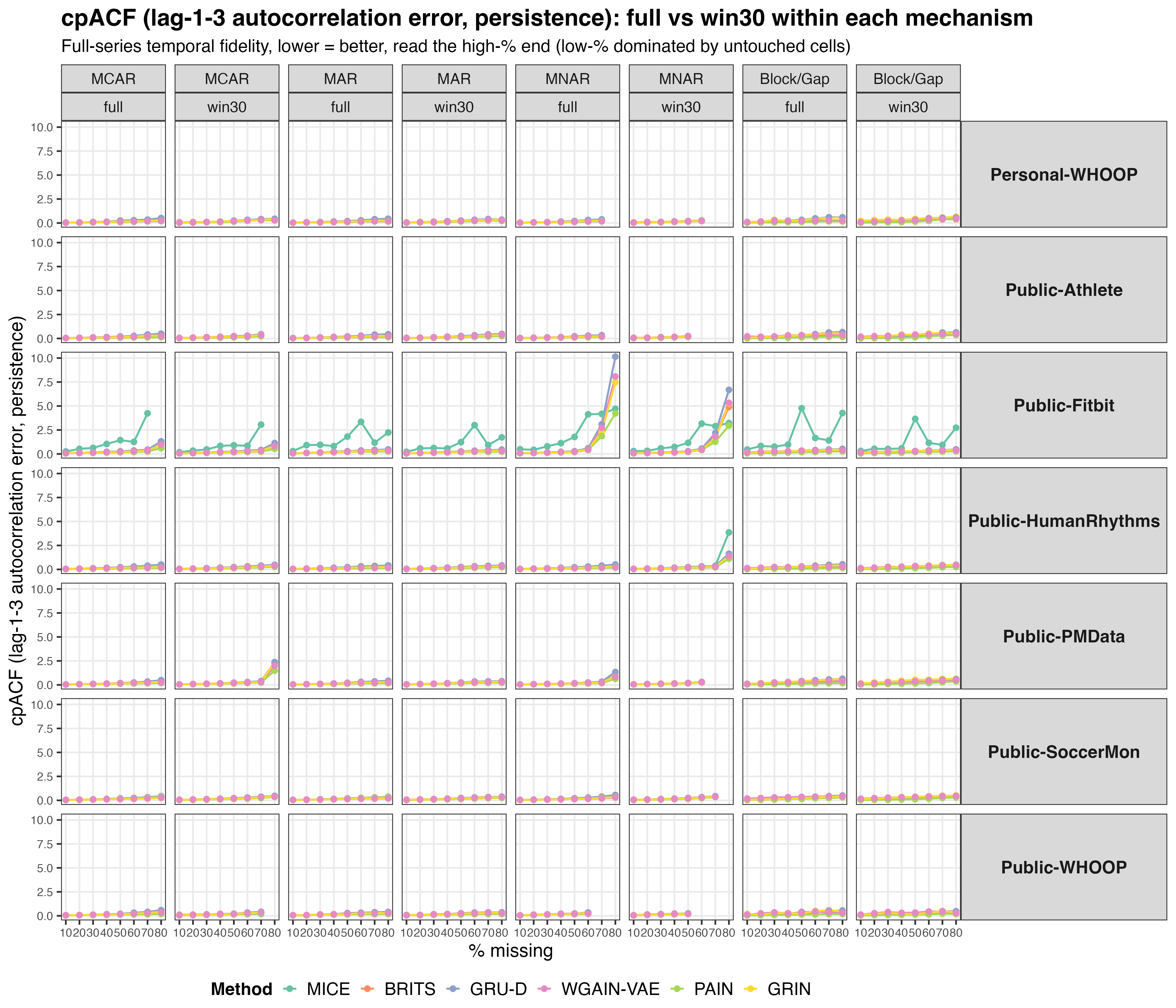} 
    \caption{cpACF (lag-1-3 autocorrelation error, persistence) by individual dataset, missingness mechanism, and window strategy.}
    \label{fig:cpacfdatasets}
\end{figure}

\begin{figure}[htbp]
    \centering
    \includegraphics[width=1\textwidth]{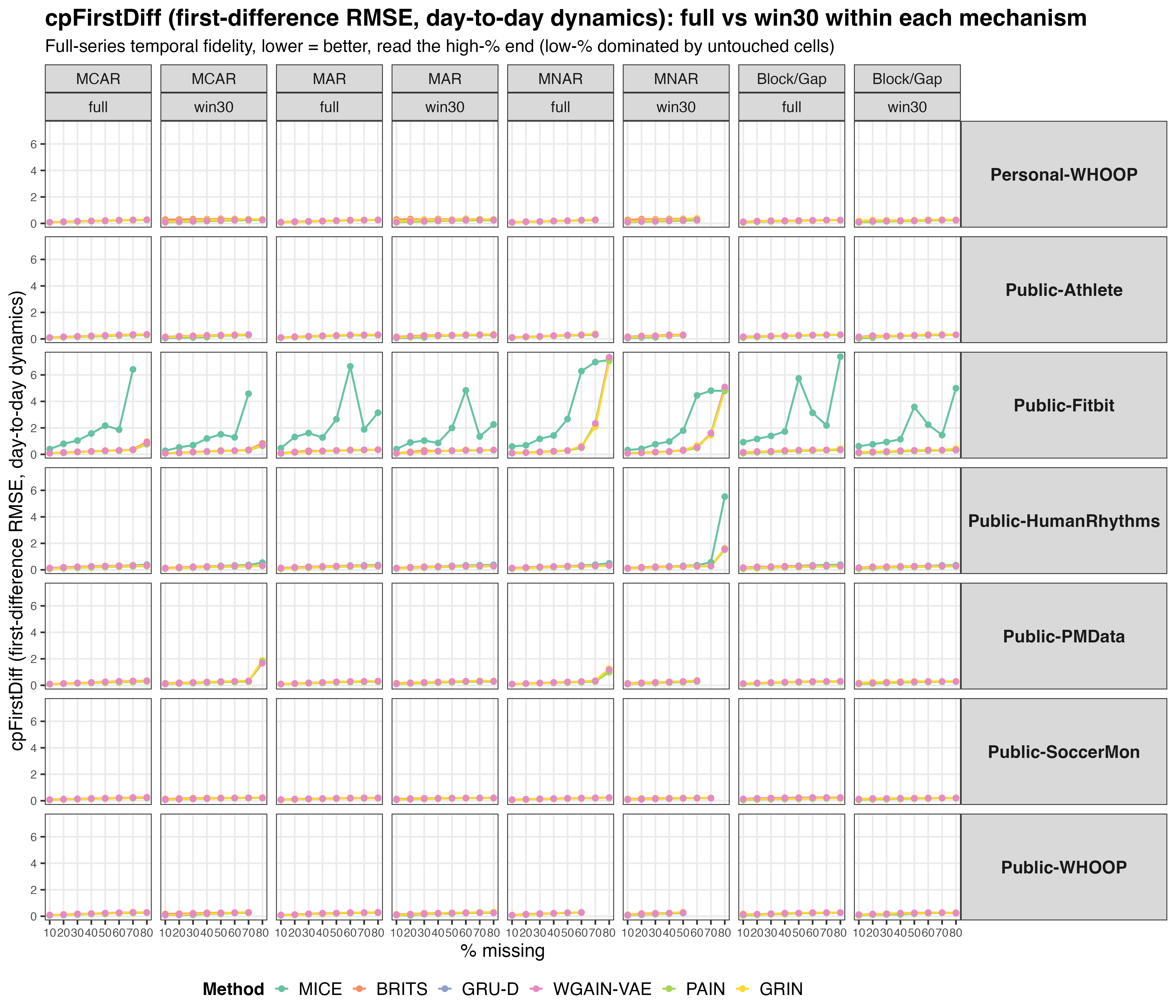} 
    \caption{cpFirstDiff (first-difference RMSE, day-to-day dynamics) by individual dataset, missingness mechanism, and window strategy.}
    \label{fig:cpfirstdiffdatasets}
\end{figure}

\begin{figure}[htbp]
    \centering
    \includegraphics[width=1\textwidth]{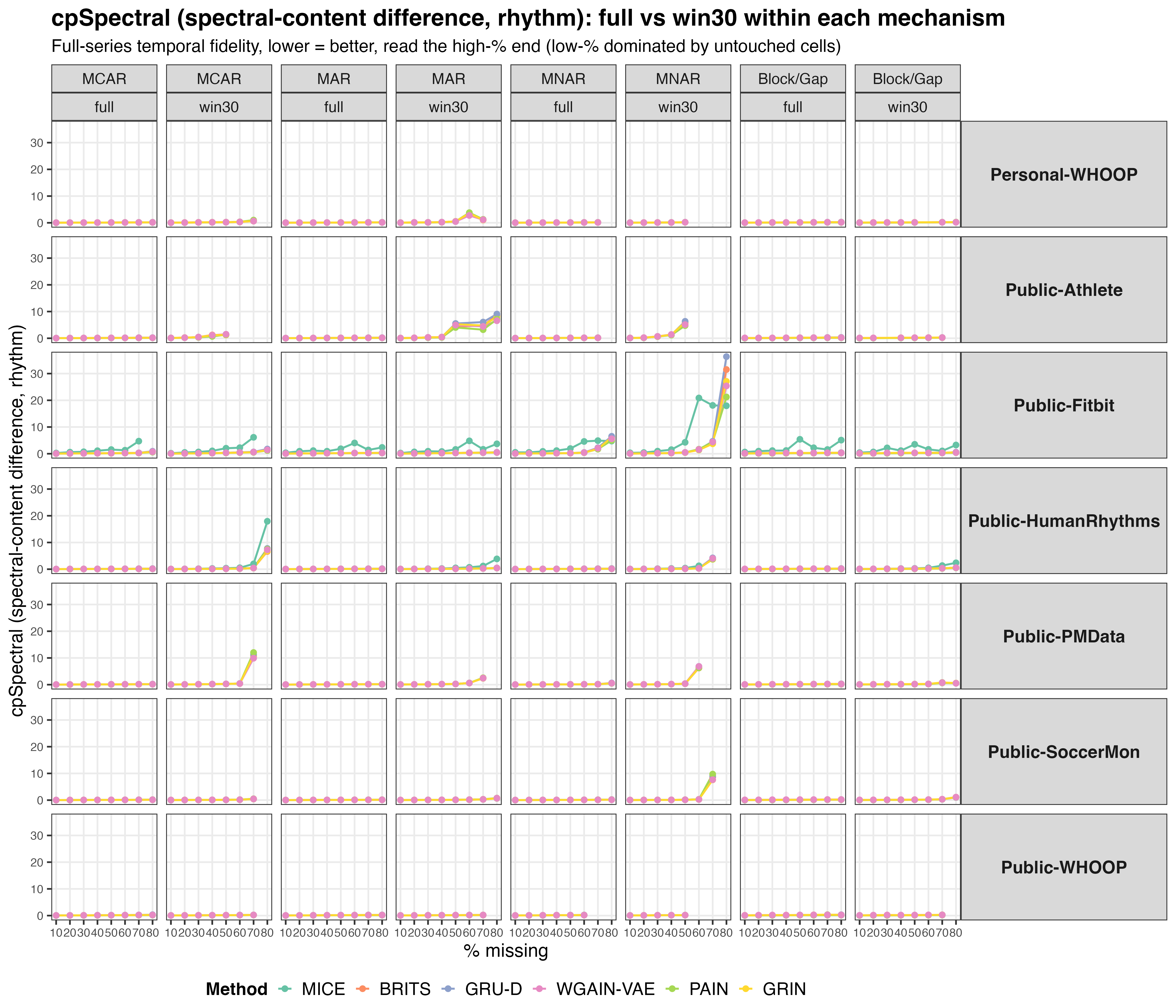} 
    \caption{cpSpectral (spectral-content difference, rhythm) by individual dataset, missingness mechanism, and window strategy.}
    \label{fig:cpspectraldatasets}
\end{figure}

The three figures below (Figure~\ref{fig:cpacfdatasets}, Figure~\ref{fig:cpfirstdiffdatasets}, and Figure~\ref{fig:cpspectraldatasets}) present the per-dataset breakdown of the three full-series temporal metrics introduced in Appendix~\ref{app:temporal_eq} (autocorrelation error, first-difference RMSE, and spectral difference), complementing the pooled view in the main text and isolating which datasets drive the MICE-outlier and MNAR/win30 breakdown patterns discussed there.

\FloatBarrier

\section{Supplementary Distributional Shape Results}
\label{app:distributional}

\begin{figure}[htbp]
    \centering
    \includegraphics[width=1\textwidth]{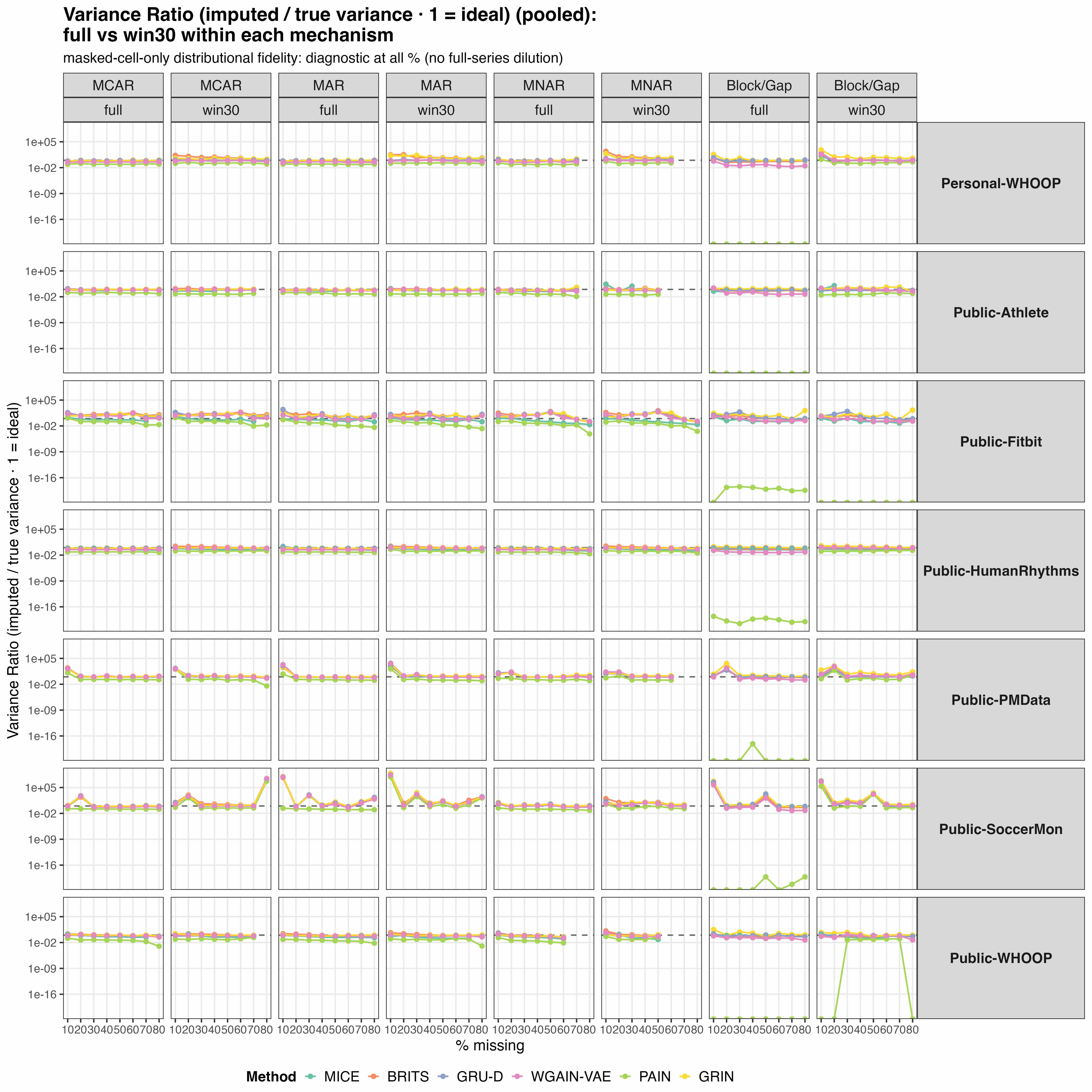}
    \caption{Variance ratio by individual dataset, missingness mechanism, and window strategy. 1 (dashed line) = ideal.}
    \label{fig:varratiodatasets}
\end{figure}

\begin{figure}[htbp]
    \centering
    \includegraphics[width=1\textwidth]{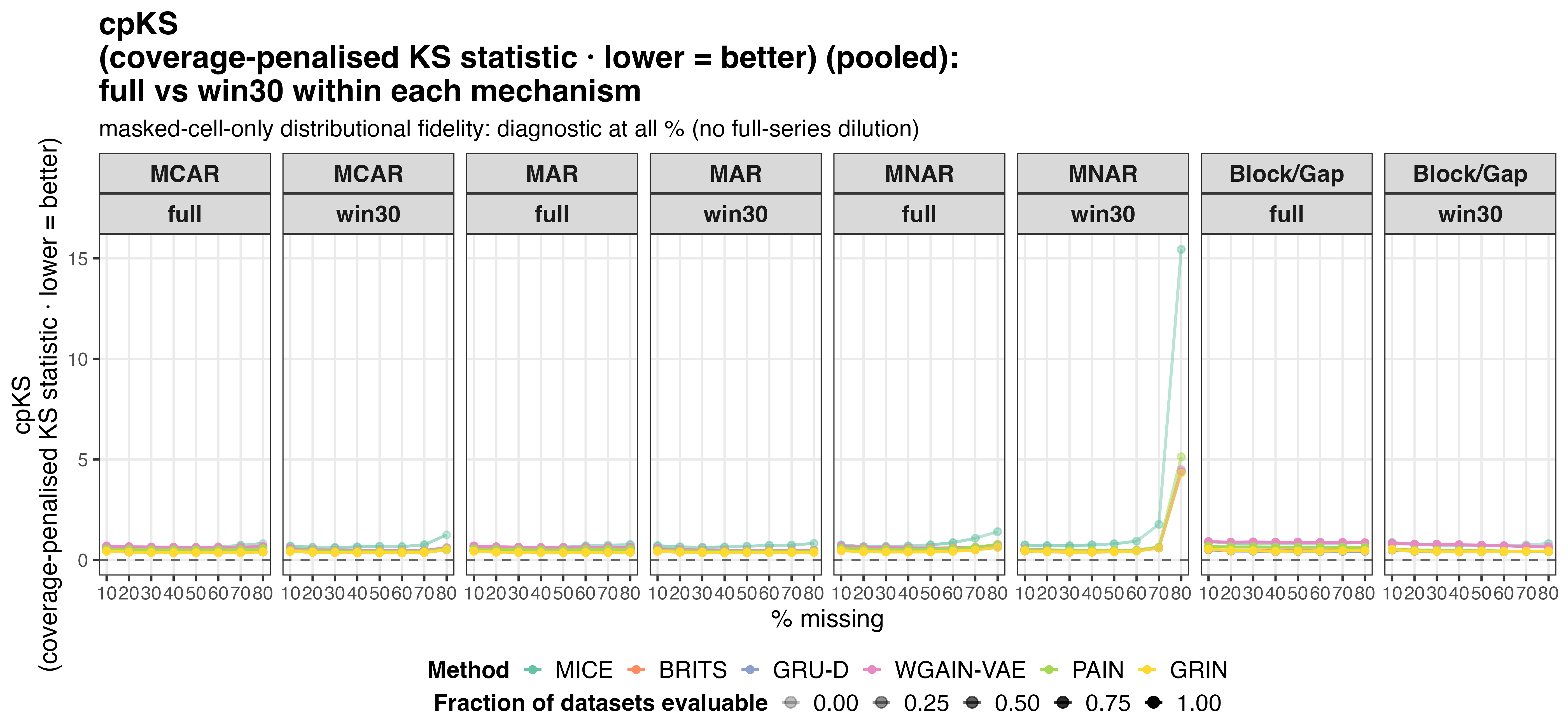} 
    \caption{Coverage-penalised KS statistic, pooled across datasets, by missingness mechanism and window strategy. Lower = better.}
    \label{fig:cpKSpooled}
\end{figure}

\begin{figure}[htbp]
    \centering
    \includegraphics[width=1\textwidth]{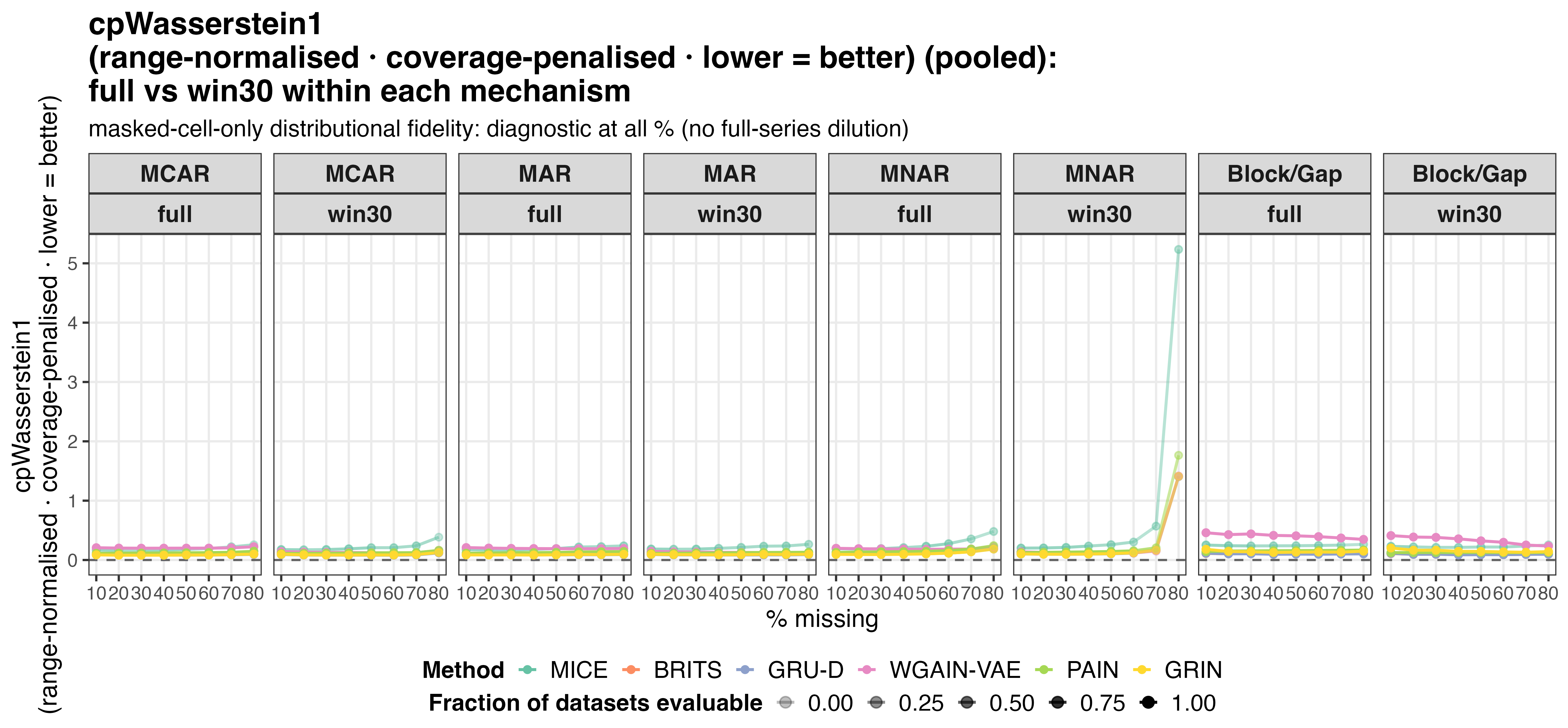} 
    \caption{Coverage-penalised, range-normalised 1-Wasserstein distance pooled over datasets.}
    \label{fig:cpW1pooled}
\end{figure}

\begin{figure}[htbp]
    \centering
    \includegraphics[width=1\textwidth]{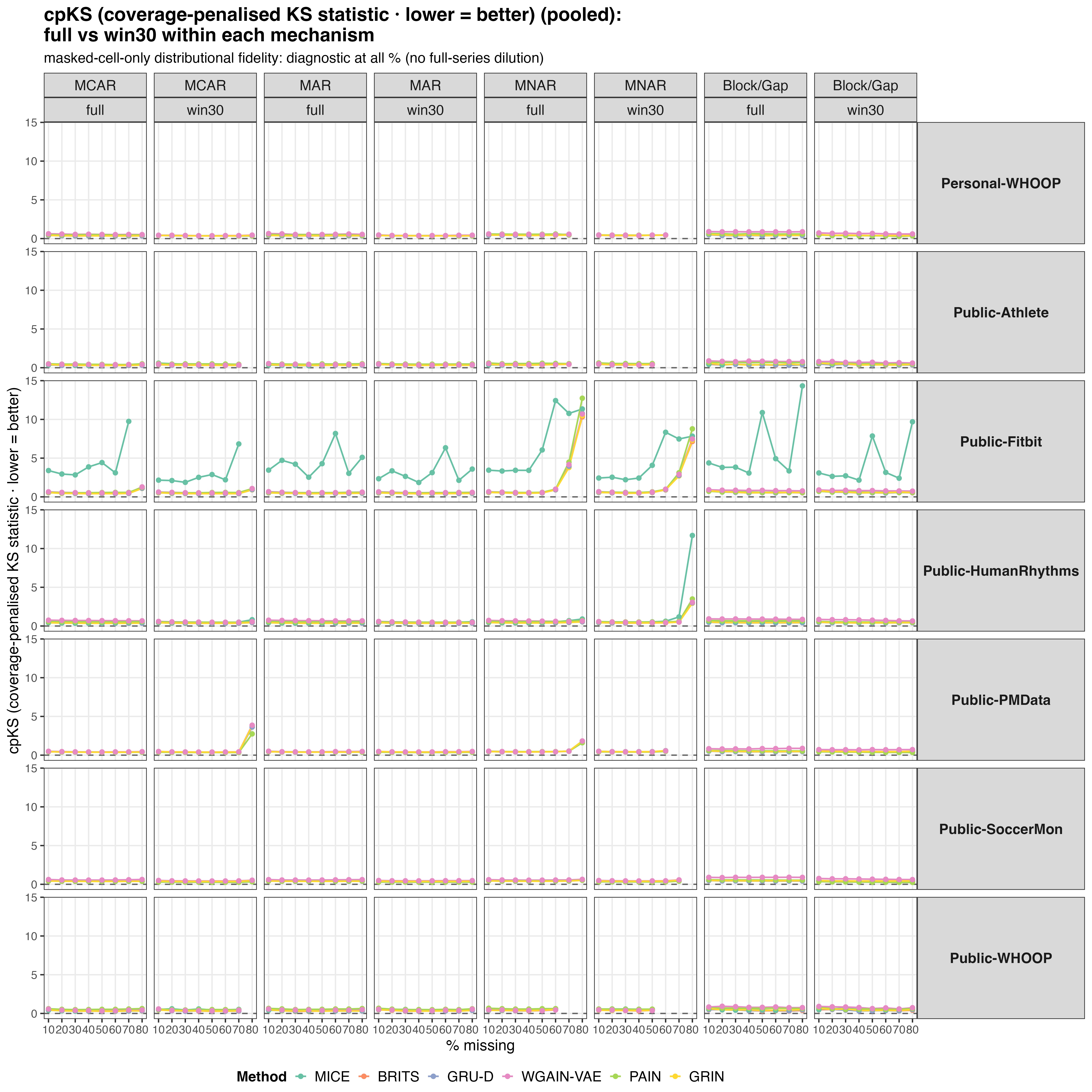} 
    \caption{Coverage-penalised KS statistic by individual dataset, missingness mechanism, and window strategy.}
    \label{fig:cpKSdatasets}
\end{figure}

\begin{figure}[htbp]
    \centering
    \includegraphics[width=1\textwidth]{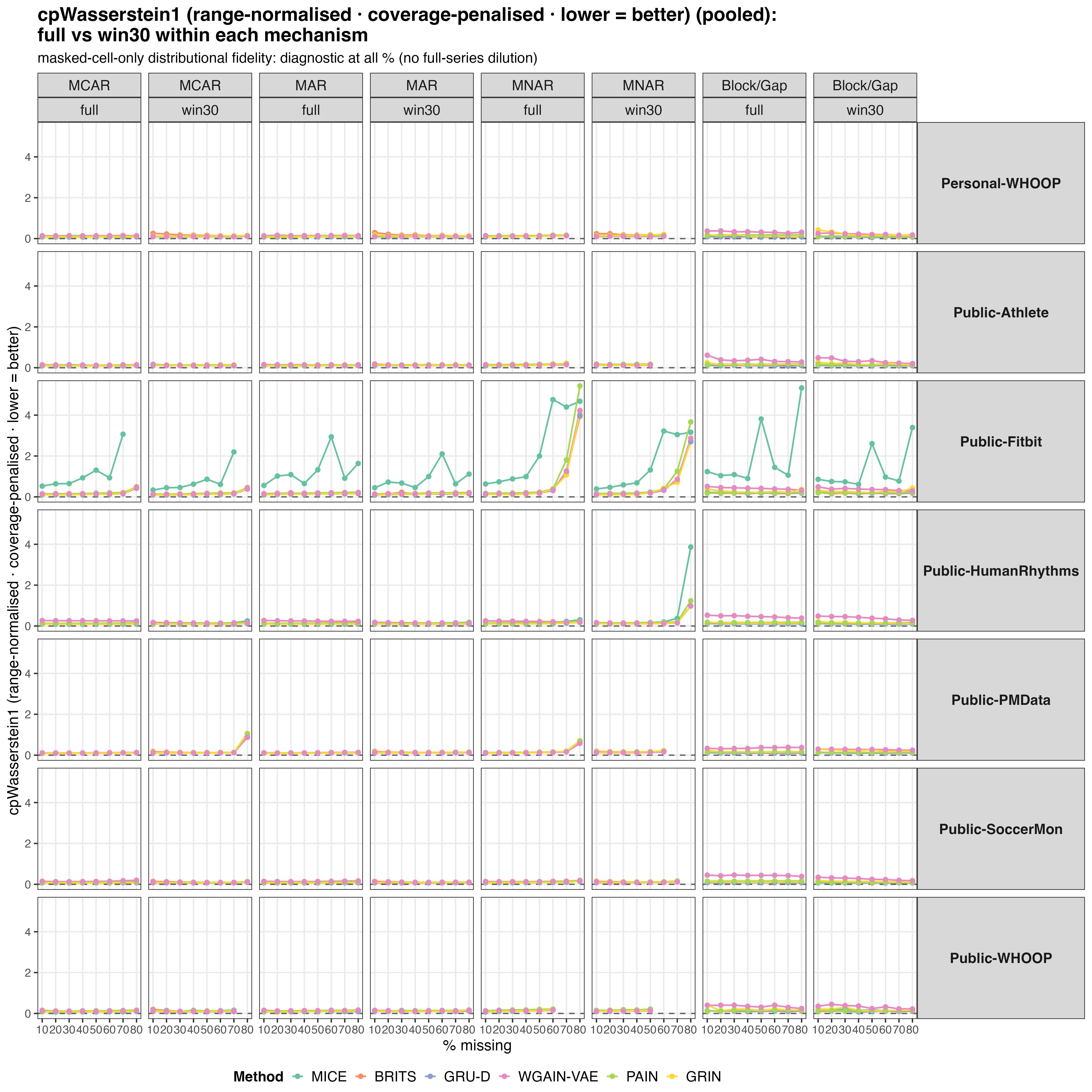} 
    \caption{Coverage-penalised, range-normalised 1-Wasserstein distance by individual dataset, missingness mechanism, and window strategy.}
    \label{fig:cpW1datasets}
\end{figure}

Figure~\ref{fig:varratiodatasets} shows the per-dataset breakdown of variance ratio, confirming that PAIN's variance collapse under Block/Gap missingness is consistent across multiple datasets rather than isolated to one. The pooled and per-dataset cpKS results below (Figure~\ref{fig:cpKSpooled} and Figure~\ref{fig:cpKSdatasets}) show MICE as the dominant outlier across mechanisms, with the DL methods sitting close to the ideal line except under the shared MNAR/win30 breakdown at high missingness. The corresponding cpWasserstein1 results (Figure~\ref{fig:cpW1pooled} and Figure~\ref{fig:cpW1datasets}) tell the same story using the full displacement between distributions rather than the single worst-point CDF gap.

\FloatBarrier

\section{Supplementary Trade-off Results: Accuracy vs. Multivariate Structure and Variance Preservation}
\label{app:tradeoffs}

\begin{figure}[htbp]
    \centering
    \includegraphics[width=1\textwidth]{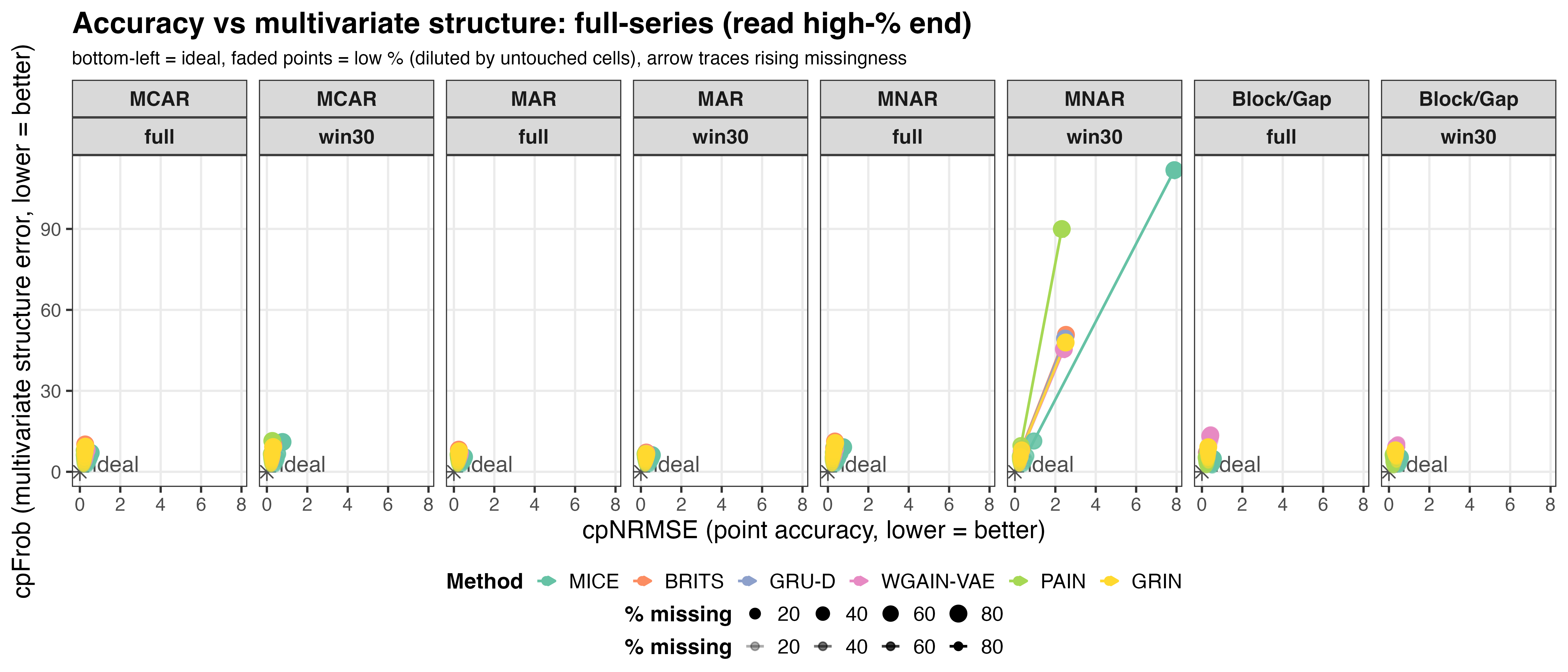} 
    \caption{Accuracy vs. multivariate structure (full-series, pooled), by missingness mechanism and window strategy. Bottom-left = ideal; faded points indicate low \% missing, where the full-series cpFrob metric is diluted by untouched observed cells; arrows trace rising missingness.}
    \label{fig:acc_vs_multivariate}
\end{figure}

\begin{figure}[htbp]
    \centering
    \includegraphics[width=1\textwidth]{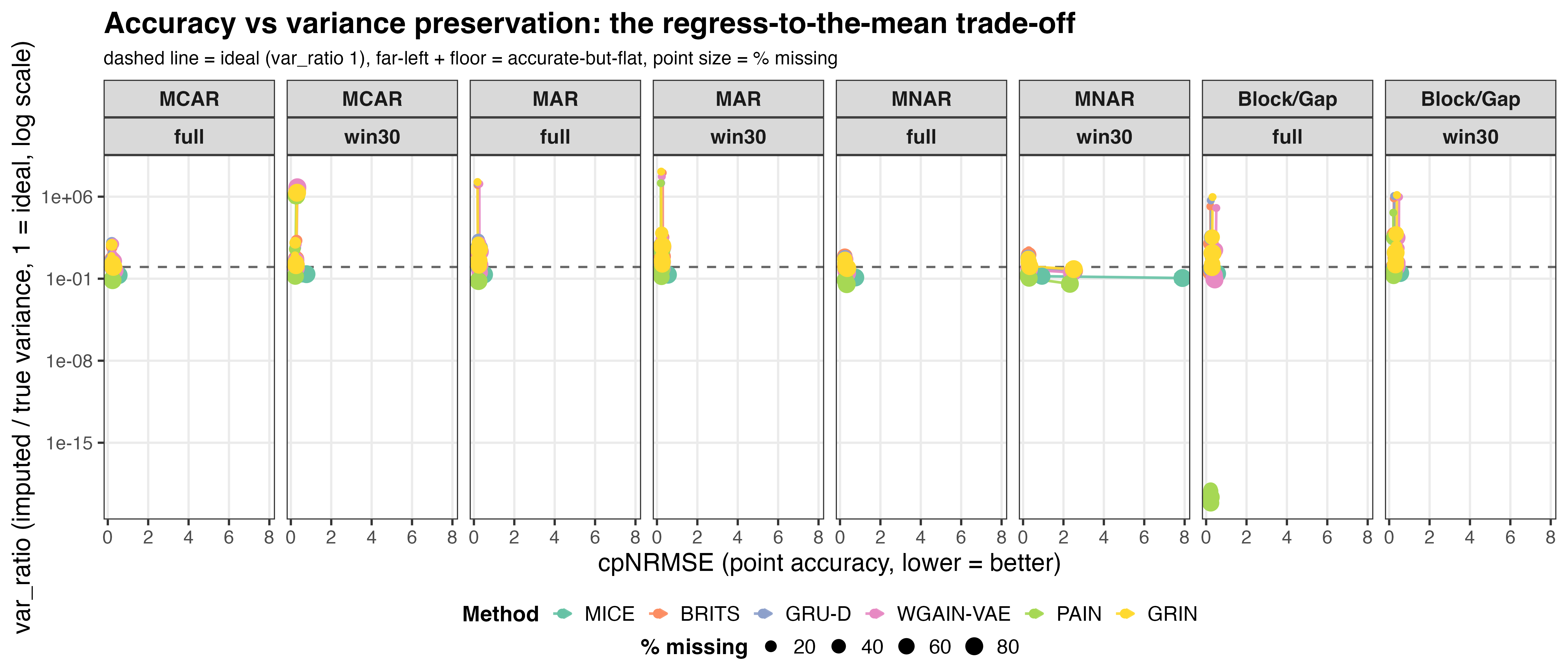} 
    \caption{Accuracy vs. variance preservation (the regress-to-the-mean trade-off), pooled across datasets, by missingness mechanism and window strategy. Dashed line = ideal (variance ratio = 1, log scale); far-left and floor = accurate-but-flat; point size indicates \% missing.}
    \label{fig:acc_vs_variance}
\end{figure}

\begin{figure}[htbp]
    \centering
    \includegraphics[width=1\textwidth]{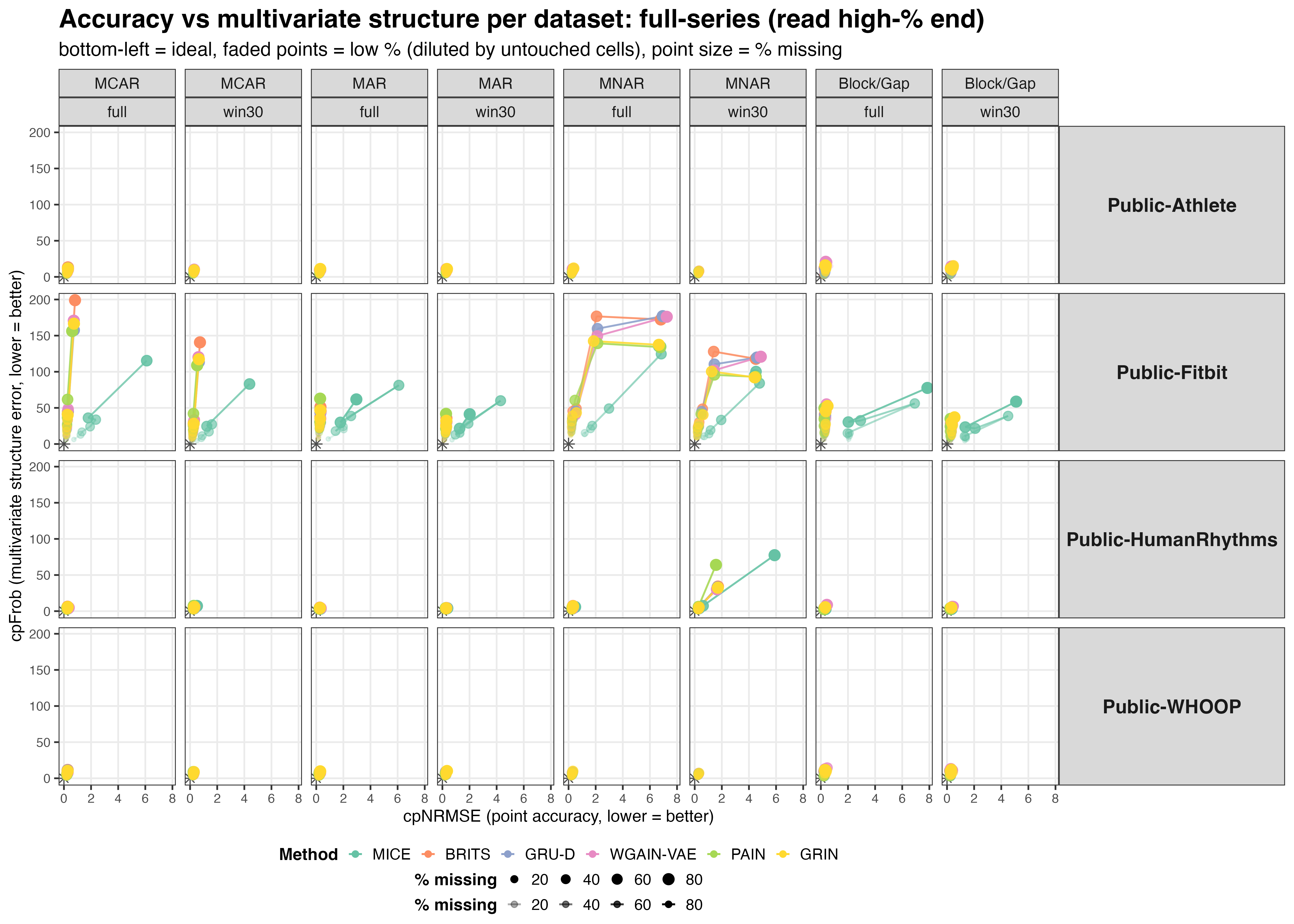} 
    \caption{Accuracy vs. multivariate structure (full-series), by individual dataset, missingness mechanism, and window strategy. Bottom-left = ideal; faded points indicate low \% missing; point size indicates \% missing.}
    \label{fig:acc_vs_multivariate_datasets}
\end{figure}

\begin{figure}[htbp]
    \centering
    \includegraphics[width=1\textwidth]{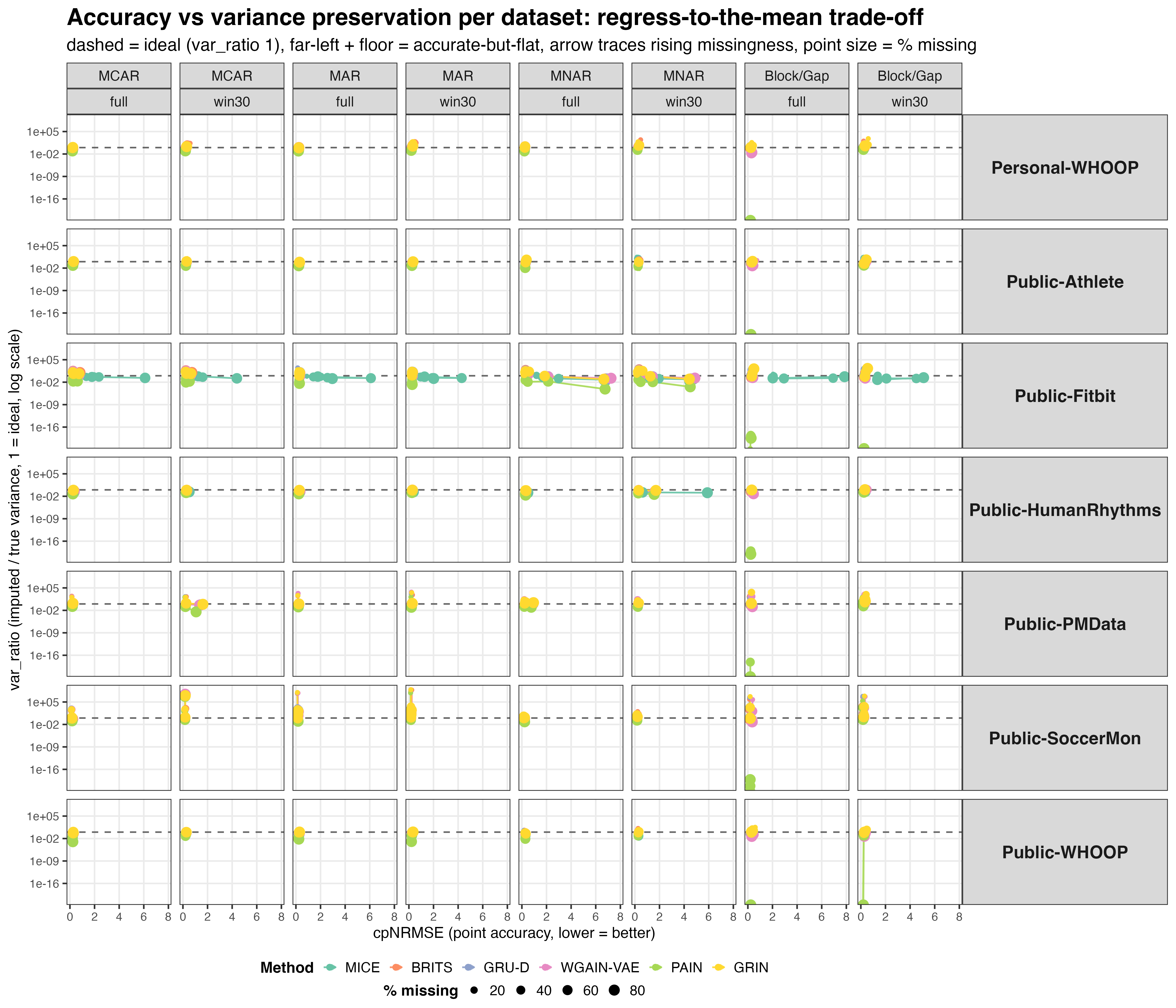} 
    \caption{Accuracy vs. variance preservation, by individual dataset, missingness mechanism, and window strategy. Dashed line = ideal (variance ratio = 1, log scale); arrows trace rising missingness; point size indicates \% missing.}
    \label{fig:acc_vs_variance_datasets}
\end{figure}

The pooled trade-off plots (Figure~\ref{fig:acc_vs_multivariate} and Figure~\ref{fig:acc_vs_variance}) below jointly visualise point accuracy against multivariate structure preservation and against variance preservation, substantiating the main text's claim that no single method dominates on both axes simultaneously. The per-dataset versions of both trade-offs (Figure~\ref{fig:acc_vs_multivariate_datasets} and Figure~\ref{fig:acc_vs_variance_datasets}) show that this pattern, in particular PAIN's Block/Gap collapse, holds consistently at the individual-dataset level rather than being an artefact of pooling.

\FloatBarrier

\subsection{Supplementary Forest Plots: Per-User cpNRMSE and cpCorr (win30)}
\label{app:forest}

\begin{figure}[p]
    \centering
    \includegraphics[width=0.75\textwidth]{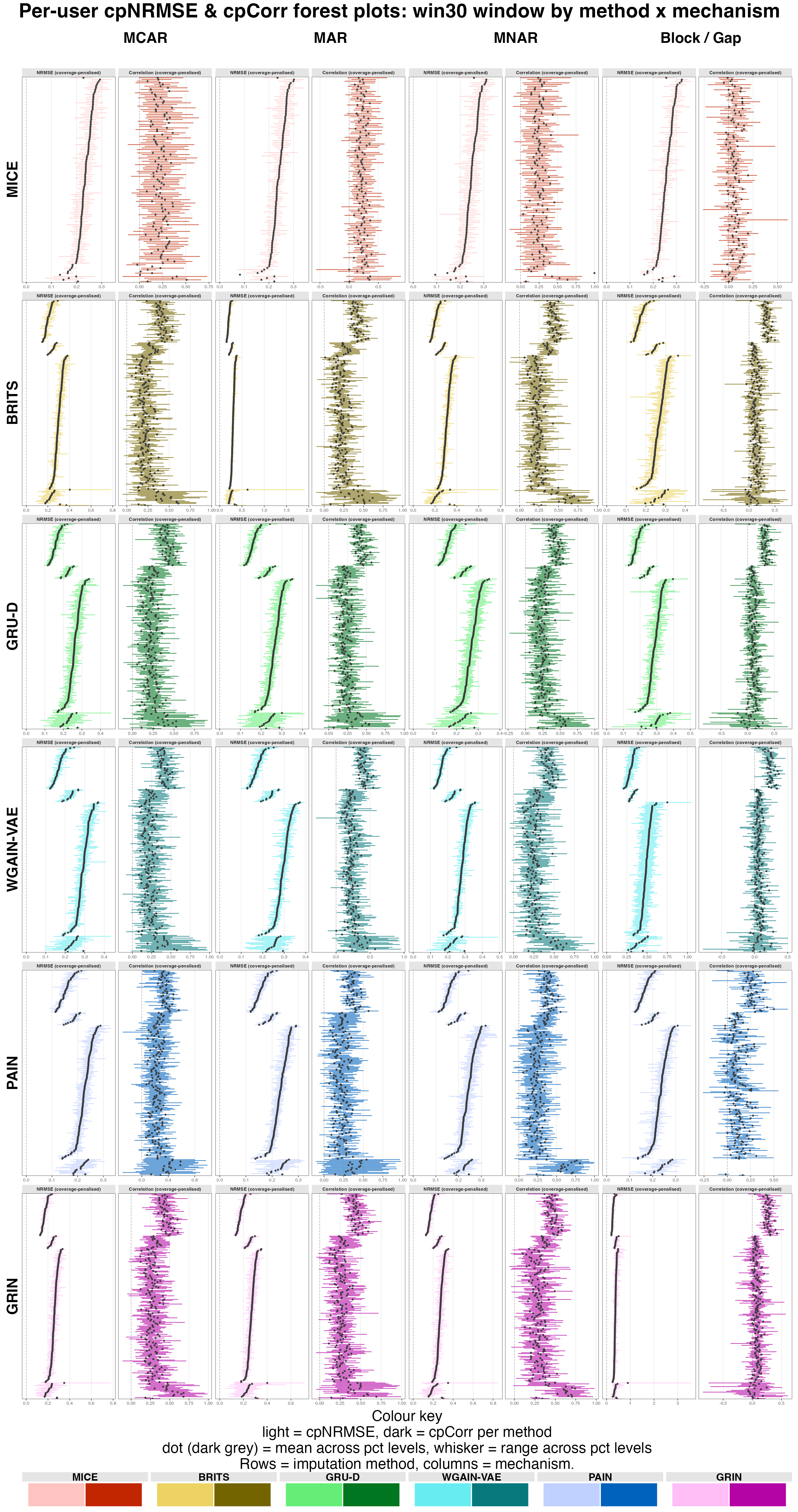} 
    \caption{Per-user cpNRMSE and cpCorr forest plots, win30 window strategy, by method (rows) and missingness mechanism (columns). Light = cpNRMSE, dark = cpCorr. Each line is one user, ordered by mean cpNRMSE within dataset; point = mean across 8 missingness percentages; whisker = range (min-max) across those percentages.}
    \label{fig:forest_win30}
\end{figure}

The forest plot Figure~\ref{fig:forest_win30} presents the win30 counterpart to the full-window forest plots in the main text, showing per-user cpNRMSE and cpCorr by method and missingness mechanism under the 30-day windowed imputation strategy.

\section{Generative AI Use Disclosure}

Generative AI Anthropic Claude Sonnet 4.5 was used to assist in testing and debugging the analysis code described in this paper, including identifying syntax errors, flagging logical inconsistencies, and suggesting diagnostic checks during development. All code logic, statistical methods, experimental design, analysis, interpretation, and written content are the sole work of the authors, who take full responsibility for the accuracy, validity, and integrity of the content presented in this paper.

\section{Analysis Code}

Analysis code and first author's personal dataset used in this study can be provided via contacting first author.

\end{document}